\documentclass[10pt, letterpaper, onecolumn, peerreview]{IEEEtran}
\usepackage{epsfig}
\usepackage{amssymb}
\usepackage[tbtags]{amsmath}
\usepackage{graphics,eepic,epic}
\usepackage{latexsym}
\usepackage{euscript}
\usepackage{graphics,eepic,epic,psfrag}

\usepackage[tbtags]{amsmath} 
\usepackage{amssymb}  
\usepackage{verbatim} 

\DeclareMathOperator*{\argmax}{arg\,max}
\DeclareMathOperator*{\argmin}{arg\,min}

\newcounter{actr}
{\begin{list}{(\alph{actr})}{\usecounter{actr}}}{\end{list}}

\newcounter{ictr}
{\begin{list}{(\roman{ictr})}{\usecounter{ictr}}}{\end{list}}

\newtheorem{thm}{Theorem}
\newtheorem{lemma}{Lemma}

\newtheorem{corol}{Corollary}

\newtheorem{defn}{Definition}

\newcommand{\qed}{\rule[0.1ex]{1.4ex}{1.6ex}}

\newcommand{\defeq}{\stackrel{\Delta}{=}}

\newcommand{\ceil}[1]{\lceil{#1}\rceil}

\newcommand{\cB}{{\mathcal{B}}}

\newcommand{\cX}{{\mathcal{X}}}

\newcommand{\cY}{{\mathcal{Y}}}

\newcounter{psctr}
\newcounter{probctr}[psctr]

\DeclareMathAlphabet{\mathbsf}{OT1}{cmss}{bx}{n}
\DeclareMathAlphabet{\mathssf}{OT1}{cmss}{m}{sl}

\DeclareSymbolFont{bsfletters}{OT1}{cmss}{bx}{n}  
\DeclareSymbolFont{ssfletters}{OT1}{cmss}{m}{n}
\DeclareMathSymbol{\bsfGamma}{0}{bsfletters}{'000}
\DeclareMathSymbol{\ssfGamma}{0}{ssfletters}{'000}
\DeclareMathSymbol{\bsfDelta}{0}{bsfletters}{'001}
\DeclareMathSymbol{\ssfDelta}{0}{ssfletters}{'001}
\DeclareMathSymbol{\bsfTheta}{0}{bsfletters}{'002}
\DeclareMathSymbol{\ssfTheta}{0}{ssfletters}{'002}
\DeclareMathSymbol{\bsfLambda}{0}{bsfletters}{'003}
\DeclareMathSymbol{\ssfLambda}{0}{ssfletters}{'003}
\DeclareMathSymbol{\bsfXi}{0}{bsfletters}{'004}
\DeclareMathSymbol{\ssfXi}{0}{ssfletters}{'004}
\DeclareMathSymbol{\bsfPi}{0}{bsfletters}{'005}
\DeclareMathSymbol{\ssfPi}{0}{ssfletters}{'005}
\DeclareMathSymbol{\bsfSigma}{0}{bsfletters}{'006}
\DeclareMathSymbol{\ssfSigma}{0}{ssfletters}{'006}
\DeclareMathSymbol{\bsfUpsilon}{0}{bsfletters}{'007}
\DeclareMathSymbol{\ssfUpsilon}{0}{ssfletters}{'007}
\DeclareMathSymbol{\bsfPhi}{0}{bsfletters}{'010}
\DeclareMathSymbol{\ssfPhi}{0}{ssfletters}{'010}
\DeclareMathSymbol{\bsfPsi}{0}{bsfletters}{'011}
\DeclareMathSymbol{\ssfPsi}{0}{ssfletters}{'011}
\DeclareMathSymbol{\bsfOmega}{0}{bsfletters}{'012}
\DeclareMathSymbol{\ssfOmega}{0}{ssfletters}{'012}

\renewcommand{\defeq}{\triangleq}

\newcommand{\svw}{w}

\newcommand{\svbw}{{\mathbf{w}}}
\newcommand{\rvx}{{\mathssf{x}}}	

\newcommand{\svx}{x}			

\newcommand{\rvbx}{{\mathbsf{x}}}

\newcommand{\svbx}{{\mathbf{\svx}}}

\newcommand{\rvy}{{\mathssf{y}}}	

\newcommand{\svy}{y}

\newcommand{\rvby}{{\mathbsf{y}}}

\newcommand{\svby}{{\mathbf{y}}}

\usepackage{fullpage}

\newcommand{\rvxtil}{\tilde{\rvx}}
\newcommand{\svxtil}{\tilde{\svx}}
\newcommand{\rvxhat}{\hat{\rvx}}
\newcommand{\svxhat}{\hat{\svx}}
\newcommand{\rvxBar}{\bar{\rvx}}
\newcommand{\svxBar}{\bar{\svx}}
\newcommand{\rvxBBar}{\bar{\bar{\rvx}}}
\newcommand{\svxBBar}{\bar{\bar{\svx}}}

\newcommand{\svbxtil}{\tilde{\svbx}}

\newcommand{\svbxBar}{\bar{\svbx}}

\newcommand{\rvytil}{\tilde{\rvy}}
\newcommand{\svytil}{\tilde{\svy}}
\newcommand{\rvyhat}{\hat{\rvy}}
\newcommand{\svyhat}{\hat{\svy}}
\newcommand{\rvyBar}{\bar{\rvy}}
\newcommand{\svyBar}{\bar{\svy}}
\newcommand{\rvyBBar}{\bar{\bar{\rvy}}}
\newcommand{\svyBBar}{\bar{\bar{\svy}}}

\newcommand{\svbytil}{\tilde{\svby}}

\newcommand{\PxyRV}{p_{\rvx \rvy}}
\newcommand{\PxCondyRV}{p_{\rvx|\rvy}}

\newcommand{\PxRV}{p_{\rvx}}
\newcommand{\PyRV}{p_{\rvy}}

\newcommand{\jointSource}[4]{\; p_{\rvx_{#1}^{#2}, \rvy_{#3}^{#4}}
    (\svx_{#1}^{#2}, \svy_{#3}^{#4})}

\newcommand{\tclass}{\mathcal{T}}

\newcommand{\PKL}{P^{k-l}}
\newcommand{\PNL}{P^{n-l}}
\newcommand{\PNK}{P^{n-k}}

\newcommand{\VKL}{V^{k-l}}
\newcommand{\VNL}{V^{n-l}}
\newcommand{\VNK}{V^{n-k}}

\newcommand{\PtilNL}{\tilde{P}^{n-l}}
\newcommand{\PtilNK}{\tilde{P}^{n-k}}

\newcommand{\VtilNK}{\tilde{V}^{n-k}}
\newcommand{\VtilKL}{\tilde{V}^{k-l}}

\newcommand{\delay}{\Delta}

\newcommand{\expML}{E_{ML}}
\newcommand{\expUniv}{E_{UN}}
\newcommand{\expMLSI}{E_{ML, SI}}
\newcommand{\expUnivSI}{E_{UN, SI}}

\newcommand{\binX}{\cB_x}
\newcommand{\binY}{\cB_y}

\newcommand{\Rx}{R_x}
\newcommand{\Ry}{R_y}
\newcommand{\Rent}{\Rx}

\newcommand{\score}{S}

\newcommand{\minEnt}{\score(\PNK, \PKL, \VNK, \VKL)}
\newcommand{\minEntTil}{\score(\PtilNK, \PKL, \VtilNK, \VtilKL)}

\newcommand{\ind}{\emph{1}}

\newcommand{\kast}{k^{\ast}(l)}
\newcommand{\BL}{N} 
\newcommand{\pf}{{\em Proof: }}

\begin{document}

\title{Lossless coding for distributed streaming sources\footnote{This
    material was presented in part at the IEEE Int Symp Inform Theory,
    Adelaide, Australia, Sept 2005.}}

\author{Cheng Chang\footnote{Department of Electrical Engineering and
    Computer Science, University of California Berkeley, Berkeley, CA
    94720}, Stark C.~Draper\footnote{Mitsubishi Electric Research Labs
    in Cambridge, MA. This work was performed while he was a postdoc
    at
    Wireless Foundations in the University of California Berkeley.}, and Anant Sahai\footnote{Wireless Foundations, Department of Electrical Engineering and Computer Science, University of California Berkeley, Berkeley, CA 94720} \\
  {\small \texttt cchang@eecs.berkeley.edu, sdraper@eecs.berkeley.edu,
    sahai@eecs.berkeley.edu} }

\maketitle

\begin{abstract} Distributed source coding is traditionally viewed in
  the block coding context --- all the source symbols are known in
  advance at the encoders. This paper instead considers a streaming
  setting in which iid source symbol pairs are revealed to the
  separate encoders in real time and need to be reconstructed at the
  decoder with some tolerable end-to-end delay using finite rate
  noiseless channels. A sequential random binning argument is used to
  derive a lower bound on the error exponent with delay and show that
  both ML decoding and universal decoding achieve the same positive
  error exponents inside the traditional Slepian-Wolf rate region. The
  error events are different from the block-coding error events and
  give rise to slightly different exponents. Because the sequential
  random binning scheme is also universal over delays, the resulting
  code eventually reconstructs every source symbol correctly with
  probability $1$.
\end{abstract}

\IEEEpeerreviewmaketitle

\section{Introduction}

Traditionally, ``lossless'' coding is considered using two distinct
paradigms: fixed block coding and variable-length
coding\footnote{There are actually four different traditional cases:
  fixed to fixed, fixed to variable, variable to fixed, and variable
  to variable. However, the last three all achieve a probability of
  error of zero and so we consider them together.}. As classically
understood, both consider that the source-symbols are known in advance
at the encoder and that they must be mapped into a string of bits
decoded by the receiver. Fixed-block coding accepts a small
probability of error and constrains the length of the bit-string,
while variable-length encoding constrains only the {\em expected}
length of the bit-string in exchange for keeping the probability of
error at zero. In the point-to-point setting, both paradigms apply
generically. In contrast, distributed source coding, has traditionally
been explored within the fixed block context. In
\cite{slepianWolf:73}, Slepian and Wolf even asked:
\begin{quotation}
What is the theory of variable-length encodings for
correlated sources?
\end{quotation}

In the classical context of source realizations known entirely in
advance, the answer is simple: there is no nontrivial sense of
variable-length encoding that applies generically while still being
interesting.\footnote{At least at sum rates close to the joint source
  entropy rate.  If the rates of communication are high enough, e.g.,
  equaling the log of the cardinalities of the source alphabets,
  zero-error communication is possible.} This is easiest to see by
example (Illustrated in Figure~\ref{fig.SWcoding} and revisited as
Example 2 in Section~\ref{sec.numerical}).  Suppose that the first
encoder observes the random vector $\rvbx$, which consists of a
sequence of $\BL$ iid uniform binary random variables.  Suppose
further that the second encoder observes $\rvby$ which is related to
$\rvbx$ via a memoryless binary symmetric channel with crossover
probability $\rho < 0.5$. The Slepian-Wolf sum-rate bound is $H(\rvx,
\rvy) = 1 + H(\rho) < 2 = H(\rvx) + H(\rvy)$.  But since the
individual encoders only see uniformly distributed binary sources,
they do not know when the sources are behaving jointly atypically.
Therefore, they have no basis on which to adjust their encoding rates
to combat joint atypicality. Since all pairs are possible when finite
blocklengths are considered, the individual encoders must use distinct
bit-strings for each of them. Since the expected length depends only
on the uniform marginal distributions, this means that the expected
length must be at least $N$. Thus, variable-length approaches do not,
in general\footnote{One should note that, in analogy to zero-error
  channel coding, there are special (non-generic) cases where
  zero-error Slepian-Wolf coding is possible~\cite{koulgiEtAl:03}
  since certain symbol pairs cannot occur.}, lead to zero-error
Slepian-Wolf codes for interesting rate-points.

\setlength{\unitlength}{1mm}
\begin{figure}
\begin{picture}(140,40)
\put(50,0){\line(1,0){20}} \put(50,10){\line(1,0){20}}
\put(50,0){\line(0,1){10}} \put(70, 0){\line(0,1){10}}
\put(52,4){Encoder $\svy$}
\put(50,30){\line(1,0){20}} \put(50,40){\line(1,0){20}}
\put(50,30){\line(0,1){10}} \put(70, 30){\line(0,1){10}}
\put(52,34){Encoder $\svx$}
\put(90,15){\line(1,0){20}} \put(90,25){\line(1,0){20}}
\put(90,15){\line(0,1){10}} \put(110, 15){\line(0,1){10}}
\put(94,19){Decoder }
\put(70, 5){\vector(3,2){20}}
\put(77, 5){$R_x$}
\put(70, 35){\vector(3,-2){20}}
\put(77, 35){$R_y$}
\put(40, 5){\vector(1,0){10}} \put(40, 35){\vector(1,0){10}}
\put(110, 20){\vector(1,0){10}}
\put(125,24){ $\hat{\rvx}_1,\hat{\rvx}_2, \ldots, \hat{\rvx}_{\BL}$}
\put(125,16){$\hat{\rvy}_1,\hat{\rvy}_2, \ldots, \hat{\rvy}_{\BL}$}
\put(18, 34){ ${\rvx}_1,{\rvx}_2, \ldots, \rvx_{\BL}$}
\put(18,4){ ${\rvy}_1,{\rvy}_2,\ldots \rvy_{\BL}$}
\put(15, 20){ $(\rvx_i,\rvy_i)\sim p_{\rvx\rvy}$}
\put(27,24){\vector(0,1){8}} \put(27,16){\vector(0,-1){8}}
\end{picture}
\caption[]{Slepian-Wolf distributed encoding and joint decoding of a pair of correlated sources.}
     \label{fig.SWcoding}
\end{figure}
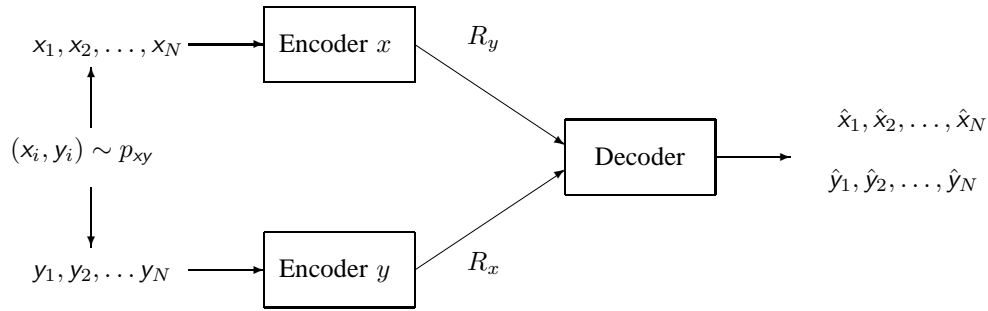

Another view of variable-length coding is as a tool that enables us to
achieve meaningful compression despite not knowing the underlying
probability distribution\footnote{In the point-to-point case, this is
  very closely related to achieving a zero-error probability. The same
  string can be an atypical realization of one source model while
  being a typical realization of another source. Encoding all the
  typical sequences correctly without knowing the underlying model
  requires getting all the possible sequences correctly for any
  specific model.} and allowing the rate used to adapt to the source.
If there is a low-rate, but reliable\footnote{It is clear that our
  techniques from \cite{sahaiSimsek:04, draperSahai:06} can also be
  adapted to make the system of \cite{draperAllerton:04} work using
  only noisy feedback channels.}, feedback link available from the
decoder to the two separate encoders, then this sense of
variable-length Slepian-Wolf coding is possible.
\cite{draperAllerton:04} gives a fixed-to-variable scheme in which the
stopping-time is chosen at the decoder and communicated back to the
encoders over a low-rate feedback link. The goal of
\cite{draperAllerton:04} is not achieving a truly zero probability of
error --- rather it is willing to accept a very small probability of
error in exchange for using a rate that is as small as possible.

To answer the question posed by Slepian and Wolf in the more classical
sense, we instead want to aim for a probability of error that goes to
zero for every source symbol, but at the cost of a variable delay.  To
do this, we propose stepping back and eliminating the modeling
assumption of encoders having access to the entire source realization
in advance. We argue that a ``streaming setting'' is required to
discern the system-level analog to variable-length source coding in
the distributed context. The streaming setting abstracts sources that
are embedded in time as well as the fact that all physically
realizable encoders/decoders must obey some form of causality. Thus
``rate'' is not just measured in bits per source symbol but in both
source symbols per second and bits per second. The source-rate
(symbols per second) is specified as a part of the problem while the
bit-rate (bits per second) is something that we get to choose. From an
engineering perspective, three desirable qualities\footnote{Of course,
  ``implementation complexity'' forms a fourth and very important
  consideration, but we will be ignoring that aspect of the problem.}
are:
\begin{itemize}
 \item Using a low rate bit-pipe(s)
 \item Low end-to-end latency
 \item Low probability of error
\end{itemize}
The theory of source-coding should tell us the tradeoffs between these
three desiderata. In addition, we will be interested in to what extent
a streaming code can be made ``universal'' over a class of probability
distributions. 

In the point-to-point streaming setting, regardless of whether block
or variable-length compression is used, the traditional initial step
is the same: group symbols into source blocks.  To compress the data
blocks, either use a fixed-rate block code, or a variable-length code.
The resulting encoding is then enqueued for transmission across the
bit-pipe.  As long as the source entropy rate is below the data-rate,
the queue will remain stable.  When block coding is used for
compression, there is a constant delay through the system, and
atypical source blocks are received in error. The probability of error
is fixed at the system's design-time and so is the end-to-end delay.

In contrast, variable-length coding induces a variable system delay.
The more unlikely the source blocks, the longer the delay experienced
at run-time. Thus, while {\em asymptotically} there are no errors when
variable-length source codes are used (assuming an infinite buffer
size), the delay till a given symbol can be decoded depends on the
random source realization. Because atypical source realizations are
large deviation events, the probability that some source symbol cannot
be reconstructed $\delay$ samples after it enters the encoder decays
exponentially\footnote{In \cite{Chang:06}, we show that variable
  length codes used in this manner actually achieve the best possible
  error exponent with delay. This is also related to the analysis of
  \cite{jelinek:68}.} in $\delay$. The choice of acceptable end-to-end
delay is left to the receiver/application. 

We show that this type of reliability {\em can} be achieved in a
generic distributed coding context --- the probability of error goes
to zero with end-to-end delay and the choice of the acceptable delay
is entirely up to the decoder. Essentially, every source symbol is
recovered correctly eventually with probability\footnote{The secret
  here is that we are considering a probability measure over infinite
  sequences. While all pairs of finite strings may be possible, most
  pairs of infinite strings collectively have probability zero.} $1$.
The only difference is that unlike the point-to-point case, the
decoder does not necessarily know when the estimate for the symbol has
converged to its final value.  Furthermore, just as in the
point-to-point setting\footnote{Sliding-window Lempel-Ziv compression
  is one example where data is naturally encoded sequentially. It is
  also universal over sources.}, both the encoding and decoding can be
made universal.

In this paper, we formally define a streaming Slepian-Wolf code, and
develop coding strategies both for situations when source statistics
are known and when they are not. The new tool is a sequential binning
argument that parallels the tree-coding arguments used to study
convolutional codes. We characterize the performance of
the streaming schemes through an error exponent analysis and
demonstrate that the exponents are equal regardless of whether the
system is informed of the source statistics (in which case we use
maximum likelihood decoding) or not (in which case we use universal
decoding). The universal decoder we design for the streaming problem
is somewhat different from those familiar from the block coding
literature, as are the nature of the error exponents.

\subsection{Potential applications and practical motivation}

In addition to our core interest in answering some basic questions
about Slepian-Wolf coding, our formulation is also motivated by the
diverse emerging application areas for distributed source coding.
Media (e.g. video-conference) sources naturally have a streaming
character.  Consequently, we are motivated to explore what sort of
streaming Slepian-Wolf technique matches naturally to such
situations.\footnote{A secondary aspect in some multimedia settings is
  a natural multi-scale nature to the source --- the high order bits
  are more important than the low order bits. To the extent that the
  high order bits can be made ``early'' and the low-order bits can be
  made ``late'', our constructions also naturally give more protection
  to the early bits as compared to the later ones. While this
  interpretation might eventually be important in practice, it is a
  bit questionable within the simplified model this paper considers.}

\subsection{Outline}

Section~\ref{sec.notation} summarizes the notation used in the paper.
Section~\ref{sec.mainresults} reviews the classical block-coding error
exponent results for Slepian-Wolf source coding and then we state the
main results of this paper: sequential error exponents for
Slepian-Wolf source coding. Section~\ref{sec.numerical} presents a
numeric study of two example sources. We observe that the sequential
error exponent is often the same as the block coding error exponent.
Sections \ref{sec.entropy}, \ref{sec.incDecSI} and \ref{sec.SW} prove
the theorems in Section~\ref{sec.mainresults}. We start with
sequential source coding for single sources in \ref{sec.entropy}.
This is the simplest case but it provides insights to the nature of
sequential source coding problem and sequential error events. We show
that the sequential error exponent is the same as the random block
source coding error exponent. Section~\ref{sec.incDecSI} moves on to
the case with decoder side-information. Finally, Section~\ref{sec.SW}
presents the proof of the main result of the paper. We derive the
sequential error exponent of distributed source coding for correlated
sources. This error exponent strictly positive everywhere inside the
achievable rate region of \cite{slepianWolf:73}. For all these three
scenarios in Sections \ref{sec.entropy}, \ref{sec.incDecSI} and
\ref{sec.SW}, both ML and universal decoding rules are studied.  The
appendix shows that the resulting error exponents are indeed the same.

\section{Notation}\label{sec.notation}

We use serifed-fonts, e.g., $\svx$ to indicate sample values, and
sans-serif, e.g., $\rvx$, to indicate random variables.  Bolded fonts
are reserved to indicate sample or random vectors, e.g., $\svbx =
\svx^n$ and $\rvbx = \rvx^n$, respectively, where the vector length
($n$ here) is understood from the context.  Subsequences, e.g.,
$\svx_l, \svx_{l+1}, \ldots, \svx_{n}$ are denoted as $\svx_l^n$ where
$\svx_i^j \defeq \emptyset$ if $i<j$.  Distributions are indicated
with lower-case $p$, e.g., $\rvx$ is distributed according to
$p_{\rvx}(\svx)$.  Sets and their elements are denoted as, e.g., $\svx
\in \cX$, and their cardinality by $|\cX|$. We use calligraphic font
to denote sets, $\mathcal{X}$, $\mathcal{F}$, $\mathcal{W}$ etc, and
reserve $\mathcal{E}$ and $\mathcal{D}$ to denote encoding and
decoding functions, respectively.  We use standard notation for types,
see, e.g., \cite{csiszarKorner}.  Let $N(a; \svbx)$ denote the number
of symbols in the length-$n$ vector $\svbx$ that take on value $a$.
Then, $\svbx$ is of type $P$ if $P(a) = N(a; \svbx)/n$.  The
type-class, or set of length-$n$ vectors of type $P$ is denoted
$\tclass_{P}$.  A sequence $\svby$ has conditional type $V$ given
$\svbx$ if $N(a,b; \svbx, \svby) = N(a; \svbx) V(b|a) = P(a) V(b|a)$
for every $a, b$. The set of sequences $\svby$ having conditional type
$V$ with respect to $\svbx$ is called the $V$-shell of $\svbx$ and is
denoted by $\tclass_{V}(\svbx)$.  When considered together, the pair
$(\svbx, \svby)$ is said to have joint type $V \times P$.  We always
use upper-case, e.g., $P$ and $V$, to denote length-$n$ types and
conditional types.  As we often discuss the types of subsequences we
add a superscript notation to remind the reader of the length of the
subsequence in question.  If, for instance, the subsequence under
consideration is $\svx_{l}^n$ we write $\svx_{l}^n \in
\tclass_{\PNL}$.  Similarly we use $\VNL$ for the conditional type of
length-$(n-l+1)$, and $\VNL \times \PNL$ for the joint type.

Given a joint type $V \times P$, entropies and conditional entropies
are denoted as $H(P)$ and $H(V|P)$, respectively. The KL divergence
between two distributions $q$ and $p$ is denoted by $D(q \| p)$.

\section{Main Results}\label{sec.mainresults}

In this section, we begin by reviewing classical results on the error
exponents of distributed block coding.  We then present the main
results of the paper: error exponents for streaming Slepian-Wolf
coding and its special cases: point-to-point coding and source coding
with decoder side information.  We analyze both maximum likelihood and
universal decoding and show that the achieved exponents are equal.
Leaving numerical examples and proofs for later sections, we here
compare the form of the streaming exponents with their block coding
counterparts.

\subsection{Block source coding and error exponents}

In the classic block-coding Slepian-Wolf paradigm, full length-$\BL$
vectors $\rvbx$ and $\rvby$ are observed by their respective encoders
before communication commences.  In this situation a rate-$(\Rx, \Ry)$
length-$\BL$ block source code consists of an encoder-decoder triplet
$(\mathcal{E}^x_{\BL},\mathcal{E}^y_{\BL}, \mathcal{D}_{\BL})$, as we
will define shortly. For the rate-region considerations, the general
case of distributed encoders can be considered by using time-sharing
among codes that alternate between sending at rates close to the
marginal entropy and those that correspond to perfectly known
side-information. However, it is easy to see that this results in a
substantial loss of error-exponent even in the block-coding case. To
get good exponents, something else is required:

\begin{defn}\label{def.SWblockCode}
  A randomized length-$\BL$ rate-$(\Rx, \Ry)$ block encoder-decoder
  triplet  $(\mathcal{E}^x_{\BL},\mathcal{E}^y_{\BL},\mathcal{D}_{\BL})$ is a
  set of maps
\begin{eqnarray*}
\begin{array}{lclcl}
\mathcal{E}^x_{\BL} &: & \mathcal{X}^{\BL} \rightarrow \{0,1\}^{ \BL R_x},
& \mbox{e.g.,} & \mathcal{E}^x_{\BL}(x^{\BL})=a^{ \BL R_x}\\
\mathcal{E}^y_{\BL} &: & \mathcal{Y}^{\BL} \rightarrow \{0,1\}^{ \BL R_y},
& \mbox{e.g.,} & \mathcal{E}^y_{\BL}(y^{\BL})=b^{ \BL R_y}\\
\mathcal{D}_{\BL} &: & \{0,1 \}^{ \BL R_x }\times \{0,1 \}^{ \BL R_y }
\rightarrow \mathcal{X}^{n}\times \mathcal{Y}^{n}, & \mbox{e.g.,}
&\mathcal{D}_{\BL}(a^{ \BL R_x }, b^{ \BL R_y })=(\hat{x}^{\BL},\hat{y}^{\BL})
\end{array}
\end{eqnarray*}
where common randomness, shared between the encoders and the decoder
is assumed.  This allows us to randomize the mappings independently of
the source sequences.
\end{defn}

The error probability typically considered in Slepian-Wolf coding is
the joint error probability, $\Pr[(\rvx^{\BL}, \rvy^{\BL})\neq
(\hat{\rvx}^{\BL},\hat{\rvy}^{\BL})]=\Pr[(\rvx^{\BL},\rvy^{\BL})\neq
\mathcal{D}_{\BL}(\mathcal{E}^x_{\BL}(\rvx^{\BL}),
\mathcal{E}^y_{\BL}(\rvy^{\BL}))]$.  This probability is taken over
the random source vectors as well as the randomized mappings.  An
error exponent $E$ is said to be achievable if there exists a family
of rate-$(\Rx, \Ry)$ encoders and decoders
$\{(\mathcal{E}^x_{\BL},\mathcal{E}^y_{\BL},\mathcal{D}_{\BL})\}$,
indexed by $\BL$,
such that 
\begin{equation}
\lim_{\BL \rightarrow \infty}-\frac{1}{\BL}\log
\Pr[(\rvx^{\BL}, \rvy^{\BL})\neq
(\hat{\rvx}^{\BL},\hat{\rvy}^{\BL})] \geq E. \label{eq.SWblockErrExp}
\end{equation}

In this paper, we study random source vectors $(\rvbx, \rvby)$ that
are iid across time but may have dependencies at any given time:
\begin{equation*}
p_{\rvx,\rvy}(\svbx,\svby)=\prod_{i=1}^{\BL}p_{\rvx,\rvy}(\svx_i,\svy_i).
\end{equation*}

For such iid sources, upper and lower bounds on the achievable error
exponents are derived in~\cite{gallagerTech:76,csiszarKorner}.  These
results are summarized by the following theorem.

\begin{thm}\label{THM.INTRO}
  (Lower bound) Given a rate pair $(\Rx, \Ry)$ such that $\Rx >
  H(\rvx|\rvy)$, $\Ry > H(\rvy|\rvx)$, $\Rx + \Ry > H(\rvx, \rvy)$.
  Then, for all
\begin{equation}
E < \min_{\rvxBar,\rvyBar} D(p_{\rvxBar,\rvyBar}\|p_{\rvx\rvy})+ \big|
\min[R_x+R_y-H(\rvxBar,\rvyBar), R_x-H(\rvxBar|\rvyBar),
 R_y-H(\rvyBar|\rvxBar) ]\big|^{+} \label{eq.SWblockLowBnd}
\end{equation}
there exists a family of randomized encoder-decoder mappings as
defined in Definition~\ref{def.SWblockCode} such
that~(\ref{eq.SWblockErrExp}) is satisfied.
In~(\ref{eq.SWblockLowBnd}) the function $|z|^{+} = z$ if $z \geq 0$
and $|z|^{+} = 0$ if $z < 0$.

(Upper bound) Given a rate pair $(\Rx, \Ry)$ such that $\Rx >
  H(\rvx|\rvy)$, $\Ry > H(\rvy|\rvx)$, $\Rx + \Ry > H(\rvx, \rvy)$. Then,
  for all
\begin{equation}
E >\min \left\{
\min_{ \rvxBar,\rvyBar: R_x<H(\rvxBar|\rvyBar)}
D(p_{\rvxBar,\rvyBar}\|p_{\rvx\rvy}) ,
\min_{ \rvxBar,\rvyBar: R_y<H(\rvyBar|\rvxBar)}
D(p_{\rvxBar,\rvyBar}\|p_{\rvx\rvy}),
\min_{ \rvxBar,\rvyBar: R_x+R_y<H(\rvxBar,\rvyBar)}
D(p_{\rvxBar,\rvyBar}\|p_{\rvx\rvy}) \right\}
\label{eq.SWblockUpBnd}
\end{equation}
there does not exists a randomized encoder-decoder mapping as defined
in Definition~\ref{def.SWblockCode} such that~(\ref{eq.SWblockErrExp}) is
satisfied.

In both bounds $(\rvxBar,\rvyBar)$ are dummy random variables
with joint distribution $p_{\rvxBar,\rvyBar}$.
\end{thm}

{\em Remark:} As long as $(R_x,R_y)$ is in the interior of the
achievable region, i.e., $R_x> H(\rvx|\rvy)$, $R_y> H(\rvy|\rvx)$ and
$R_x+R_y> H(\rvx, \rvy)$ then the lower-bound~(\ref{eq.SWblockLowBnd})
is positive.  The achievable region is illustrated in
Fig~\ref{fig.SW_region_intro}.  As shown in \cite{csiszarKorner},
the upper and lower bounds~(\ref{eq.SWblockUpBnd})
and~(\ref{eq.SWblockLowBnd}) match when the rate pair $(R_x,R_y)$ is
achievable and close to the boundary of the region. This is analogous
to the high rate regime in channel coding where the random coding
bound (analogous to~(\ref{eq.SWblockLowBnd})) and the sphere packing
bound (analogous to~(\ref{eq.SWblockUpBnd})) agree.

Theorem~\ref{THM.INTRO} can also be used to generate bounds on the
exponent for source coding with decoder side information (i.e.,
$\rvby$ observed at the decoder), and for source coding without side
information (i.e., $\rvby$ is a constant).  These corollaries will
prove useful as a basis for comparison as we build up to the complete
solution for streaming Slepian-Wolf coding.

\begin{corol}\label{thm.blockSI}
  (Source coding with decoder side information) Consider a
  Slepian-Wolf problem where $\rvby$ is known by the decoder. Given a
  rate $\Rx$ such that $\Rx > H(\rvx|\rvy)$, then for all
\begin{equation}
E < \min_{\rvxBar,\rvyBar} D(p_{\rvxBar,\rvyBar}\|p_{\rvx\rvy}) +
|R_x-H(\rvxBar|\rvyBar)|^{+}, \label{eq.SIblockLowBnd}
\end{equation}
there exists a family of randomized encoder-decoder mappings as
defined in Definition~\ref{def.SWblockCode} such
that~(\ref{eq.SWblockErrExp}) is satisfied.
\end{corol}

The proof of Corollary~\ref{thm.blockSI} follows from
Theorem~\ref{THM.INTRO} by letting $\Ry$ be arbitrarily large.
Similarly, by letting $\rvby$ be deterministic so that $H(\rvx|\rvy)
= H(\rvx)$ and $H(\rvy) = 0$, we get the following random-coding
bound for the point-to-point case of a single source $\rvbx$.

\begin{corol}\label{thm.blockEnt} (point-to-point)
  Consider a Slepian-Wolf problem where $\rvby$ is deterministic,
  i.e., $\rvby = \svby$.  Given a rate $\Rx$ such that $\Rx >
  H(\rvx)$, for all
\begin{equation}
E < \min_{\rvxBar} D(p_{\rvxBar}\|p_{\rvx})+ |R_x-H(\rvxBar)|^{+}
=E_x(R_x) \label{eq.EntblockLowBnd}
\end{equation}
there exists a family of randomized encoder-decoder triplet as defined
in Definition~\ref{def.SWblockCode} such that~(\ref{eq.SWblockErrExp}) is
satisfied.
\end{corol}

\setlength{\unitlength}{1mm}
 \begin{figure}[htbp]
   \begin{center}
     \leavevmode
 \begin{picture}(130,80)

\put(40, 10){\vector(1,0){55}}
 \put(40, 10){\vector(0,1){55}}

\put(40,70){$R_y$} \put(100,10){$R_x$}

\put(28,40){$H(\rvy)$} \put(28,31){$H(\rvy| \rvx)$}

\put(68,5){$H(\rvx)$} \put(56, 5){$H(\rvx| \rvy)$}

\put(28,60){$\log|\mathcal{Y}|$}   \put(87,5){$\log|\mathcal{X}|$}

 \put(90,60){\line(0,-1){29}}  \put(90,31){\line(-1,0){20}}
  \put(70,31){\line(-1, 1){9}}
   \put(90,60){\line(-1,0){29}}  \put(61,60){\line(0,-1){20}}

 \put(64, 52){Achievable} \put(66, 48){Region}

 \put(47,
25){$R_x+R_y=H(\rvx,\rvy)$} \put(60, 28){\vector(1,1){6.5}}

\end{picture}
\caption{ Achievable region for Slepian-Wolf source coding }
\label{fig.SW_region_intro}
\end{center}
\end{figure}
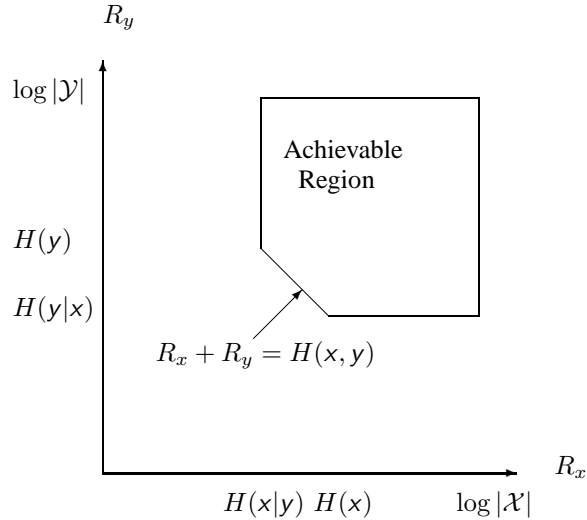

\subsection{Sequential Distributed Source Coding}

We now state our main results for streaming encoding, and contrast
them with the block-coding results of the last section.  To begin, we
define a streaming encoder.

\begin{defn}
\label{def.seqn_coding}
A randomized sequential encoder-decoder triplet
$\mathcal{E}^x,\mathcal{E}^y,\mathcal{D}$ is a sequence of mappings,
$\{\mathcal{E}^x_j\},j=1,2,...$, $\{\mathcal{E}^y_j\},j=1,2,...$ and
$\{\mathcal{D}_j\},j=1,2,...$:

\begin{equation}
\begin{array}{lclcl}
\mathcal{E}^x_j & : & \mathcal{X}^{j} \longrightarrow
    \{0,1\}^{ R_x }, & \mbox{e.g.,} & \mathcal{E}^x_j(x^j)=a_{ (j-1)R_x +1}^{
      jR_x }, \\
\mathcal{E}^y_j & : & \mathcal{Y}^{j}
    \longrightarrow \{0,1\}^{ R_y }, & \mbox{e.g.,} & \mathcal{E}^y_j(y^j)=b_{
      (j-1)R_y +1}^{ jR_y }.
\end{array}
\label{eq.xEnc}
\end{equation}
Common randomness, shared between encoders and decoder, is assumed.
This allows us to randomize the mappings independently of the source
sequence.
\end{defn}

In this paper, the sequential encoding maps will always work by
assigning random ``parity bits'' in a causal manner to the observed
source sequence.  That is, the $\Rx$ (or $\Ry$) bits generated at each
time in~(\ref{eq.xEnc}), are iid Bernoulli-$(0.5)$.\footnote{We assume
  that $\Rx$ and $\Ry$ are integer.  To justify this assumption note
  that we can always group sets of $\alpha$ successive symbols into
  super-symbols.  These larger symbols can be encoded at an average
  rate $\alpha \Rent$.  Generally, if we group $\alpha$ symbols
  together, and transmit $\beta$ bits per super-symbol, we can realize
  an average rate $\alpha/ \beta$, i.e., a rational rate.  If desired,
  non-integer average rates are easily implemented by a time-varying
  transmission rate.  For example, say we want to implement an average
  encoding rate of $5/4$ bits per source symbol.  Say we generate one
  new parity bit per symbol for each symbol observed except for the
  fourth symbol, eighth symbol, etc, when we generate two.  The
  average encoding rate is $5/4$.  As long as the decoding delay
  $\delay$ we target is long enough so that the decoder received an
  ``average'' number of encoded bits -- $\delta \Rent$ -- before we
  must make an estimate (e.g., if $\delay \gg 1 / \Rent$), these
  small-scale issues even out.  In particular, they do not effect the
  exponents.}  Since parity bits are assigned causally, if two source
sequences share the same length-$l$ prefix, then their first $l
{\Rent}$ parity bits must match.  Subsequent parities are drawn
independently. Such a sequential coding strategy is the source-coding
parallel to tree and convolutional codes used for channel coding
\cite{Forney:74}. In fact, we call these ``parity bits'' as they can
be generated using an infinite constraint-length time-varying random
convolutional code.

\begin{defn}
The decoder mapping
\begin{eqnarray}
  &&\mathcal{D}_j: \{0,1 \}^{ jR_x }\times\{0,1 \}^{ jR_y }
  \longrightarrow \mathcal{X}^j \times \mathcal{Y}^j \nonumber\\
  &&\mathcal{D}_j(a^{ jR_x },b^{ jR_y
  })=(\svxhat_{1}^{j}(j),\svyhat_{1}^{j}(j))\nonumber
\end{eqnarray}
At each time $j$ the decoder $\mathcal{D}_j$ outputs estimates of all
the source symbols that have entered the encoder by time $j$.
\end{defn}

{\em Remark:} While we state Definition~\ref{def.seqn_coding} only for
Slepian-Wolf coding, it immediately specializes to source coding with
decoder side information (dropping the $\mathcal{E}_y$ and revealing
$\rvy^n$ to the decoder), and source coding without side information
(dropping the $\mathcal{E}_y$).  We present results for both these
situations as well.

In this paper we study two error probabilities.  We define the pair of
source estimates at time $n$ as $(\hat{\rvx}^n, \hat{\rvy}^n) =
\mathcal{D}_n(\prod_{j=1}^n \mathcal{E}^x_j, \prod_{j=1}^n
\mathcal{E}^y_j)$, where $\prod_{j=1}^n \mathcal{E}^x_j$ indicates the
full $n \Rx$ bit stream from encoder $x$ up to time $n$.  We use
$(\hat{\rvx}^{n - \delay}, \hat{\rvy}^{n - \delay})$ to indicate the
first $n - \delay$ symbols of each estimate, where for conciseness of
notation both the estimate time, $n$, and the decoding delay,
$\delay$, are indicated in the superscript.  With these definitions
the two error probabilities we study are
\begin{align}
\Pr[\rvxhat^{n - \delay} \neq \rvx^{n - \delay}] \;\; \mbox{and} \;\;
\Pr[\rvyhat^{n - \delay} \neq \rvy^{n - \delay}]. \nonumber
\end{align}
A pair of exponents $E_x > 0$ and $E_y > 0$ is said to be achievable
if there exists a family of rate-$(\Rx, \Ry)$ encoders and decoders
$\{(\mathcal{E}_j^x, \mathcal{E}_j^y, \mathcal{D}_j)\}$ such that
\begin{align}
\lim_{\delay \rightarrow \infty} \lim_{n \rightarrow \infty}
- \frac{1}{\delay} \log \Pr[\hat{\rvx}^{n - \delay} \neq \rvx^{n - \delay}]
&\geq E_x \label{eq.errExpX}\\
\lim_{\delay \rightarrow \infty} \lim_{n \rightarrow \infty}
- \frac{1}{\delay} \log \Pr[\hat{\rvy}^{n - \delay} \neq \rvy^{n - \delay}]
&\geq E_y \label{eq.errExpY}
\end{align}

{\em Remarks:} In contrast to~(\ref{eq.SWblockErrExp}) the error
exponent we look at is in the delay, $\delay$, rather than total
observation time, $n$. The order of the limits is important since the
total time-period $n$ is allowed to go to infinity faster than the
delay $\delay$. While the definitions
of~(\ref{eq.errExpX})--(\ref{eq.errExpY}) and
of~(\ref{eq.SWblockErrExp}) are asymptotic in nature, the results hold
for finite block-lengths and delays as well.  Finally, we note that
while in~(\ref{eq.SWblockErrExp}) the error exponent of a joint error
event on either $\rvbx$ or $\rvby$ is considered, we provide a refined
analysis specifying potentially different exponents on either
decision.  The results for joint errors are found by taking the
minimum of the individual exponents, i.e.,
\begin{equation*}
\lim_{\delay \rightarrow \infty} \lim_{n \rightarrow \infty}
- \frac{1}{\delay} \log \Pr[(\hat{\rvx}^{n-\delay}, \hat{\rvy}^{n - \delay})
\neq (\rvx^{n-\delay}, \rvy^{n - \delay})]  \geq
\min\{E_x, E_y\}.
\end{equation*}

\subsection{Streaming source coding}

Our first results concern streaming coding in the point-to-point
setting.  The first theorem we state gives random coding error
exponents for maximum likelihood decoding where the source statistics
are known, and the second exponents for universal decoding, where they
are not.
\begin{thm} \label{thm.entCodeML}
  Given a rate $\Rent > H(\PxRV)$, there exists a randomized streaming
  encoder and maximum likelihood decoder pair (per
  Definition~\ref{def.seqn_coding}) such that for all $E < \expML(\Rent)$
  there is a constant $K > 0$ such that $\Pr[\rvxhat^{n - \delay} \neq
  \rvx^{n - \delay}] \leq K \exp\{- \delay \expML(\Rent)\}$ for all
  $n, \delay \geq 0$ where
\begin{equation}
\expML(\Rent) = \sup_{0 \leq \rho \leq 1} \rho \Rent - (1 + \rho) \log
\left( \sum_{\svx} \PxRV(\svx)^{\frac{1}{1 + \rho}} \right).
\label{eq.errExpML}
\end{equation}
\end{thm}

\begin{thm} \label{thm.entCodeUniv}
  Given a rate $\Rent > H(\PxRV)$, there exists a randomized streaming
  encoder and universal decoder pair (per Definition~\ref{def.seqn_coding})
  such that for all $E < \expUniv(\Rent)$ there is a constant $K > 0$
  such that $\Pr[\rvxhat^{n - \delay} \neq \rvx^{n - \delay}] \leq K
  \exp\{- \delay E\}$ for all $n, \delay \geq 0$ where
\begin{equation}
\expUniv(\Rent) = \inf_q D(q \| \PxRV) + |\Rent - H(q)|^{+},
\label{eq.errExpUniv}
\end{equation}
where $q$ is an arbitrary probability distribution on $\cX$ and where
$|z|^{+} = z$ if $z \geq 0$ and $|z|^{+} = 0$ if $z < 0$.
\end{thm}

{\em Remark:} The error exponents of Theorems~\ref{thm.entCodeML}
and~\ref{thm.entCodeUniv} both equal their respective random
block-coding exponents for ML and universal decoders.
For example, compare~(\ref{eq.errExpUniv})
with~(\ref{eq.EntblockLowBnd}).  The main difference in the
formulation is that the error probability decays with delay $\delay$
rather than block length $\BL$.  Furthermore, it is known
that~(\ref{eq.errExpML}) and~(\ref{eq.errExpUniv}) are equal --- see
\cite{csiszarKorner} exercise $13$ on page $44$.  Such equality is
required by the formal definition of a universal scheme, i.e., for the
same source statistics and coding rates, the universal decoder should
asymptotically achieve the same error exponent as the maximum
likelihood decoder.  See~\cite{lapidothNarayan:98} for a detailed
discussion of universal versus maximum likelihood decoding in the
context of channel coding.

\subsection{Streaming distributed source coding with decoder side information}

This section summarizes our results for distributed streaming source
coding when the side information is observed at the decoder, but not
the encoder:

\begin{thm} \label{thm.decSIML}
  Given a rate $\Rent > H(\rvx|\rvy)$, there exists a randomized
  encoder decoder pair (per Definition~\ref{def.seqn_coding}) such that for
  all $E < \expMLSI(\Rent)$ there is a constant $K > 0$ such that
  $\Pr[\rvxhat^{n-\delay} \neq \rvx^{n-\delay}] \leq K \exp\{- \delay
  E\}$ for all $n, \delay \geq 0$ where
\begin{equation}
\expMLSI(\Rent) = \sup_{0 \leq \rho \leq 1} \rho \Rx - \log \Big[
\sum_{\svy} \Big[ \sum_{\svx}
p_{\rvx\rvy}(\svx,\svy)^{\frac{1}{1+\rho}} \Big]^{1+\rho} \Big].
\label{eq.errExpMLSI}
\end{equation}
\end{thm}

\begin{thm} \label{thm.decSIUniv}
  Given a rate $\Rent > H(\rvx|\rvy)$, there exists a randomized
  encoder decoder pair (per Definition~\ref{def.seqn_coding} ) such that for
  all $E < \expUnivSI(\Rent)$ there is a constant $K > 0$ such that
  $\Pr[\rvxhat^{n-\delay} \neq \rvx^{n-\delay}] \leq K \exp\{- \delay
  E\}$ for all $n, \delay \geq 0$ where
\begin{equation}
\expUnivSI(\Rent)
=\inf_{\rvxtil, \rvytil} D(p_{\rvxtil, \rvytil} \| \PxyRV) +
|\Rent - H(\rvxtil | \rvytil)|^{+}, \label{eq.errExpUnivSI}
\end{equation}
and $(\rvxtil, \rvytil)$ are random variables with joint distribution
$p_{\rvxtil, \rvytil}$, $H(\rvxtil | \rvytil)$ is their conditional
entropy, and where $|z|^{+} = z$ if $z \geq 0$ and $|z|^{+} = 0$ if $z
< 0$.
\end{thm}

{\em Remark:} Similar to the point-to-point case, the error exponents
of Theorems~\ref{thm.decSIML} and~\ref{thm.decSIUniv} both equal their
respective random block-coding exponents.  For example,
compare~(\ref{eq.errExpUnivSI}) with~(\ref{eq.SIblockLowBnd}).
Similarly, (\ref{eq.errExpMLSI}) and~(\ref{eq.errExpUnivSI}) can be
shown to be equal.

\subsection{Streaming Slepian-Wolf coding}

In contrast to streaming point-to-point coding and streaming source
coding with decoder side information, the general case of streaming
Slepian-Wolf coding with two distributed encoders results in error
exponents that differ from their block coding counterparts.  In the
streaming setting, fundamentally different error events dominate as
compared to the block setting.

\begin{thm} \label{thm.jointCodeML}

  Let $(\Rx, \Ry)$ be a rate pair such that $\Rx > H(\rvx|\rvy)$, $\Ry
  > H(\rvy|\rvx)$, $\Rx + \Ry > H(\rvx, \rvy)$.  Then, there exists a
  randomized encoder pair and maximum likelihood decoder triplet (per
  Definition~\ref{def.seqn_coding}) that satisfies the following three
  decoding criteria.

  (i) For all $E < E_{ML,SW,x}(\Rx, \Ry)$, there is a constant $K > 0$
  such that
$\Pr[ \rvxhat^{n-\delay} \neq \rvx^{n-\delay}] \leq K
  \exp\{- \delay E\}$ for all $n, \delay \geq 0$ where
\begin{equation}
E_{ML,SW, x}(\Rx,  \Ry) = \min \Bigg\{ \inf_{\gamma \in [0,1]}
E_x^{ML}(\Rx, \Ry, \gamma), \inf_{\gamma \in [0,1]}
\frac{1}{1-\gamma} E_y^{ML}(\Rx, \Ry, \gamma) \Bigg\}.\nonumber
\end{equation}

(ii) For all $E < E_{ML,SW,y}(\Rx, \Ry)$ there is a constant $K > 0$
such that $\Pr[\rvyhat^{n-\delay} \neq \rvy^{n-\delay}] \ \leq K
\exp\{- \delay E\}$ for all $n, \delay \geq 0$ where
\begin{equation}
E_{ML,SW, y}(\Rx,  \Ry) = \min \Bigg\{ \inf_{\gamma \in [0,1]}
\frac{1}{1-\gamma} E_x^{ML}(\Rx, \Ry, \gamma), \inf_{\gamma \in
[0,1]} E_y^{ML}(\Rx, \Ry, \gamma) \Bigg\}.\nonumber
\end{equation}

(iii) For all $E < E_{ML,SW,xy}(\Rx, \Ry)$ there is a constant $K >
0$ such that $\Pr[(\rvxhat^{n-\delay}, \rvyhat^{n-\delay}) \neq
(\rvx^{n-\delay}, \rvy^{n-\delay})] \ \leq K \exp\{- \delay E\}$ for
all $n, \delay \geq 0$ where
\begin{equation}
E_{ML,SW, xy}(\Rx,  \Ry) = \min \Bigg\{ \inf_{\gamma \in [0,1]}
E_x^{ML}(\Rx, \Ry, \gamma), \inf_{\gamma \in [0,1]} E_y^{ML}(\Rx,
\Ry, \gamma) \Bigg\}.\nonumber
\end{equation}

In definitions (i)--(iii),
\begin{equation}
\begin{array}{lll}
E_x^{ML}(\Rx, \Ry, \gamma) & = & \sup_{\rho \in [0,1]} [ \gamma
E_{x|y}(\Rx, \rho) + (1-\gamma) E_{xy}(\Rx, \Ry, \rho)]
\vspace{1ex} \\
E_y^{ML}(\Rx, \Ry, \gamma) & = & \sup_{\rho \in [0,1]} [ \gamma
E_{y|x}(\Rx, \rho) + (1-\gamma) E_{xy}(\Rx, \Ry, \rho)]
\end{array} \label{eq.compoundExp}
\end{equation}
and
\begin{equation}
\begin{array}{lll}
E_{xy}(\Rx, \Ry, \rho) & = & \rho (\Rx + \Ry) - \log \Big[ \sum_{\svx, \svy}
p_{\rvx\rvy}(\svx,\svy)^{\frac{1}{1+\rho}} \Big]^{1 + \rho} \vspace{1ex}\\
E_{x|y}(\Rx, \rho) & =  & \rho \Rx - \log \Big[ \sum_{\svy}
\Big[ \sum_{\svx}
p_{\rvx\rvy}(\svx,\svy)^{\frac{1}{1+\rho}} \Big]^{1+\rho} \Big]\vspace{1ex}\\
E_{y|x}(\Ry, \rho) & = & \rho \Ry - \log \Big[ \sum_{\svx} \Big[
\sum_{\svy} p_{\rvx\rvy}(\svx,\svy)^{\frac{1}{1+\rho}}
\Big]^{1+\rho} \Big] \vspace{1ex}
\end{array}\label{eq.defBasicExp}
\end{equation}
\end{thm}

\begin{thm} \label{thm.jointCode}

  Let $(\Rx, \Ry)$ be a rate pair such that $\Rx > H(\rvx|\rvy)$, $\Ry
  > H(\rvy|\rvx)$, $\Rx + \Ry > H(\rvx, \rvy)$.  Then, there exists a
  randomized encoder pair and universal decoder triplet (per
  Definition~\ref{def.seqn_coding}) that satisfies the following three
  decoding criteria.

  (i) For all $E < E_{UN,SW,x}(\Rx, \Ry)$, there is a constant $K >
  0$ such that $\Pr[ \rvxhat^{n-\delay} \neq \rvx^{n-\delay}] \leq K
  \exp\{- \delay E\}$ for all $n, \delay \geq 0$ where
\begin{equation}
E_{UN,SW, x}(\Rx,  \Ry) = \min \Bigg\{ \inf_{\gamma \in [0,1]}
E_x^{UN}(\Rx, \Ry, \gamma), \inf_{\gamma \in [0,1]}
\frac{1}{1-\gamma} E_y^{UN}(\Rx, \Ry, \gamma) \Bigg\}.
\end{equation}

(ii) For all $E < E_{UN,SW,y}(\Rx, \Ry)$, there is a constant $K >
0$ such that $\Pr[ \rvyhat^{n-\delay} \neq \rvy^{n-\delay}] \leq K
\exp\{- \delay E\}$ for all $n, \delay \geq 0$ where
\begin{equation}
E_{UN,SW, y}(\Rx,  \Ry) = \min \Bigg\{ \inf_{\gamma \in [0,1]}
\frac{1}{1-\gamma} E_x^{UN}(\Rx, \Ry, \gamma), \inf_{\gamma \in
[0,1]} E_y^{UN}(\Rx, \Ry, \gamma) \Bigg\}.
\end{equation}

(iii) For all $E < E_{UN,SW,xy}(\Rx, \Ry)$, there is a constant $K >
0$ such that $\Pr[ (\rvxhat^{n-\delay}, \rvxhat^{n-\delay}) \neq
(\rvx^{n-\delay}, \rvy^{n-\delay})] \leq K \exp\{- \delay E\}$ for
all $n, \delay \geq 0$ where
\begin{equation}
E_{UN,SW, xy}(\Rx,  \Ry) = \min \Bigg\{ \inf_{\gamma \in [0,1]}
 E_x^{UN}(\Rx, \Ry, \gamma), \inf_{\gamma \in
[0,1]} E_y^{UN}(\Rx, \Ry, \gamma) \Bigg\}.
\label{eq.expSWUnivJoint}
\end{equation}

In definitions (i)--(iii),
\begin{align}
& E_{x}^{UN}(\Rx, \Ry, \gamma) \nonumber \\
& = \inf_{\tiny \rvxtil, \rvytil,
\rvxBar, \rvyBar} \gamma D(p_{\rvxtil, \rvytil} \| \PxyRV) +
(1-\gamma) D(p_{\rvxBar, \rvyBar} \| \PxyRV)
+ \left|\gamma [\Rx - H(\rvxtil | \rvytil )]
+ (1-\gamma)  [\Rx + \Ry - H(\rvxBar, \rvyBar)]\right|^{+}
\nonumber \\
& E_{y}^{UN}(\Rx, \Ry, \gamma) \nonumber \\
& = \inf_{\tiny \rvxtil, \rvytil,
\rvxBar, \rvyBar} \gamma D(p_{\rvxtil, \rvytil}\| \PxyRV) +
(1-\gamma) D(p_{\rvxBar, \rvyBar}  \| \PxyRV)
+ \left|\gamma [\Ry - H(\rvytil|\rvxtil)]
+ (1-\gamma) [\Rx + \Ry - H(\rvxBar, \rvyBar)]\right|^{+}
\end{align}
where the random variables $(\rvxtil, \rvytil)$ and $(\rvxBar,
\rvyBar)$ have joint distributions $p_{\rvxtil, \rvytil}$ and
$p_{\rvxBar, \rvyBar}$, respectively.  The function $|z|^{+} = z$ if
$z \geq 0$ and $|z|^{+} = 0$ if $z < 0$.
\end{thm}

{\em Remark:} Definitions (i) and (ii) in
Theorems~\ref{thm.jointCodeML} and~\ref{thm.jointCode} concern
individual decoding error events which might be useful in applications
where the $\rvbx$ and $\rvby$ streams are decoded jointly, but
utilized individually.  The more standard joint error event is given
by (iii).

{\em Remark:} We can compare the joint error event for block and
streaming Slepian-Wolf coding, c.f.~(\ref{eq.expSWUnivJoint})
with~(\ref{eq.SWblockLowBnd}).  The streaming exponent differs by the
extra parameter $\gamma$ that must be minimized over. If the
minimizing $\gamma = 1$, then the block and streaming exponents are
the same. The minimization over $\gamma$ results from a fundamental
difference in the types of error-causing events that can occur in
streaming Slepian-Wolf as compared to block Slepian-Wolf.

{\em Remark:} The error exponents of maximum likelihood and universal
decoding in Theorems~\ref{thm.jointCodeML} and~\ref{thm.jointCode} are
the same.  However, because there are new classes of error events
possible in streaming, this needs proof. The equivalence is
summarized in the following theorem.

\begin{thm}  \label{THM:Universal_ML_SW}
  Let $(\Rx, \Rx)$ be a rate pair such that $\Rx > H(\rvx|\rvy)$, $\Ry
  > H(\rvy|\rvx)$, and $\Rx + \Ry > H(\rvx, \rvy)$.  Then,
\begin{equation}
E_{ML, SW, x}(\Rx, \Ry) = E_{UN, SW, x}(\Rx, \Ry),
\end{equation}
and
\begin{equation}
E_{ML, SW, x}(\Rx, \Ry) = E_{UN, SW, x}(\Rx, \Ry).
\end{equation}
\end{thm}

Theorem~\ref{THM:Universal_ML_SW} follows directly from the
following lemma, shown in the  appendix.
\begin{lemma}
For all $\gamma \in [0,1]$
\begin{equation}
E^{ML}_x(R_x,R_y,\gamma)=E^{UN}_x(R_x,R_y,\gamma),
\end{equation}
and
\begin{equation}
E^{ML}_y(R_x,R_y,\gamma)=E^{UN}_y(R_x,R_y,\gamma).
\end{equation}.
\end{lemma}

{\em Remark:} This theorem allows us to simplify notation.  For
example, we can define $E_x(R_x,R_y,\gamma)$ as
$E_x(R_x,R_y,\gamma)=E^{ML}_x(R_x,R_y,\gamma)=E^{UN}_x(R_x,R_y,\gamma)$,
and can similarly define $E_y(R_x,R_y,\gamma)$.  Further, since the
ML and universal exponents are the same for the whole rate region we
can define $E_{SW,x}(\Rx, \Ry)$ as $E_{SW,x}(\Rx, \Ry) =
E_{ML,SW,x}(\Rx, \Ry) = E_{UN,SW,x}(\Rx, \Ry)$, and can similarly
define $E_{SW,y}(R_x,R_y)$.

\section{Numerical Results}\label{sec.numerical} To build insight
into the differences between the sequential error exponents of Theorem
\ref{thm.entCodeML} - \ref{THM:Universal_ML_SW} and block-coding error
exponents, we give some examples of the exponents for binary sources.

For the point-to-point case, the error exponents of random sequential
and block source coding are identical everywhere in the achievable
rate region as can be seen by comparing Theorem~\ref{thm.entCodeUniv}
and Corollary~\ref{thm.blockEnt}. The same is true for source coding
with decoder side information (cf.~Theorem~\ref{thm.decSIUniv} and
Corollary~\ref{thm.blockSI}). For distributed Slepian-Wolf source
coding however, the sequential and block error exponents can be
different. The reason for the discrepancy is that a new type of error
event can be dominant in Slepian-Wolf source coding. This is reflected
in Theorem~\ref{thm.jointCodeML} by the minimization over $\gamma$.
Example $2$ illustrates the impact of this $\gamma$ term.

For Slepian-Wolf source coding at very high rates, where $\Rx >
H(\rvx)$, the decoder can ignore any information from encoder $y$ and
still decode $x$ with with a positive error exponent.  However, the
decoder could also choose to decode source $x$ and $y$ jointly.
Fig~\ref{fig.numerical1}.a and \ref{fig.numerical1}.b illustrate that
joint decoding may or surprisingly {\em may not} help decoding source
$x$. This is seen by comparing the error exponent when the decoder
ignores the side information from encoder $y$ (the dotted curves) to
the joint error exponent (the lower solid curves). It seems that when
the rate for source $y$ is low, atypical behaviors of source $y$ can
cause joint decoding errors that end up corrupting $x$ estimates.
This holds for both block and sequential coding.

\subsection{Example 1: symmetric source with uniform marginals}

\begin{figure}
\begin{center}
\begin{picture}(100,70)
\put(40, 10){\vector(1,0){55}} \put(40, 10){\vector(0,1){55}}
\put(40,70){$\Ry$} \put(97,10){$\Rx$}
\put(90,60){\line(0,-1){27}} \put(90,33){\line(-1,1){27}}
\put(90,60){\line(-1,0){27}} \multiput(40,45)(2,0){25}{$.$}
\put(30,45 ){$0.49$   } \multiput(40,59)(2,0){25}{$.$}  \put(30,59
){$0.67$ } \put(72, 52){Achievable } \put(78, 48){Region } \put(57,
35){$\Rx+\Ry=H(\rvx,\rvy)$} \put(72, 37){\vector(1,1){6.5}}
\end{picture}
\caption{Rate region for the example 1 source, we focus on the error
exponent on source $x$ for fixed encoder $y$ rates: $R_y=0.49$
and $R_y=0.67$ } \label{fig.SWregion1}
\end{center}
\end{figure}
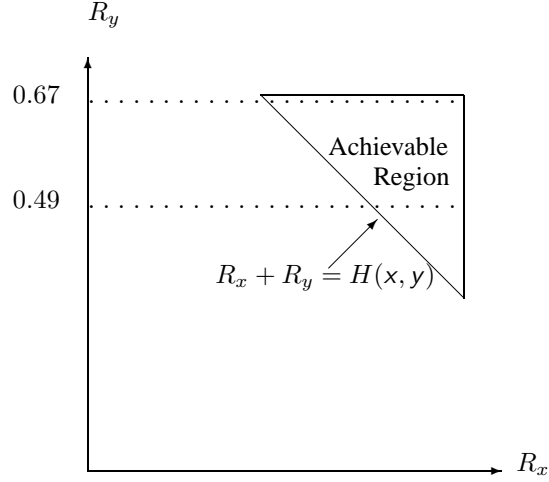

Consider a symmetric source where $|\mathcal{X}|=|\mathcal{Y}|=2$,
$p_{\rvx\rvy}(0,0)=0.45$, $p_{\rvx\rvy}(0,1)= p_{\rvx\rvy}(1,0)=0.05$
and $p_{\rvx\rvy}(1,1)=0.45$. This is a marginally-uniform source:
$\rvx$ is Bernoulli(1/2), $\rvy$ is the output from a BSC with input
$\rvx$, thus $\rvy$ is Bernoulli(1/2) as well. For this source
$H(\rvx)=H(\rvy)=\log(2)$, $H(\rvx|\rvy)=H(\rvy|\rvx)=0.32$,
$H(\rvx,\rvy)=1.02$. The achievable rate region is the triangle shown
in Figure(\ref{fig.SWregion1}).

For this source, as will be shown later, the dominant sequential
error event is on the diagonal line in Fig~\ref{fig.twoD2}. This is
to say that:
\begin{equation}
E_{SW,x}(\Rx, \Ry)= E_{SW,x}^{BLOCK}(\Rx, \Ry)=  E^{ML}_x(\Rx, \Ry,
0) = \sup_{\rho \in [0,1]} [ E_{xy}(\Rx, \Ry, \rho)].
\end{equation}

Where $E_{SW,x}^{BLOCK}(\Rx, \Ry)=\min\{E^{ML}_x(\Rx, \Ry,
0),E^{ML}_x(\Rx, \Ry, 1)\} $ as shown in \cite{gallagerTech:76}.

Similarly for source $y$:
\begin{equation}
E_{SW,y}(\Rx, \Ry)= E_{SW,y}^{BLOCK}(\Rx, \Ry)=  E^{ML}_y(\Rx, \Ry,
0) = \sup_{\rho \in [0,1]} [ E_{xy}(\Rx, \Ry, \rho)].
\end{equation}

We   first show that for this source $\forall \rho\geq 0$, $
E_{x|y}(\Rx, \rho) \geq  E_{xy}(\Rx, \Ry, \rho)$. By definition:
\begin{eqnarray}
E_{x|y}(\Rx, \rho)- E_{xy}(\Rx, \Ry, \rho) & = & \rho \Rx - \log
\Big[ \sum_{\svy} \Big[ \sum_{\svx}
p_{\rvx\rvy}(\svx,\svy)^{\frac{1}{1+\rho}} \Big]^{1+\rho}
\Big]\nonumber\\
&&-\Big(\rho (\Rx + \Ry) - \log \Big[ \sum_{\svx, \svy}
p_{\rvx\rvy}(\svx,\svy)^{\frac{1}{1+\rho}} \Big]^{1 + \rho}
\Big)\nonumber\\
& = & -\rho \Ry - \log \Big[ 2 \Big[ \sum_{\svx}
p_{\rvx\rvy}(\svx,0)^{\frac{1}{1+\rho}} \Big]^{1+\rho} \Big] + \log
\Big[ 2\sum_{\svx }
p_{\rvx\rvy}(\svx,0)^{\frac{1}{1+\rho}} \Big]^{1 + \rho}  \nonumber\\
& = & -\rho \Ry - \log \Big[ 2   \Big] + \log \Big[ 2 \Big]^{1 + \rho}  \nonumber\\
& =& \rho (\log [2] -\Ry)\nonumber\\
& \geq & 0\nonumber
\end{eqnarray}

The last inequality is true because we only consider the problem when
$\Ry \leq \log|\mathcal{Y}|$. Otherwise, $y$ is better viewed as
perfectly known side-information. Now

\begin{eqnarray}
E^{ML}_x(\Rx, \Ry, \gamma) &=& \sup_{\rho \in [0,1]} [ \gamma
E_{x|y}(\Rx, \rho) + (1-\gamma) E_{xy}(\Rx, \Ry, \rho)]\nonumber\\
&\geq & \sup_{\rho \in [0,1]} [   E_{xy}(\Rx, \Ry, \rho)]\nonumber\\
&=&   E^{ML}_x(\Rx, \Ry, 0) \nonumber
\end{eqnarray}

Similarly $E^{ML}_y(\Rx, \Ry, \gamma) \geq   E^{ML}_y(\Rx, \Ry, 0)=
E^{ML}_x(\Rx, \Ry, 0)$. Finally,
\begin{eqnarray}
E_{SW, x}(\Rx,  \Ry)& =& \min \Bigg\{ \inf_{\gamma \in [0,1]}
E_x(\Rx, \Ry, \gamma), \inf_{\gamma \in [0,1]}
\frac{1}{1-\gamma} E_y(\Rx, \Ry, \gamma) \Bigg\}\nonumber\\
&=& E^{ML}_x(\Rx, \Ry, 0)\nonumber
\end{eqnarray}

Particularly $E_x(\Rx, \Ry, 1) \geq E_x(\Rx, \Ry, 0)$, so
\begin{eqnarray}
E_{SW,x}^{BLOCK}(\Rx, \Ry) &=& \min\{E^{ML}_x(\Rx, \Ry,
0),E^{ML}_x(\Rx, \Ry, 1)\}\nonumber\\
&=& E^{ML}_x(\Rx, \Ry, 0)\nonumber
\end{eqnarray}
The same proof holds for source $y$.

\begin{figure}[htbp]
\begin{center}
\leavevmode
\includegraphics[width=100mm]{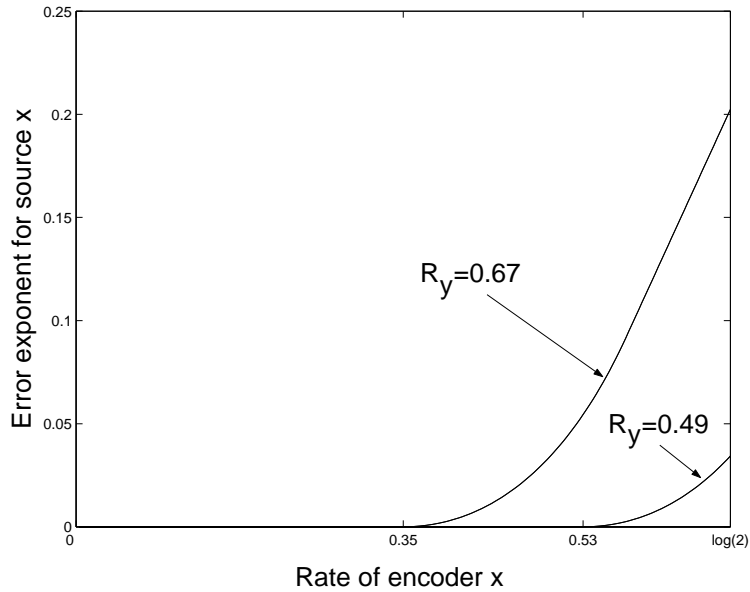}
\caption[]{ Error exponents plot:  $E_{SW,x}(\Rx, \Ry)$ plotted for $R_y=0.49$ and $R_y=0.67$\\
$E_{SW,x}(\Rx, \Ry)= E_{SW ,x}^{BLOCK}(\Rx,
  \Ry)=E_{SW,y}(\Rx, \Ry)= E_{SW,y}^{BLOCK}(\Rx, \Ry)$ and
  $E_{x}(\Rx)=0$  }
\label{fig.numerical3}
\end{center}
\end{figure}

In Fig~\ref{fig.numerical3} we plot the joint sequential/block coding
error exponents $E_{SW,x}(\Rx, \Ry)=E_{SW,x}^{BLOCK}(\Rx, \Ry)$, the
error exponents are positive iff $\Rx> H(\rvx\rvy)-\Ry=1.02-\Ry$.

\subsection{Example 2: non-symmetric source}

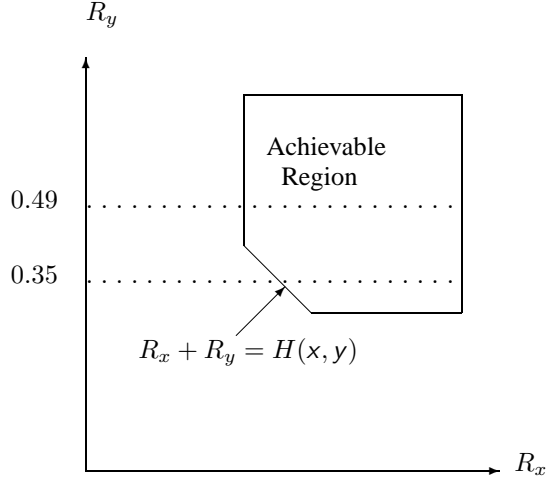
\begin{figure}
\begin{center}
\begin{picture}(100,70)
\put(40, 10){\vector(1,0){55}} \put(40, 10){\vector(0,1){55}}
\put(40,70){$\Ry$} \put(97,10){$\Rx$}
\put(90,60){\line(0,-1){29}} \put(90,31){\line(-1,0){20}}
\put(70,31){\line(-1, 1){9}} \put(90,60){\line(-1,0){29}}
\put(61,60){\line(0,-1){20}}
\multiput(40,35)(2,0){25}{$.$}
\multiput(40,45)(2,0){25}{$.$}
\put(30,35 ){$0.35$   }
\put(30,45 ){$0.49$ }
\put(64, 52){Achievable} \put(66, 48){Region}

\put(47, 25){$\Rx+\Ry=H(\rvx,\rvy)$} \put(60, 28){\vector(1,1){6.5}}
\end{picture}
\caption{ Rate region for the example 2 source, we focus on the error
exponent on source $x$ for  fixed encoder $y$ rates: $R_y=0.35$
and $R_y=0.49$  } \label{fig.SWregion}
\end{center}
\end{figure}

\begin{figure}[htbp]
\begin{center}
\leavevmode
\includegraphics[width=140mm]{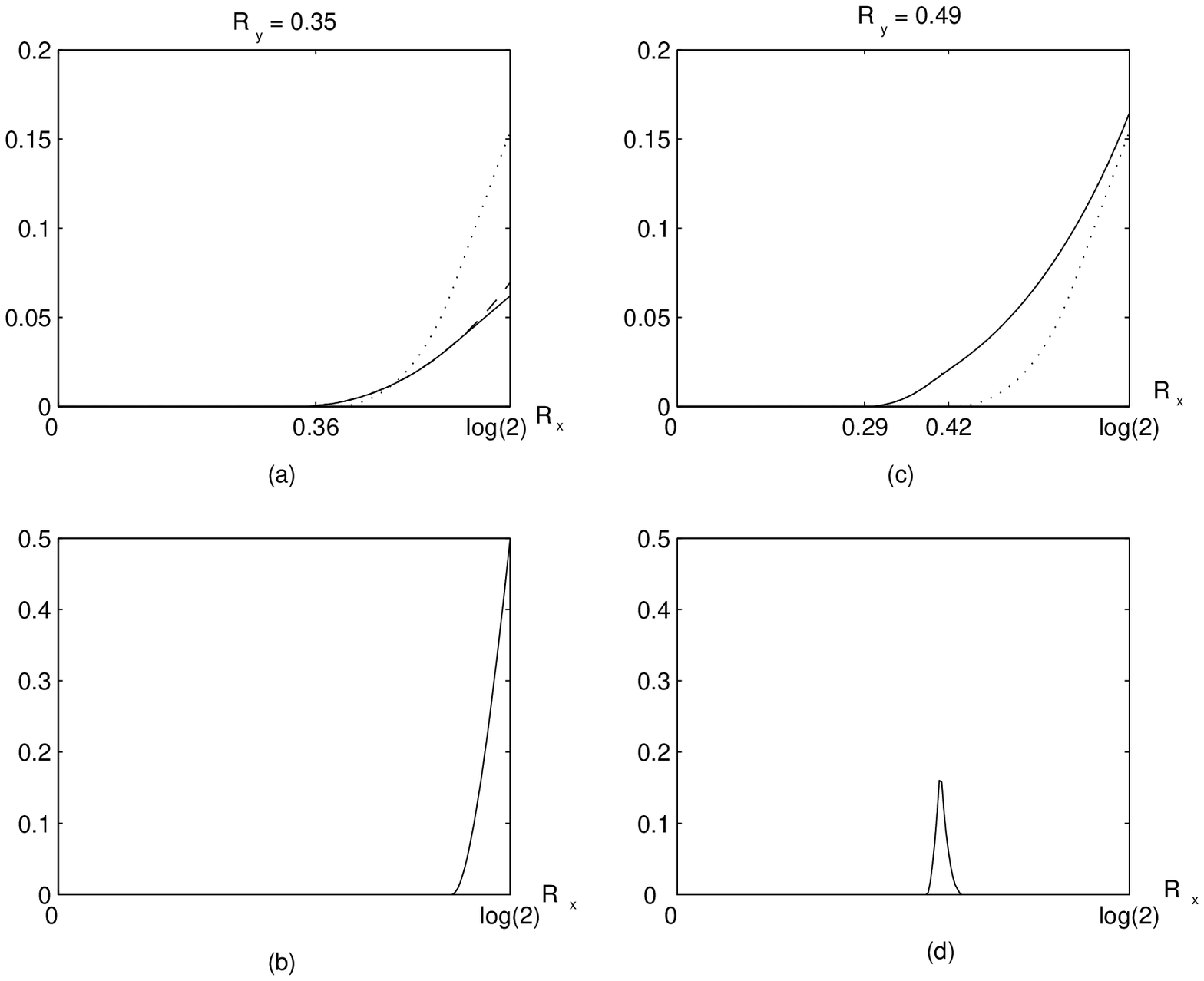}
\caption[]{ Error exponents plot for source $x$ for fixed $\Ry$ as $\Rx$ varies:\\
$\Ry=0.35$:\\(a) Solid curve: $E_{SW,x}(\Rx, \Ry)$, dashed curve
  $ E_{SW,x}^{BLOCK}(\Rx, \Ry)$ and dotted
  curve: $E_{x}(\Rx)$,  notice that $E_{SW,x}(\Rx, \Ry)\leq
  E_{SW,x}^{BLOCK}(\Rx, \Ry)$ but the difference is  small.\\(b) $10
  \log_{10}(\frac{E_{SW,x}^{BLOCK}(\Rx, \Ry)}{E_{SW,x}(\Rx, \Ry)})$. This shows the difference is there at high rates.\\
  $\Ry=0.49$:\\(c) Solid curve $E_{SW,x}(\Rx, \Ry)$, dashed
  curve $ E_{SW,x}^{BLOCK}(\Rx, \Ry)$ and
  dotted curve: $E_{x}(\Rx)$, again $E_{SW,x}(\Rx, \Ry)\leq
  E_{SW,x}^{BLOCK}(\Rx, \Ry)$ but the difference is extremely small.\\(d) $10
  \log_{10}(\frac{E_{SW,x}^{BLOCK}(\Rx, \Ry)}{E_{SW,x}(\Rx, \Ry)})$. This shows the difference is there at intermediate low rates. }
\label{fig.numerical1}
\end{center}
\end{figure}

\begin{figure}[htbp]
\begin{center}
\leavevmode
\includegraphics[width=140mm]{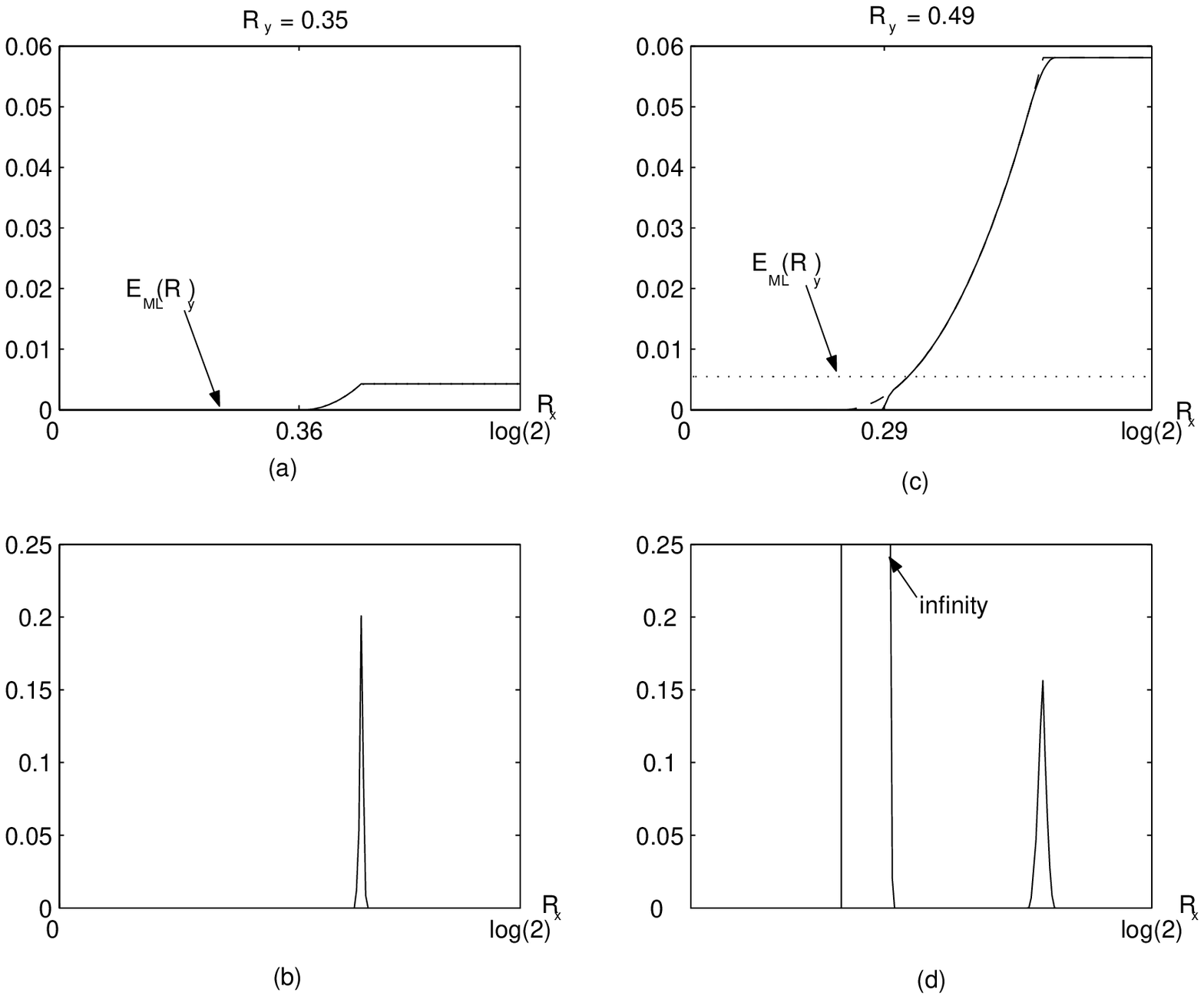}
\caption[]{ Error exponents plot for source $y$ for fixed $\Ry$ as $\Rx$ varies:\\
$\Ry=0.35$:

(a) Solid curve:  $E_{SW,y}(\Rx, \Ry)$
  and dashed curve $E_{SW,y}^{BLOCK}(\Rx, \Ry)$, $E_{SW,y}(\Rx, \Ry)\leq
  E_{SW,y}^{BLOCK}(\Rx, \Ry)$, the difference is extremely small.
  $E_{y}(\Ry)$ is $0$ because $R_y=0.35< H(\rvy)$. (b) $ 10 \log_{10}(\frac{E_{SW,y}^{BLOCK}(\Rx,
    \Ry)}{E_{SW,y}(\Rx, \Ry)})$. This shows the two exponents are not identical everywhere. \\$\Ry=0.49$:\\(c) Solid curves:
  $E_{SW,y}(\Rx, \Ry)$, dashed curve $ E_{SW,y}^{BLOCK}(\Rx, \Ry)$ and  $E_{SW,y}(\Rx,
  \Ry)\leq E_{SW,y}^{BLOCK}(\Rx, \Ry)$ and $E_{y}(\Ry)$ is constant
  shown in a dotted line.\\(d) $ 10 \log_{10}(\frac{E_{SW,y}^{BLOCK}(\Rx,
    \Ry)}{E_{SW,y}(\Rx, \Ry)})$. Notice how the gap goes to infinity when we leave the Slepian-Wolf region. }
\label{fig.numerical2}
\end{center}
\end{figure}

Consider a non-symmetric source where $|\mathcal{X}|=|\mathcal{Y}|=2$,
$p_{\rvx\rvy}(0,0)=0.1$, $p_{\rvx\rvy}(0,1)= p_{\rvx\rvy}(1,0)=0.05$
and $p_{\rvx\rvy}(1,1)=0.8$.  For this source $H(\rvx)=H(\rvy)=0.42$,
$H(\rvx|\rvy)=H(\rvy|\rvx)=0.29$ and $H(\rvx,\rvy)=0.71$. The
achievable rate region is shown in Fig~\ref{fig.SWregion}.  In
Fig~\ref{fig.numerical1}.a, \ref{fig.numerical1}.b,
\ref{fig.numerical1}.c and \ref{fig.numerical1}.d, we compare the
joint sequential error exponent $E_{SW,x}(\Rx, \Ry)$ the joint block
coding error exponent $E_{SW,x}^{BLOCK}(\Rx, \Ry)=\min\{E_x(\Rx, \Ry,
0),E_x(\Rx, \Ry, 1)\} $ as shown in \cite{gallagerTech:76} and the
individual error exponent for source $X$, $E_{x}(\Rx)$ as shown in
Corollary~\ref{thm.blockEnt}. Notice that $E_{x}(\Rx)>0$ only if $\Rx>
H(\rvx)$. In Fig~\ref{fig.numerical2}, we compare the sequential error
exponent for source $y$: $E_{SW,y}(\Rx, \Ry)$ and the block coding
error exponent for source $y$: $E_{SW,y}^{BLOCK}(\Rx, \Ry)
=\min\{E_y(\Rx, \Ry, 0),E_y(\Rx, \Ry, 1)\}$ and $E_{y}(\Ry)$ which is
a constant since we fix $\Ry$.

For $\Ry=0.35$ as shown in Fig~\ref{fig.numerical1}.a.b and
\ref{fig.numerical2}.a.b, the difference between the block coding and
sequential coding error exponents is very small for both source $x$
and $y$. More interestingly, as shown in Fig~\ref{fig.numerical1}.a,
because the rate of source $y$ is low, i.e. it is more likely to get a
decoding error due to the atypical behavior of source $y$. So as $\Rx$
increases, it is sometimes better to ignore source $y$ and decode $x$
individually. This is evident as the dotted curve is above the solid
curves.

For $\Ry=0.49$ as shown in Fig~\ref{fig.numerical1}.c.d and
\ref{fig.numerical2}.c.d, since the rate for source $y$ is high
enough, source $y$ can be decoded with a positive error exponent
individually as shown in Fig~\ref{fig.numerical2}.c. But as the rate
of source $x$ increases, joint decoding gives a better error exponent.
When $\Rx$ is very high, then we observe the saturation of the error
exponent on $y$ as if source $x$ is known perfectly to the decoder!
This is illustrated by the flat part of the solid curves in
Fig~\ref{fig.numerical2}.c.

\section{Streaming point-to-point coding via sequential random binning}
\label{sec.entropy}

In this section we prove Theorems~\ref{thm.entCodeML}
and~\ref{thm.entCodeUniv}.  While the emphasis of the paper is on
distributed source coding, the basic causal random binning ideas and
analysis techniques can be more easily developed in the point-to-point
context. 

\subsection{Maximum-likelihood decoding}
\label{sec.MLent}

To show Theorems~\ref{thm.entCodeML} and~\ref{thm.entCodeUniv}, we
first develop the common core of the proof in the context of ML
decoding.  The proof strategy is as follows.  A decoding error can
only occur if there is some spurious source sequence $\svxtil^n$ that
satisfies three conditions: (i) it must be in the same bin (share the
same parities) as $\svx^n$, i.e., $\svxtil^n \in \binX(\svx^n)$, (ii)
it must be more likely than the true sequence, i.e.,
$p_{\rvbx}(\svxtil^n) > p_{\rvbx}(\svx^n)$, and (iii) $\svxtil_{l}
\neq \svx_{l}$ for some $l \leq n - \delay$.

\begin{figure}
\setlength{\unitlength}{1mm}
 \begin{picture}(100,20)
\multiput(30,10)(5,0){14}{\circle*{1.5}}
\multiput(30,10)(5,0){9}{\oval(2,3)}
 \thicklines
 \put(75,10){\oval(2,3)}
 \thinlines
 \put(25,10){\vector(1,0){85}}
 \put(115,9){$l$}
 \put(29,5){$1$}
 \put(94,5){$n$}  \put(70,5){$n-\Delta$}
  \end{picture}
\caption{Decoding error probability at $n-\delay$ can be union
bounded by the sum of probabilities of first decoding error at $l$,
$1\leq l\leq n-\delay$. The dominant error event $p_n(n-\delay)$ is
the one in the highlighted oval(shortest delay).} \label{fig.oneD1}
\end{figure}
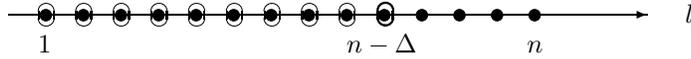

The error probability is
\begin{align}
\Pr [ \rvxhat^{n-\delay} \neq \rvx^{n-\delay}] 
= &  \sum_{\svx^n} \Pr [\rvxhat^{n-\delay} \neq \svx^{n-\delay} |
\rvx^n = \svx^n]
p_\rvbx(\svx^n) \label{eq.condSS} \\
= & \sum_{\svx^n} \sum_{l=1}^{n- \delay} \Pr \big[ \exists \;
\svxtil^n \in \mathcal{B}_x(\svx^n)\cap \mathcal{F}_n(l, \svx^n) \;
\mbox{s.t.} \; p_{\rvbx} (\svxtil^n) \geq p_{\rvbx}(\svx^n)  \big]
p_\rvbx(\svx^n) \label{eq.decomp}\displaybreak[2]\\
= & \sum_{l=1}^{n- \delay} \Big\{ \sum_{\svx^n}  \Pr \big[ \exists
\; \svxtil^n \in \mathcal{B}_x(\svx^n)\cap \mathcal{F}_n(l, \svx^n)
\; \mbox{s.t.} \; p_{\rvbx} (\svxtil^n) \geq p_{\rvbx}(\svx^n)  \big]
p_\rvbx(\svx^n) \Big\} \nonumber \\
 =&\sum_{l=1}^{n- \delay} p_n(l).
\label{eq.sufDec}
\end{align}
After conditioning on the realized source sequence
in~(\ref{eq.condSS}), the remaining randomness is only in the binning.
In~(\ref{eq.decomp}) we decompose the error event into a number of
mutually exclusive events (see Fig~\ref{fig.oneD1}) by partitioning
all source sequences $\svxtil^n$ into sets $\mathcal{F}_n(l,\svx^n)$
defined by the time $l$ of the first sample in which they differ from
the realized source $\svx^n$,
\begin{equation}
\mathcal{F}_n(l,\svx^n) =\{\svxtil^n\in
\mathcal{X}^n|\svxtil^{l-1} = \svx^{l-1}, \svxtil_{l} \neq
\svx_{l}\}, \label{eq.partition}
\end{equation}
and define $\mathcal{F}_n(n+1,\svx^n)=\{\svx^n\}$. Finally,
in~(\ref{eq.sufDec}) we define
\begin{equation}
p_n(l)= \sum_{\svx^n}  \Pr \big[ \exists
\; \svxtil^n \in \mathcal{B}_x(\svx^n)\cap \mathcal{F}_n(l, \svx^n)
\; \mbox{s.t.} \; p_{\rvbx} (\svxtil^n) \geq p_{\rvbx}(\svx^n)  \big]
p_\rvbx(\svx^n).
\label{eq.errTimeL}
\end{equation}

We now upper bound $p_n(l)$ using a Chernoff bound argument similar to
\cite{gallagerTech:76}.
\begin{lemma}\label{Lemma.indivupperbound}
$p_n(l)\leq \exp\{-(n-l+1)\expML(\Rent)\}$.
\end{lemma}

\pf 
\begin{align}
p_n(l) =&\sum_{\svx^n} \Pr \big[ \exists \; \svxtil^n \in
\mathcal{B}_x(\svx^n)\cap \mathcal{F}_n(l, \svx^n) \; \mbox{s.t.} \;
p_{\rvbx} (\svxtil^n) \geq p_{\rvbx}(\svx^n)  \big]
p_\rvbx(\svx^n)  \nonumber\\
\leq &  \sum_{\svx^n} \min \Big[1, \hspace{-1.5em} \sum_{\tiny
\begin{array}{c} \svxtil^n\in \mathcal{F}_n(l,\svx^n) \mbox{s.t.}  \\ p_{\rvbx}
(\svx^n)\leq p_{\rvbx} (\svxtil^n) \end{array}} \hspace{-1.5em} \Pr[
\svxtil^n \in \binX(\svx^n)] \Big] p_\rvbx(\svx^n)
\label{eq.setBoundViaScore} \displaybreak[2]\\
= &  \sum_{\svx^{l-1}, \svx_{l}^n}
\min \Big[1, \hspace{-1em}
\sum_{\tiny \begin{array}{c} \svxtil_{l}^n \; \mbox{s.t.}  \\
p_{\rvx }(\svx_{l}^n) < p_{\rvx }(\svxtil_{l}^n)
\end{array}}
\exp\{-(n-l + 1) \Rent\} \Big] p_\rvbx(\svx^{l-1}) p_\rvbx(\svx_l^{n})
\label{eq.randBin} \\
= &  \sum_{\svx_{l}^n}
\min \Big[1, \hspace{-1em}
\sum_{\tiny \begin{array}{c} \svxtil_{l}^n \; \mbox{s.t.}  \\
p_{\rvx }(\svx_{l}^n) < p_{\rvx }(\svxtil_{l}^n)
\end{array}}
\exp\{-(n-l +1) \Rent\} \Big]
 p_\rvbx(\svx_l^{n})\nonumber\\
= &
\sum_{\svx_{l}^n}
\min \Big[ 1, \sum_{\tiny \svxtil_{l}^n } \ind[ p_{\rvx
}(\svxtil_{l}^n) >  p_\rvbx(\svx_l^{n})] \exp\{-(n-l+1) \Rent\} \Big]
 p_\rvbx(\svx_l^{n}) \label{eq.indicator}\\
\leq &  \sum_{\svx_{l}^n}
\min \left[1, \sum_{\tiny \svxtil_{l}^n } \min \left[ 1,
\frac{p_{\rvx }(\svxtil_{l}^n)}{p_{\rvx }(\svx_{l}^n)} \right]
\exp\{-(n-l +1) \Rent\} \right]
 p_\rvbx(\svx_l^{n}) \nonumber \displaybreak[2]\\
\leq &  \sum_{\svx_{l}^n}
\left[ \sum_{\tiny \svxtil_{l}^n } \left[ \frac{p_{\rvx
}(\svxtil_{l}^n)}{p_{\rvx }(\svx_{l}^n)} \right]^{\frac{1}{1+\rho}}
\exp\{-(n-l+1) \Rent\} \right]^{\rho}
 p_\rvbx(\svx_l^{n}) \label{eq.limOnRho} \displaybreak[2]\\
= &  \sum_{\svx_{l}^n}  p_\rvbx(\svx_l^{n})^{\frac{1}{1+\rho}}
\left[ \sum_{\tiny \svxtil_{l}^n } \left[ p_{\rvx }(\svxtil_{l}^n)
\right]^{\frac{1}{1+\rho}}\right]^{\rho}
\exp\{-(n-l +1) \rho \Rent\} \nonumber \displaybreak[2] \\
= &  \left[\sum_{\svx} \PxRV(x)^{\frac{1}{1+\rho}}\right]^{(n-l+1)}
\left[
\sum_{\svx} \PxRV(\svx)^{\frac{1}{1+\rho}}\right]^{(n-l+1)\rho}
\exp\{-(n-l +1) \rho \Rent\} \label{eq.iid} \displaybreak[2]\\
= &  \left[ \sum_{\svx}
\PxRV(\svx)^{\frac{1}{1+\rho}}\right]^{(n-l+1)(1+\rho)}
\exp\{-(n-l+1) \rho \Rent\} \nonumber\\
= & \exp\left\{-(n-l+1) \left[\rho \Rent - (1+\rho) \ln
\left(\sum_{\svx} \PxRV(\svx)^{\frac{1}{1+\rho}}\right) \right]
\right\}. \label{eq.rhoBnd}
\end{align}

In~(\ref{eq.setBoundViaScore}) the union bound is applied.
In~(\ref{eq.randBin}) we use the fact that after the first symbol in
which two sequences differ, the remaining parity bits are independent,
and the fact that only the likelihood of the differing suffixes
matter.  That is, if $\svx^{l-1} = \svxtil^{l-1}$, then $p_{\rvbx}
(\svx^n)< p_{\rvbx} (\svxtil^n)$ if and only if $p_{\rvbx}(\svx_{l}^n)
< p_{\rvbx}(\svxtil_{l}^n)$.  In~(\ref{eq.indicator}) $\ind(\cdot)$ is
the indicator function, taking the value one if the argument is true,
and zero if it is false. We get~(\ref{eq.limOnRho}) by limiting $\rho$
to the range $0 \leq \rho \leq 1$ since the arguments of the
minimization are both positive and upper-bounded by one. We use the
iid property of the source, exchanging sums and products to
get~(\ref{eq.iid}).  The bound in~(\ref{eq.rhoBnd}) is true for all
$\rho$ in the range $0 \leq \rho \leq 1$. Maximizing~(\ref{eq.rhoBnd})
over $\rho$ gives $p_n(l)\leq \exp\{-(n-l+1)\expML(\Rent)\}$ where
$\expML(\Rent)\}$ is defined in Theorem~\ref{thm.entCodeML}, in
particular~(\ref{eq.errExpML}). \hfill$\blacksquare$

Using Lemma~\ref{Lemma.indivupperbound} in~(\ref{eq.sufDec}) gives
\begin{align}
\Pr [ \rvxhat^{n-\delay} \neq \rvx^{n-\delay}] 
\leq & \sum_{l=1}^{n-\delay} \exp\{- (n-l+1) E_{ML}(\Rent)\}
\label{eq.delayTerm}\\
= & \sum_{l=1}^{n-\delay} \exp\{- (n-l+1-\delay) E_{ML}(\Rent)\}
\exp\{- \delay E_{ML}(\Rent)\} \nonumber \\
\leq & K_0 \exp\{- \delay E_{ML}(\Rent)\} \label{eq.pullOutExp}
\end{align}
In~(\ref{eq.pullOutExp}) we pull out the exponent in $\delay$.  The
remaining summation is a sum over decaying exponentials, can thus
can be bounded by some constant $K_0$. This proves Theorem~\ref{thm.entCodeML}.

\subsection{Error events and sequential decoding}
\label{sec.entMLseq}

To better understand the dominant error event in the
sum~(\ref{eq.delayTerm}), consider constructing the ML estimate in a
symbol-by-symbol sequential manner.  The decoder starts by first
identifying as candidates those sequences whose parities match the
received bit stream up to time $n$.  If the encoder observes the
length-$n$ sequence $\rvbx = \svbx$, this is $\{ \svbxBar \;
\mbox{s.t.} \; \svbxBar \in \binX(\svbx)\}$.  The $l$th symbol of the
estimate, $\rvxhat_l$, is defined as
\begin{equation}
\svxhat_l = \svw_l \;\;\; \mbox{where} \;\;\;
\svbw = \argmax_{\svbxBar \in \binX(\svbx) \;\; \mbox{s.t.} \;\;
\svxBar^{l-1} = \svxhat^{l-1}} p_{\rvx_{l}^n}(\svxBar_{l}^n).
\label{eq.defSeqDec}
\end{equation}
The estimate thus produced is the maximum likelihood estimate because
the decision regarding which pair of sequences is more likely depends
only on which one's suffix is more likely.

This is a decision-directed decoder. Semi-hard\footnote{Decisions are
  only ``hard'' for computational time. As soon as the next set of
  parities arrive and real-time advances, all the computations are
  done again.} estimate are made sequentially for each symbol.  These
estimates are then fixed, and taken as true when estimating subsequent
symbols.  Each such hard-decision is analogous to a classic
block-coding Slepian-Wolf problem. This is because we only need to
decide between sequences that start to differ in the symbol we are
trying to estimate---previous symbols have been fixed, and subsequent
symbols are not yet in question.  Thus, all sequences that could lead
to different estimates of symbol $l$ are binned independently for the
remainder of the block.  This is why the error exponent we derive
in~(\ref{eq.pullOutExp}) equals Gallager's block coding
exponent~\cite{gallagerTech:76}.  Since the error exponent for each
block-decoding problem is the same, the dominant error event is the
hard-decision with the shortest block-length.  This symbol is the last
symbol we need to estimate.  Its block-length equals the estimation
delay $\delay$. We revisit this story in Section~\ref{sec.SW} when we
consider Slepian-Wolf coding.  In that context the dominant error
event has some features that do not arise in block coding.

\subsection{Universal decoding}
\label{sec.univEnt}

In this section we prove Theorem~\ref{thm.entCodeUniv}.  We use the
sequential decoder introduced in Section~\ref{sec.entMLseq}, but with
minimum-entropy, rather than maximum-likelihood, decoding.  That is,
\begin{equation}
\svxhat_l = \svw_l[l] \;\;\; \mbox{where} \;\;\; \svw^n[l] =
\argmin_{\svxBar^n \in \binX(\svx^n) \;\; \mbox{s.t.} \;\;
\svxBar^{l-1} = \svxhat^{l-1}} H(\svxBar_{l}^n).
\label{eq.defSeqUniv}
\end{equation}
We term this a minimum suffix-entropy decoder.  The reason for using
this decoder instead of the standard minimum block-entropy decoder is
that the block-entropy decoder has a polynomial term in $n$ (resulting
from summing over the type classes) that multiplies the exponential
decay in $\delay$.  For $n$ large, this polynomial can dominate.
Using the minimum suffix-entropy decoder results in a polynomial term
in $\delay$.

With this decoder, errors can only occur if there is some sequence
$\svxtil^n$ such that (i) $\svxtil^n \in \binX(\svx^n)$, (ii)
$\rvxtil^{l-1} = \rvx^{l-1}$, and $\rvxtil_l \neq \rvx_l$, for some $l
\leq n-\delay$, and (iii) the empirical suffix entropy of
$\svxtil_l^n$ is such that $H(\rvxtil_{l}^n) < H(\svx_l^n)$.  Building
on the common core of the
achievability~(\ref{eq.condSS})--(\ref{eq.sufDec}) with the
substitution of universal decoding in the place of maximum likelihood
results in the following definition of $p_n(l)$ (cf.~(\ref{eq.pnUniv})
with~(\ref{eq.errTimeL}),

\begin{align}
p_n(l)=\sum_{\svx^n}  \Pr \big[ \exists \; \svxtil^n \in
\mathcal{B}_x(\svx^n)\cap \mathcal{F}_n(l, \svx^n) \; \mbox{s.t.} \;
 H(\svxtil_{l}^n) \leq H(\svx_{l}^n)  \big]
p_\rvbx(\svx^n) \label{eq.pnUniv}
\end{align}

The following lemma gives a bound on $p_n(l)$.
\begin{lemma}\label{Lemma.indivUniv}
  For minimum suffix-entropy decoding, $p_n(l)\leq (n-l+2)^{2|\cX|}
  \exp\{-(n - l +1) E_{UN}(\Rent)\}.$
\end{lemma}

\pf We define $\PNL$ to be the type of length-$(n-l+1)$ sequence
$x_{l}^n$, and $\tclass_{\PNL}$ to be the corresponding type class so
that $x_{l}^n \in \tclass_{\PNL}$. Analogous definitions hold for
$\PtilNL$ and $\tilde{x}_{l}^n$.  We rewrite the constraint
$H(\svxtil_{l}^n) < H(\svxtil_{l}^n)$ as $H(\PtilNL) < H(\PNL)$.
Thus,
\begin{align}
 p_n(l)=&\sum_{\svx^n} \Pr \big[ \exists \; \svxtil^n \in
\mathcal{B}_x(\svx^n)\cap \mathcal{F}_n(l, \svx^n) \; \mbox{s.t.} \;
 H(\svxtil_{l}^n) \leq H(\svx_{l}^n)  \big]
p_\rvbx(\svx^n)\nonumber\\
\leq & %
\sum_{\svx_{1}^n}
\min \Big[1, \hspace{-1em}
\sum_{\tiny \begin{array}{c} \svxtil_{1}^n \in
\mathcal{F}_n(l,\svx^n)\; \mbox{s.t.}  \\
H(\svxtil_{l}^n) \leq H(\svx_{l}^n) \end{array}}
 \Pr[\svxtil_{1}^n\in \mathcal{B}_x(\svx_{1}^n)] \Big]
p_\rvbx(\svx^n)\nonumber\\
=& %
\sum_{\svx_{1}^{l-1},\svx_{l}^{n}}
\min \Big[1, \hspace{-1em}
\sum_{\tiny \begin{array}{c} \svxtil_{l}^n  \; \mbox{s.t.}  \\
H(\svxtil_{l}^n) \leq H(\svx_{l}^n) \end{array}}
 \exp\{-(n-l +1) \Rent\}  \Big] p_\rvbx(\svx^{l-1})p_\rvbx(\svx_l^{n})
\nonumber\\
= & %
\sum_{\svx_{l}^n}
\min \Big[1, \hspace{-1em}
\sum_{\tiny \begin{array}{c} \svxtil_{l}^n \; \mbox{s.t.}  \\
H(\svxtil_{l}^n) \leq H(\svx_{l}^n) \end{array}}
\exp\{-(n-l +1) \Rent\} \Big] p_\rvbx(\svx_l^{n})\label{eq.nonBlock}\\
=&
\sum_{\PNL}
\sum_{\tiny \svx_{l}^n \in \tclass_{\PNL}}
\min \Big[ 1, \hspace{-2em}
\sum_{\tiny \begin{array}{c}\PtilNL \; \mbox{s.t.}\\
H (\PtilNL) \leq H(\PNL) \end{array}}
%
\sum_{\tilde{x}_{l}^n \in \tclass_{\PtilNL}} \exp\{-(n-l+1) \Rent\}
\Big] p_\rvbx(\svx_l^{n})
\label{eq.tildeType} \displaybreak[2]\\
\leq &
\sum_{\PNL}
\sum_{\tiny \svx_{l+1}^n \in \tclass_{\PNL}}
\min \Big[ 1, (n-l+2)^{|\cX|}
%
\exp\{-(n-l) [\Rent - H(\PNL)]\} \Big]
p_\rvbx(\svx_l^{n})\label{eq.entBnd} \displaybreak[2]\\
\leq&    (n-l+2)^{|\cX|}
\sum_{\PNL} \sum_{\svx_{l}^n \in \tclass_{\PNL}}
\exp\{-(n-l+1) [ |\Rent \! - \! H(\PNL)|^{+} ]\} \nonumber\\
&\hspace{1in} \exp\{-(n-l+1) [D(\PNL \| \PxRV) + H(\PNL)]\}
\label{eq.incExp} \displaybreak[2]\\
\leq&  (n-l+2)^{|\cX|}
\sum_{\PNL}
\exp\{-(n-l+1)
\inf_{q}[D(q \| \PxRV)  + |\Rent -
H(q)|^{+}]\} \label{eq.optErrExp} \displaybreak[2]\\
\leq&  (n-l+2)^{2|\cX|}
\exp\{-(n - l +1) E_{UN}(\Rent)\} \label{eq.defEr} \displaybreak[2]
\end{align}
In going from~(\ref{eq.tildeType}) to~(\ref{eq.entBnd}) first note
that the argument of the inner-most summation (over $\svxtil_{l}^n$)
does not depend on $\svbx$.  We then use the following relations: (i)
$\sum_{\svxtil_{l}^n \in \tclass_{\PtilNL}} = |\tclass_{\PtilNL}| \leq
\exp\{(n-l+1) H(\PtilNL)\}$, which is a standard bound on the size of
the type class, (ii) $H(\PtilNL) \leq H(\PNL)$ by the
minimum-suffix-entropy decoding rule, and (iii) the polynomial bound
on the number of types, $|\{\PtilNL\}| \leq (n-l+2)^{|\cX|}$.
In~(\ref{eq.incExp}) we recall the function definition $|\cdot|^+
\defeq \max\{0, \cdot\}$.  We pull the polynomial term out of the
minimization and use $p_\rvbx(\svx_l^{n}) = \exp\{-(n-l+1) [ D(\PNL \|
\PxRV) + H(\PNL)]\}$ for all $p_\rvbx(\svx_l^{n}) \in \tclass_{\PNL}$.
It is also in~(\ref{eq.incExp}) that we see why we use a minimum
suffix-entropy decoding rule instead of a minimum entropy decoding
rule.  If we had not marginalized out over $\svx^{l-1}$ in
~(\ref{eq.nonBlock}) then we would have a polynomial term out front in
terms of $n$ rather than $n-l$, which for large $n$ could dominate the
exponential decay in $n-l$.  As the expression in~(\ref{eq.optErrExp})
no longer depends on $\svx_{l}^n$, we simplify by using
$|\tclass_{\PNL}| \leq \exp\{(n-l+1) H(\PNL)\}$.  In~(\ref{eq.defEr})
we use the definition of the universal error exponent $E_{UN}(\Rent)$
from~(\ref{eq.errExpUniv}) of Theorem~\ref{thm.entCodeUniv}, and the
polynomial bound on the number of types.  \hfill $\blacksquare$

Lemma~\ref{Lemma.indivUniv} and  $\Pr [   \rvxhat^{n-\delay} \neq
\rvx^{n-\delay}]\leq \sum_{l=1}^{n-\delay} p_n(l)$ imply that:
\begin{align}
\Pr [   \rvxhat^{n - \delay} \neq \rvx^{n - \delay}]  \leq&
\sum_{l=1}^{n-\delay}  (n-l+2)^{2|\cX|}
 \exp\{-(n - l +1) E_{UN}(\Rent)\}\nonumber\\
 \leq&
\sum_{l=1}^{n-\delay} K_1
\exp\{-(n -l + 1 ) [E_{UN}(\Rent) - \gamma]\}
\label{eq.polyIntoExp} \displaybreak[2]\\
\leq & K_2 \exp\{ - \delay [E_{UN}(\Rent) - \gamma] \}
\label{eq.entErrExp}
\end{align}
 In~(\ref{eq.polyIntoExp}) we
incorporate the polynomial into the exponent. Namely, for all $a
>0$, $b>0$, there exists a $C$ such that $z^a \leq C \exp \{b (z
-1)\}$ for all $z \geq 1$.

We then   make explicit the delay-dependent term.  Pulling out the
exponent in $\delay$, the remaining summation is a sum over decaying
exponentials, and can be bounded by a constant. Together with $K_1$,
this gives the constant $K_2$ in~(\ref{eq.entErrExp}). This proves
Theorem~\ref{thm.entCodeUniv}.  Note that the $\gamma$
in~(\ref{eq.entErrExp}) does not enter the optimization because
$\gamma > 0$ can be picked equal to any constant.  The choice of
$\gamma$ effects the constant $K$ in Theorem~\ref{thm.entCodeUniv}.


\section{Streaming source coding with side information at the
decoder}
\label{sec.incDecSI}

If a random sequence $\rvy^n$, related to the source $\rvx^n$ through
a discrete memoryless channel, is observed at the decoder, then this
side information can be used to reduce the rate of the source code.
In this model $p_{\rvbx, \rvby}(\svx^n, \svy^n) = \prod_{i=1}^n \PxyRV
(\svx_i, \svy_i) = \prod_{i=1}^n \PxCondyRV (\svx_i | \svy_i)
\PyRV(\svy_i)$.  The source $\rvx^n$ is observed at the encoder, and
the decoder, which observes $\rvy^n$ and a bit stream from the
encoder, wants to estimate each source symbol $\rvx_i$ with a
probability of error that decreases exponentially in the decoding
delay $\delay$.

We can apply the analysis of Section~\ref{sec.entropy} to this problem
with a few minor modifications. For ML decoding, we need to pick the
sequence with the maximum conditional probability given $\rvy^n$. The
error exponent can be derived using a similar Chernoff bounding
argument as in section ~\ref{sec.entropy}.  For universal decoding,
the only change is that we now use a minimum suffix
conditional-entropy decoder that compares sequence pairs $(\svxBar^n,
\svy^n)$ and $(\svxBBar^n, \svy^n)$. In terms of the analysis, one
change enters in~(\ref{eq.condSS}) where we must also sum over the
possible side information sequences. And in~(\ref{eq.tildeType}) the
entropy condition in the summation over $\svbxtil$ changes to
$H(\svxtil_{l+1}^n|\svy_{l+1}^n) < H(\svx_{l+1}^n| \svy_{l+1}^n)$ (or
the equivalent type notation).  Since there is no ambiguity in the
side information, since $\rvy^n$ is observed at the decoder, this
condition is equivalent to $H(\svxtil_{l+1}^n, \svy_{l+1}^n) <
H(\svx_{l+1}^n, \svy_{l+1}^n)$.

These results are summarized in Theorems~\ref{thm.decSIML}
and~\ref{thm.decSIUniv}.  We do not include the full derivation of
these theorems as no new ideas are required.

\section{Streaming Slepian-Wolf source coding}
\label{sec.SW}

In this section we provide the proofs of
Theorems~\ref{thm.jointCodeML} and~\ref{thm.jointCode}, which consider
the two-user\footnote{The multiuser case is essentially the same, just
  with a lot more notation and minimization parameters
  $\gamma_1,\gamma_2,\ldots$.} Slepian-Wolf problem. As with the
proofs of Theorems~\ref{thm.entCodeML} and~\ref{thm.entCodeUniv} in
Sections~\ref{sec.MLent} and~\ref{sec.univEnt}, we start by developing
the common core of the proof in the context of maximum likelihood
decoding.  This allows us to develop the results for universal
decoding more quickly and transparently.  Furthermore, as shown in
Theorem~\ref{THM:Universal_ML_SW}, maximum likelihood decoding and
universal decoding provide the same reliability with delay.

\subsection{Maximum Likelihood Decoding}
\label{sec.MLSW}

In Theorems~\ref{thm.jointCodeML} and~\ref{thm.jointCode} three error
events are considered: (i) $\Pr[\rvx^{n - \delay} \neq
\rvxhat^{n-\delay}]$, (ii) $\Pr[\rvy^{n - \delay} \neq
\rvyhat^{n-\delay}]$, and (iii) $\Pr[(\rvx^{n - \delay}, \rvy^{n -
  \delay}) \neq (\rvxhat^{n-\delay}, \rvyhat^{n-\delay})]$.  We
develop the error exponent for case (i).  The error exponent for case
(ii) follows from a similar derivation, and that of case (iii) from an
application of the union bound resulting in an exponent that is the
minimum of the exponents of cases (i) and (ii).

To lead to the decoding error $\Pr[\rvx^{n - \delay} \neq
\rvxhat^{n-\delay}]$ there must be some spurious source pair
$(\svxtil^n, \svytil^n)$ that satisfies three conditions: (i)
$\svxtil^n \in \binX(\svx^n)$ and $\svytil^n \in \binY(\svy^n)$, (ii)
it must be more likely than the true pair $p_{\rvbx, \rvby}(\svxtil^n,
\svytil^n) > p_{\rvbx, \rvby}(\svx^n, \svy^n)$, and (iii) $\svxtil_{l}
\neq \svx_{l}$ for some $l \leq n - \delay$.

The error probability is
\begin{align}
\Pr[&\rvxhat^{n-\delay} \neq \rvx^{n-\delay}] 
= \sum_{\svx^n, \svy^n} \Pr [\rvxhat^{n-\delay} \neq \svx^{n-\delay}
| \rvx^n = \svx^n, \rvy^n = \svy^n]
p_{\rvbx,\rvby}(\svx^n, \svy^n) \nonumber\\
&\leq \sum_{\svx^n, \svy^n} p_{\rvbx,\rvby}(\svx^n, \svy^n)\Big\{
 \sum_{l=1}^{n - \delay}
\sum_{k=1}^{n+1}   \nonumber \\
&   \hspace{0.75in}
 \Pr \big[ \exists \; (\svxtil^n, \svytil^n) \in
\binX(\svx^n) \times \binY(\svy^n)\cap
\mathcal{F}_n(l,k,\svx^n, \svy^n) \; \mbox{s.t.} \;
  p_{\rvbx,\rvby}(\svxtil^n, \svytil^n) \geq
p_{\rvbx,\rvby}(\svx^n, \svy^n)\big]  \Big\}
\label{eq.diffTime} \\
& = \sum_{l=1}^{n - \delay} \sum_{k=1}^{n+1} \Big\{ \sum_{\svx^n,
\svy^n} p_{\rvbx,\rvby}(\svx^n, \svy^n)\nonumber \\
&   \hspace{0.75in}
 \Pr \big[ \exists \; (\svxtil^n, \svytil^n)\in
\binX(\svx^n)\times \binY(\svy^n)\cap \mathcal{F}_n(l,k,\svx^n,
\svy^n) \; \mbox{s.t.} \;
    p_{\rvbx,\rvby}(\svxtil^n, \svytil^n) \geq
p_{\rvbx,\rvby}(\svx^n, \svy^n) \big]  \Big\}
\nonumber\\
= & \sum_{l=1}^{n - \delay} \sum_{k=1}^{n+1} p_n(l,k).
\label{eq.defPn}
\end{align}
In~(\ref{eq.diffTime}) we decompose the error event into a number of
mutually exclusive events by partitioning all source pairs
$(\svxtil^n, \svytil^n)$ into sets $\mathcal{F}_n(l, k,\svx^n,
\svy^n)$ defined by the times $l$ and $k$ at which $\svxtil^n$ and
$\svytil^n$ diverge from the realized source sequences.  The set
$\mathcal{F}_n(l, k,\svx^n, \svy^n)$ is defined as
\begin{equation}
 \mathcal{F}_n(l,k,x^n,y^n)=\{(\svxBar^n,\svytil^n)\in
\mathcal{X}^{n} \times\mathcal{Y}^{n} \; \mbox{s.t.} \;
\svxBar^{l-1} = x^{l-1},\svxBar_l \neq x_l,\svyBar^{k-1}= y^{k-1},
\svyBar_k\neq y_k\}, \label{eq.jointPart}
\end{equation}
In contrast to streaming point-to-point or side-information coding
(cf.~(\ref{eq.jointPart}) with~(\ref{eq.partition})), the partition is
now doubly-indexed.  To find the dominant error event, we must search
over both indices.  Having two dimensions to search over results in an
extra minimization when calculating the error exponent (and leads to
the infimum over $\gamma$ in Theorem~\ref{thm.jointCodeML}).

Finally, to get~(\ref{eq.defPn}) we define $p_n(l,k)$ as
\begin{eqnarray*}
& & p_n(l,k) \\
&=& \sum_{\svx^n, \svy^n} p_{\rvbx,\rvby}(\svx^n, \svy^n) \Pr \Big[
\exists \; (\svxtil^n, \svytil^n)\in \binX(\svx^n)\times
\binY(\svy^n)\cap \mathcal{F}_n(l,k,\svx^n, \svy^n) \; \mbox{s.t.} \;
    p_{\rvbx,\rvby}(\svxtil^n, \svytil^n) \geq
p_{\rvbx,\rvby}(\svx^n, \svy^n)\Big]. 
\end{eqnarray*}
The following lemma provides an upper bound on $p_n(l,k)$:
\begin{lemma} \label{lemm.jointPn}
\begin{equation}
\begin{array}{lllll}
p_n(l,k) & \leq & \exp\{-(n-l+1) E_x(\Rx, \Ry, \frac{k-l}{n-l+1})\} &
\mbox{if} & l \leq k, \vspace{1ex} \\
p_n(l,k) & \leq & \exp\{-(n-k+1) E_y(\Rx, \Ry, \frac{l-k}{n-k+1})\} &
\mbox{if} & l \geq k,
\end{array} \label{eq.mlSWbnd}
\end{equation}
where $E_x(\Rx, \Ry, \gamma)$ and $E_y(\Rx, \Ry, \gamma)$ are
defined in ~(\ref{eq.compoundExp}) and~(\ref{eq.defBasicExp})
respectively. Notice that $l,k \leq n$, for $l\leq k$: $
\frac{k-l}{n-l+1}\in [0,1]$ serves as $\gamma$ in the error exponent
$E_x(\Rx, \Ry, \gamma)$. Similarly for  $l\geq k$.
\end{lemma}

\pf The bound depends on whether $l \leq k$ or $l \geq k$. Consider
the case for $l \leq k$,
\begin{align}
& p_n(l,k) \nonumber \\
&=\sum_{\svx^n, \svy^n} p_{\rvbx,\rvby}(\svx^n, \svy^n) \Pr[
\exists \; (\svxtil^n, \svytil^n)\in \binX(\svx^n)\times
\binY(\svy^n)\cap \mathcal{F}_n(l,k,\svx^n, \svy^n) \; \mbox{s.t.} \;
    p_{\rvbx,\rvby}(\svx^n, \svy^n) < p_{\rvbx,\rvby}(\svxtil^n,
\svytil^n)]\nonumber\\
&\leq  \sum_{\svx^n, \svy^n}
\min\Big[1, \sum_{\tiny \begin{array}{c} (\svxtil^n, \svytil^n) \in
\mathcal{F}_n(l,k,\svx^n, \svy^n)\; \\
  p_{\rvbx,\rvby}(\svx^n, \svy^n) <
p_{\rvbx,\rvby}(\svxtil^n, \svytil^n)
\end{array}}
\Pr[ \svxtil^n \in \binX(\svx^n), \svytil^n \in \binY(\svy^n)]\Big]
p_{\rvbx,\rvby}(\svx^n, \svy^n)  \label{eq.enumJoint} \displaybreak[2]\\
&\leq     \sum_{\svx_l^n, \svy_l^n}
\min \Big[1, \sum_{\tiny \begin{array}{c} (\svxtil_l^n,
    \svytil_l^n) \; \mbox{s.t.} \;   \svytil^{k-1}=\svy^{k-1}  \; \\
    p_{\rvbx,\rvby}(\svx_l^n, \svy_l^n) < p_{\rvbx,\rvby}(\svxtil_l^n,
    \svytil_l^n)
\end{array}}
\exp\{-(n-l +1) \Rx -(n-k+1) \Ry\} \Big]
p_{\rvbx,\rvby}(\svx_l^n,\svy_l^n)  \label{eq.indepBin} \\
&= \sum_{\svx_l^n, \svy_l^n}
\min \Big[1, \sum_{\svxtil_l^n, \svytil_k^n}
\exp\{-(n-l+1) \Rx -(n-k+1) \Ry\} \nonumber \\\
& \hspace{0.75in} \ind [ p_{\rvbx,\rvby}(\svxtil_l^{k-1}, \svy_l^{k-1})
p_{\rvbx,\rvby}(\svxtil_k^{n}, \svytil_k^{n}) > p_{\rvbx,\rvby}
(\svx_l^{n}, \svy_l^{n})]
\Big] p_{\rvbx,\rvby}(x_l^n,y_l^n) \nonumber \\
&\leq
\sum_{\svx_l^n, \svy_l^n}
\min  \Bigg[1, \sum_{\svxtil_l^n, \svytil_k^n}
\exp\{-(n-l+1) \Rx -(n-k+1) \Ry\} \nonumber \\\
& \hspace{0.5in}  \min \Bigg[1, \frac{p_{\rvbx,\rvby}(\svxtil_l^{k-1},
\svy_l^{k-1}) p_{\rvbx,\rvby} (\svxtil_k^{n}, \svytil_k^{n})}{
p_{\rvbx,\rvby}(\svx_l^{n}, \svy_l^{n})} \Bigg] \Bigg]
p_{\rvbx,\rvby}(\svx_l^{n}, \svy_l^{n}) \nonumber
\displaybreak[2] \\
&\leq
\sum_{\svx_l^n, \svy_l^n}
\Bigg[\sum_{\svxtil_l^n, \svytil_k^n}
e^{-(n-l+1) \Rx -(n-k+1) \Ry}
 \Bigg[
\frac{p_{\rvbx,\rvby}(\svxtil_l^{k-1}, \svy_l^{k-1}) p_{\rvbx,\rvby}
(\svxtil_k^{n}, \svytil_k^{n})}{ p_{\rvbx,\rvby}(\svx_l^{n},
\svy_l^{n})} \Bigg]^{\frac{1}{1+\rho}} \Bigg]^{\rho}
p_{\rvbx,\rvby}(\svx_l^{n}, \svy_l^{n})
\displaybreak[2] \label{eq.gallagerRho} \\
&=
e^{-(n-l+1) \rho \Rx -(n-k+1) \rho \Ry}
\sum_{\svx_l^n, \svy_l^n}
\Bigg[\sum_{\svxtil_l^n, \svytil_k^n}
[p_{\rvbx,\rvby}(\svxtil_l^{k-1}, \svy_l^{k-1}) p_{\rvbx,\rvby}
(\svxtil_k^{n}, \svytil_k^{n}) ]^{\frac{1}{1+\rho}} \Bigg]^{\rho}
p_{\rvbx,\rvby}(\svx_l^n,\svy_l^n)^{\frac{1}{1+\rho}}
\nonumber \displaybreak[2]\\
&= e^{-(n-l+1) \rho \Rx -(n-k+1) \rho \Ry} \sum_{\svy_l^{k-1}}
\Big[ \sum_{\svx_l^{k-1}}
p_{\rvbx,\rvby}(\svx_l^{k-1},\svy_l^{k-1})^{\frac{1}{1+\rho}}\Big]
\Big[\sum_{\svxtil_l^{k-1}}
p_{\rvbx,\rvby} (\svxtil_l^{k-1}, \svy_l^{k-1})^{\frac{1}{1+\rho}}
\Big]^{\rho}
\nonumber \\
& \hspace{0.5in} \Big[ \sum_{\svxtil_k^n, \svytil_k^n} p_{\rvbx,\rvby}
(\svxtil_k^{n}, \svytil_k^{n})^{\frac{1}{1+\rho}} \Big]^{\rho}
\sum_{\svx_k^n, \svy_k^n} p_{\rvbx,\rvby}(\svx_k^{n},
\svy_k^{n})^{\frac{1}{1+\rho}}
\nonumber \displaybreak[2] \\
&= e^{-(n-l+1) \rho \Rx -(n-k+1) \rho \Ry}
\Bigg[\sum_{\svy_l^{k-1}} \Big[ \sum_{\svx_l^{k-1}}
p_{\rvbx,\rvby}(\svx_l^{k-1}, \svy_l^{k-1})^{\frac{1}{1+\rho}} \Big]^{1
+ \rho} \Bigg]
\Big[\sum_{\svx_k^n, \svy_k^n}
p_{\rvbx,\rvby}(\svx_k^{n}, \svy_k^{n})^{\frac{1}{1+\rho}} \Big]^{1
+\rho}
\nonumber \displaybreak[2]\\
&= e^{-(n-l+1) \rho \Rx -(n-k+1) \rho \Ry} \Bigg[\sum_{\svy} \Big[
\sum_{\svx} p_{\rvx,\rvy}(\svx,\svy)^{\frac{1}{1+\rho}} \Big]^{1 +
\rho} \Bigg]^{k-l}
\Big[\sum_{\svx, \svy}
p_{\rvx,\rvy}(\svx,\svy)^{\frac{1}{1+\rho}} \Big]^{(1 +\rho)(n-k+1)}
\label{eq.rearranging}\displaybreak[2] \\
&= \exp\left\{-(k-l) \Bigg[ \rho \Rx - \log \Big[ \sum_{\svy} \Big[
\sum_{\svx} p_{\rvx,\rvy}(\svx,\svy)^{\frac{1}{1+\rho}}
\Big]^{1+\rho} \Big] \Bigg]
\right\} \nonumber \\
& \hspace{0.5in} \exp\left\{ -(n-k+1) \Bigg[ \rho (\Rx + \Ry) -
(1+\rho) \log \Big[ \sum_{\svx, \svy}
p_{\rvx,\rvy}(\svx,\svy)^{\frac{1}{1+\rho}} \Big] \Bigg] \right\}
\nonumber \displaybreak[2] \\
&= \exp\left\{-(k-l) E_{x|y}(\Rx, \rho)
          -(n-k+1) E_{xy}(\Rx, \Ry, \rho) \right\} \label{eq.defElk}\\
&= \exp \left\{ -(n-l+1) \Big[ \frac{k-l}{n-l+1} E_{x|y}(\Rx,\rho) +
\frac{n-k+1}{n-l+1} E_{xy}(\Rx, \Ry, \rho)\Big] \right\} \label{eq.defEl2}\\
&\leq \exp \left\{ -(n-l+1) \sup_{\rho \in [0,1]}
\Big[ \frac{k-l}{n-l+1} E_{x|y}(\Rx,\rho) +
\frac{n-k+1}{n-l+1} E_{xy}(\Rx, \Ry, \rho)\Big] \right\}
\label{eq.jointMLoptRho}\\
&= \exp \left\{ -(n-l+1) E_{x}^{ML} \left(\Rx, \Ry,
    \frac{k-l}{n-l+1}\right) \right\} = \exp \left\{ -(n-l+1)
  E_{x}(\Rx, \Ry, \frac{k-l}{n-l+1}) \right\}.
\label{eq.subDefsEx}
\end{align}

In~(\ref{eq.enumJoint}) we explicitly indicate the three conditions
that a suffix pair $(\svxtil_{l}^n, \svytil_{k}^n)$ must satisfy to
result in a decoding error.  In~(\ref{eq.indepBin}) we sum out over
the common prefixes $(\svx^{l-1}, \svy^{l-1})$, and use the fact that
the random binning is done independently at each encoder, see
Definition.~\ref{def.seqn_coding}.  We get~(\ref{eq.gallagerRho}) by
limiting $\rho$ to the interval $0 \leq \rho \leq 1$, as
in~(\ref{eq.limOnRho}). Getting~(\ref{eq.rearranging})
from~(\ref{eq.gallagerRho}) follows by a number of basic
manipulations.  In~(\ref{eq.rearranging}) we get the single letter
expression by again using the memoryless property of the sources.
In~(\ref{eq.defElk}) we use the definitions of $E_{x|y}$ and $E_{xy}$
from~(\ref{eq.defBasicExp}) of Theorem~\ref{thm.jointCodeML}.  Noting
that the bound holds for all $\rho \in [0,1]$ optimizing over $\rho$
results in~(\ref{eq.jointMLoptRho}).  Finally, using the definition
of~(\ref{eq.compoundExp}) and the remark following
Theorem~\ref{THM:Universal_ML_SW} that the maximum-likelihood and
universal exponents are equal gives~(\ref{eq.subDefsEx}).  The bound
on $p_n(l,k)$ when $l > k$, is developed in an analogous
fashion.\hfill $\blacksquare$

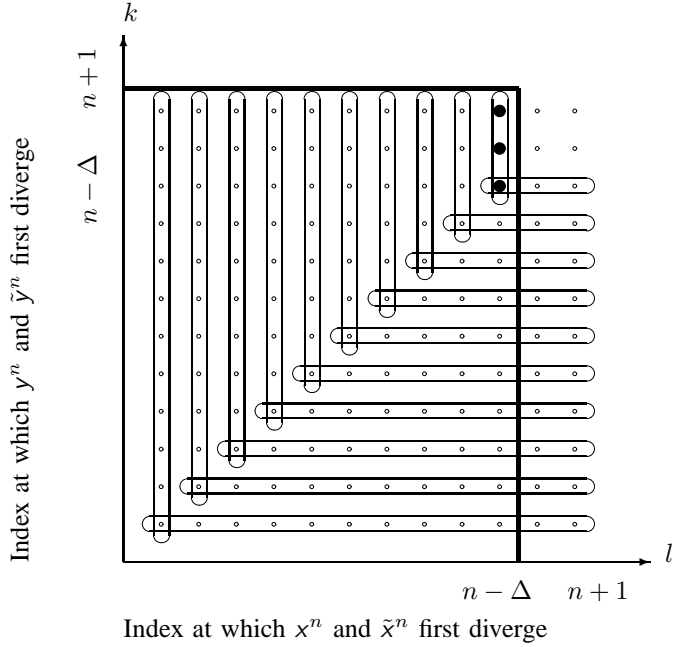
\begin{figure}
 \begin{picture}(100,100)
\multiput(50,20)(5,0){12}{\multiput(0,0)(0,5){12}{\circle{0.5}}}
\put(45, 15){\vector(1,0){70}} \put(45, 15){\vector(0,1){70}}
\put(45,87){$k$} \put(117,15){$l$}

\put(39,74) {\rotatebox{90}{ $n+1$}}
  \put(39,60){\rotatebox{90}{$n-\Delta$}}

\put(30,15) {\rotatebox{90}{Index at which $ \rvy^n$ and $
\rvytil^n$ first diverge}}

\put(45,5)  {Index at which $ \rvx^n$ and $ \rvxtil^n$ first
diverge}
 \put(104,10){$n+1$}
\put(90,10){$n-\Delta$}

\linethickness{0.5mm}\put(97.5,15){\line(0,1){63}} 
\put(45,78){\line(1,0){52.5}} 

\multiput(95,65)(0, 5){ 3 }{\circle*{1.5}}

\thinlines \put(50,47.5){\oval(2,60)} \put(55,50){\oval(2,55)}
\put(60,52.5){\oval(2,50)} \put(65,55){\oval(2,45)}
\put(70,57.5){\oval(2,40)} \put(75,60){\oval(2,35)}
\put(80,62.5){\oval(2,30)} \put(85,65){\oval(2,25)}
\put(90,67.5){\oval(2,20)} 
\put(95,70){\oval(2,15)} \thinlines \put(77.5,20){\oval(60,2)}
\put(80,25){\oval( 55,2)} \put(82.5,30){\oval(50,2)}
\put(85,35){\oval(45,2)} \put(87.5,40){\oval(40,2)}
\put(90,45){\oval( 35,2)} \put(92.5,50){\oval(30,2)}
\put(95,55){\oval( 25,2)} \put(97.5,60){\oval(20,2)}
\put(100,65){\oval( 15,2)} \thinlines
 \end{picture}
\caption{Two dimensional plot of the error probabilities $p_n(l,k)$,
corresponding to error events $(l,k)$,
  contributing to $\Pr[ \rvxhat^{n - \delay} \neq \rvx^{n-\delay}]$ in
  the situation where $\inf_{\gamma \in [0,1] } E_y(\Rx, \Ry, \rho,
  \gamma) \geq {\inf_{\gamma \in [0,1] } E_x(\Rx, \Ry, \rho,\gamma)}$.
  }   \label{fig.twoD2}
\end{figure}

We use Lemma~\ref{lemm.jointPn} together with~(\ref{eq.defPn}) to
bound $\Pr[\rvxhat^{n-\delay} \neq \rvx^{n-\delay}]$ for two distinct
cases.  The first, simpler case, is when $\inf_{\gamma \in [0,1] }
E_y(\Rx, \Ry, \gamma) > {\inf_{\gamma \in [0,1] } E_x(\Rx, \Ry,
  \gamma)}$.  To bound $\Pr[\rvxhat^{n-\delay} \neq \rvx^{n-\delay}]$
in this case, we split the sum over the $p_n(l,k)$ into two terms,
as visualized in Fig~\ref{fig.twoD2}.  There are $(n+1)\times
(n-\delay)$ such events to account for
  (those inside the box).  The probability of the event within each oval are
  summed together to give an upper bound on $\Pr[ \rvxhat^{n - \delay} \neq \rvx^{n-\delay}]$.
  We add extra probabilities outside of the box but within the ovals
  to make the summation symmetric thus simpler. Those extra
  error events do not impact the error exponent because $\inf_{\gamma \in [0,1] } E_y(\Rx, \Ry, \rho,
  \gamma) \geq {\inf_{\gamma \in [0,1] } E_x(\Rx, \Ry, \rho,\gamma)}$.
  The possible dominant   error events are highlighted in Figure \ref{fig.twoD2} . Thus,
\begin{align}
  & \Pr[ \rvxhat^{n - \delay} \neq \rvx^{n-\delay}] \leq \sum_{l=1}^{n
    - \delay} \sum_{k=l}^{n+1} p_n(l,k) + \sum_{k=1}^{n - \delay}
    \sum_{l=k}^{n+1} p_n(l,k)  \label{eq.twoTerms} \\
&\leq \sum_{l=1}^{n - \delay}  \sum_{k=l}^{n+1} \exp\{ -(n-l+1)
\inf_{\gamma \in [0,1]} E_x(\Rx, \Ry, \gamma) \}
 + \sum_{k=1}^{n-\delay}\sum_{l=k}^{n+1}
\exp\{-(n-k+1) \inf_{\gamma \in [0,1]} E_y(\Rx, \Ry, \gamma)\}
 \label{eq.usinglemma} \\
& =   \sum_{l=1}^{n - \delay} \Big[ (n-l+2) \exp\{ -(n-l+1)
\inf_{\gamma \in [0,1]} E_x(\Rx, \Ry, \gamma) \} \nonumber \\
 & \ \ \ \ \ +  \sum_{k=1}^{n - \delay} \Big[ (n-k+2) \exp\{ -(n-k+1)
\inf_{\gamma \in [0,1]}
E_y(\Rx, \Ry, \gamma) \} \nonumber\\
& \leq 2 \sum_{l=1}^{n - \delay} \Big[ (n-l+2) \exp\{ -(n-l+1)
\inf_{\gamma \in [0,1]} E_x(\Rx, \Ry, \gamma) \}
 \label{eq.sumTerms} \\
& \leq \sum_{l=1}^{n - \delay}   C_1 \exp\{ -(n-l+2)[ \inf_{\gamma
\in [0,1]} E_x(\Rx, \Ry, \gamma) -\alpha]\}
  \label{eq.smallerExp}\\
& \leq C_2 \exp\{-\delay [ \inf_{\gamma \in [0,1]} E_x(\Rx, \Ry,
\gamma) -\alpha]\} \label{eq.boundEyBigEx}
\end{align}

Equation (\ref{eq.twoTerms}) follows directly from (\ref{eq.defPn}),
in the first  term  $l\leq k $, in the second
 term $l\geq k$.  In~(\ref{eq.usinglemma}), we use Lemma~\ref{lemm.jointPn}. In~(\ref{eq.sumTerms}) we
use the assumption that $\inf_{\gamma \in [0,1] } E_y(\Rx, \Ry,
\gamma) > \inf_{\gamma \in [0,1] } E_x(\Rx, \Ry, \gamma)$.
In~(\ref{eq.smallerExp}) the $\alpha > 0$ results from incorporating
the polynomial into the first exponent, and can be chosen as small
as desired.  Combining terms and summing out the decaying
exponential yield the bound~(\ref{eq.boundEyBigEx}).

\begin{figure}[t]
\begin{picture}(100,100)

\multiput(50,20)(5,0){12}{\multiput(0,0)(0,5){12}{\circle{0.5}}}
\put(45, 15){\vector(1,0){70}} \put(45, 15){\vector(0,1){70}}
\put(45,87){$k$} \put(117,15){$l$}

\put(39,74) {\rotatebox{90}{ $n+1$}}
  \put(39,60){\rotatebox{90}{$n-\Delta$}}

\put(30,15) {\rotatebox{90}{Index at which $ \rvy^n$ and $
\rvyhat^n$ first diverge}}

\put(45,5)  {Index at which $ \rvx^n$ and $ \rvxhat^n$ first
diverge}

   \put(110,49){$k^*(n-\Delta)-1$}
\put(104,10){$n+1$}  \put(90,10){$n-\Delta$}
 \multiput(95,55)(0,5){
5 }{\circle*{1.5}}

\linethickness{0.5mm}\put(97.5,15){\line(0,1){63}} 
\put(45,78){\line(1,0){52.5}} 

 \thinlines
\put(105,75){\line(-1,-1){58}}  \dottedline{1}(105,75)(77.5,20)

\dottedline{1}(77.5,20)(50,20)

\put(50,47.5){\oval(2,60)} \put(55,50){\oval(2,55)}
\put(60,52.5){\oval(2,50)} \put(65,55){\oval(2,45)}
\put(70,57.5){\oval(2,40)} \put(75,60){\oval(2,35)}
\put(80,62.5){\oval(2,30)} \put(85,65){\oval(2,25)}
\put(90,67.5){\oval(2,20)}
 \put(95,70){\oval(2,15)}
\put(55,20){\oval(2,5)} \put(60,22.5){\oval(2,10)}
\put(65,25){\oval(2,15)} \put(70,27.5){\oval(2,20)}
\put(75,30){\oval(2,25)} \put(80,35){\oval(2,25)}
\put(85,42.5){\oval(2,20)} \put(90,50){\oval(2,15)}   \put(95,57
){\oval(2,10)}  \put(92.5,20){\oval( 30,2)} \put(95,25){\oval(
25,2)} \put(95,30){\oval(25,2)} \put(97.5,35){\oval(20,2)}
\put(97.5,40){\oval(20,2)} \put(100,45){\oval( 15,2)}
\put(100,50){\oval(15,2)} \thinlines \put(61,56){Region
1}\put(65,27){Region 2} \put(96,27){Region 3} \put(108,
61){$k^*(l)$} \put(107, 62){\vector(-1,0){8.5}}
\end{picture}
\caption{Two dimensional plot of the error probabilities $p_n(l,k)$,
corresponding to error events $(l,k)$,
  contributing to $\Pr[ \rvxhat^{n - \delay} \neq \rvx^{n-\delay}]$ in
  the situation where $\inf_{\gamma \in
    [0,1] } E_y(\Rx, \Ry,   \gamma) < {\inf_{\gamma \in [0,1] }
    E_x(\Rx, \Ry,  \gamma)}$. }\label{fig.errEvents}
\end{figure}
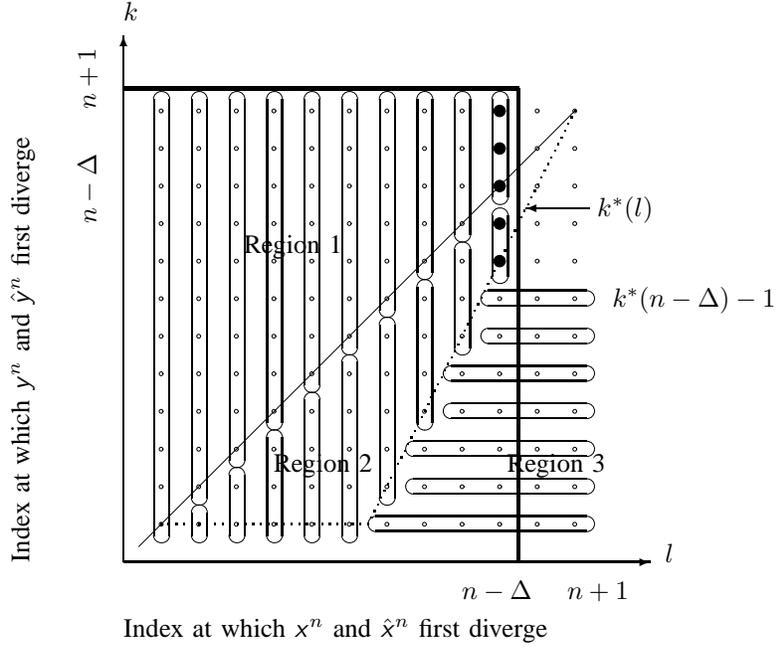

The second, more involved case, is when $\inf_{\gamma \in [0,1] }
E_y(\Rx, \Ry, \rho, \gamma) < {\inf_{\gamma \in [0,1] } E_x(\Rx, \Ry,
  \rho, \gamma)}$.  To bound $\Pr[ \rvxhat^{n - \delay} \neq
\rvx^{n-\delay}]$, we could use the same bounding technique used in
the first case. This gives the error exponent $\inf_{\gamma \in [0,1]
} E_y(\Rx, \Ry, \gamma)$ which is generally smaller than what we can
get by dividing the error events in a new scheme as shown in Figure
\ref{fig.errEvents}. In this situation we split~(\ref{eq.defPn}) into
three terms, as visualized in Fig~\ref{fig.errEvents}.  Just as in the
first case shown in Fig~\ref{fig.twoD2}, there are $(n+1)\times
(n-\delay)$ such events to account for (those inside the box). The
error events are partitioned into 3 regions. Region 2 and 3 are
separated by $k^*(l)$ using a dotted line.  In region 3, we add extra
probabilities outside of the box but within the ovals to make the
summation simpler.  Those extra error events do not affect the error
exponent as shown in the proof.  The possible dominant error events
are highlighted shown in Fig~\ref{fig.errEvents}. Thus,
\begin{equation}
\Pr[ \rvxhat^{n - \delay} \neq \rvx^{n-\delay}] \leq \sum_{l=1}^{n -
\delay} \sum_{k=l}^{n+1} p_n(l,k) + \sum_{l=1}^{n - \delay}
\sum_{k=k^{\ast}(l)}^{l-1} p_n(l,k) + \sum_{l=1}^{n - \delay}
\sum_{k=1}^{k^{\ast}(l)-1} p_n(l,k) \label{eq.threeTerms}
\end{equation}
Where  $\sum_{k=1}^{0} p_k=0$. The lower boundary of Region 2 is
$k^{\ast}(l) \geq 1$ as a function of $n$ and $l$:
\begin{equation}
\kast = \max\left\{1, n +1-  \ceil{\frac{ \inf_{\gamma \in [0,1]}
E_x(\Rx, \Ry, \gamma)}{ \inf_{\gamma \in [0,1] } E_y(\Rx, \Ry,
\gamma)}} (n+1-l )\right\} = \max\left\{1, n+1 - G
(n+1-l)\right\}\label{eq.kast}
\end{equation}
where we use $G$ to denote the ceiling of the ratio of exponents.
Note that when $\inf_{\gamma \in [0,1] } E_y(\Rx, \Ry,  \gamma)
> {\inf_{\gamma \in [0,1] } E_x(\Rx, \Ry,   \gamma)}$ then $G =
1$ and region two of Fig.~\ref{fig.errEvents} disappears.  In other
words, the middle term of~(\ref{eq.threeTerms}) equals zero.  This
is the first case considered.  We now consider the cases when $G
\geq 2$ (because of the ceiling function $G$ is a positive integer).

The first term of~(\ref{eq.threeTerms}), i.e., region one in
Fig.~\ref{fig.errEvents} where $ l\leq k$, is bounded in the same
way that the first term of~(\ref{eq.twoTerms}) is, giving
\begin{equation}
\sum_{l=1}^{n - \delay} \sum_{k=l}^{n+1} p_n(l,k) \leq
C_2 \exp\{-\delay [ \inf_{\gamma \in [0,1]} E_x(\Rx, \Ry, \gamma) -\alpha]\}.
\label{eq.firstTerm}
\end{equation}

In Fig.~\ref{fig.errEvents}, region two is upper bounded by the
45-degree line, and lower bounded by $k^{\ast}(l)$. The second term
of~(\ref{eq.threeTerms}), corresponding to this region where $ l\geq
k$,
\begin{align}
 \sum_{l=1}^{n - \delay} \sum_{k=\kast}^{l-1} p_n(l,k)
&\leq \sum_{l=1}^{n - \delay}\sum_{k=\kast}^{l-1} \exp\{-(n-k+1)
  E_y(\Rx, \Ry, \frac{l-k}{n-k+1})\}
\nonumber \\
  &= \sum_{l=1}^{n - \delay}\sum_{k=\kast}^{l-1}
\exp\{-(n-k+1) \frac{n-l+1}{n-l+1} E_y(\Rx, \Ry,
\frac{l-k}{n-k+1})\}
 \label{eq.gammaInv} \\
 & \leq \sum_{l=1}^{n - \delay}\sum_{k=\kast}^{l-1}
 \exp\{ -(n-l+1) \inf_{\gamma \in [0,1]}
\frac{1}{1-\gamma} E_y(\Rx, \Ry, \gamma)\}
\label{eq.defGamma}\\
 & = \sum_{l=1}^{n - \delay} (l-\kast)
 \exp\{ -(n-l+1) \inf_{\gamma \in [0,1]}
\frac{1}{1-\gamma} E_y(\Rx, \Ry, \gamma)\} \label{eq.secTerm}
\end{align}
In~(\ref{eq.gammaInv}) we note that $l\geq k $, so define
$\frac{l-k}{n-k+1}=\gamma$ as in~(\ref{eq.defGamma}). Then
$\frac{n-k+1}{n-l+1} = \frac{1}{1-\gamma}$.

The third term of~(\ref{eq.threeTerms}), i.e., the intersection of
region three and the ``box'' in Fig.~\ref{fig.errEvents} where $
l\geq k$, can be bounded as,
\begin{align}
 \sum_{l=1}^{n - \delay}\sum_{k = 1}^{\kast-1}
 p_n(l,k) &\leq \sum_{l=1}^{n + 1}\sum_{k = 1}^{\min\{l, k^*(n-\delay)-1\}}
 p_n(l,k) \label{eq.changeOrder}\\
 &=\sum_{k=1}^{k^*(n-\Delta)-1}
\sum_{l=k}^{n+1}p_n(l,k) \label{eq.changeOrder1}\\
&\leq \sum_{k=1}^{k^*(n-\Delta)-1}
\sum_{l=k}^{n+1}\exp\{-(n-k+1)E_y(R_x,R_y,\frac{l-k}{n-k+1})\}\nonumber\\
&\leq \sum_{k=1}^{k^*(n-\Delta)-1}
\sum_{l=k}^{n+1}\exp\{-(n-k+1)\inf_{\gamma\in[0,1]}E_y(R_x,R_y,\gamma)\}
\nonumber\\
&\leq \sum_{k=1}^{k^*(n-\Delta)-1}(n-k+2)
\exp\{-(n-k+1)\inf_{\gamma\in[0,1]}E_y(R_x,R_y,\gamma)\}
\label{eq.thirdTerm}
\end{align}

In (\ref{eq.changeOrder}) we note that $l\leq n-\delay$ thus
$k^*(n-\delay) -1 \geq \kast -1$, also $l\geq 1$, so $l\geq
\kast-1$. This can be visualized in Fig~\ref{fig.errEvents} as we
extend the summation from the intersection of the ``box'' and region
3 to the whole region under the diagonal line and  the horizontal
line $k=k^*(n-\delay)-1$. In (\ref{eq.changeOrder1}) we simply
switch the order of the summation.

Finally when $G \geq 2$, we substitute~(\ref{eq.firstTerm}),
(\ref{eq.secTerm}), and~(\ref{eq.thirdTerm})
into~(\ref{eq.threeTerms}) to give
\begin{align}
\Pr[\rvxhat^{n - \delay} \neq \rvx^{n-\delay}] &\leq C_2
  \exp\{-\delay [ \inf_{\gamma \in [0,1]} E_x(\Rx, \Ry, \gamma)
  -\alpha]\} \nonumber \\
  &+ \sum_{l=1}^{n - \delay} (l-\kast) \exp\{ -(n-l+1) \inf_{\gamma
    \in [0,1]} \frac{1}{1-\gamma} E_y(\Rx, \Ry, \gamma)\}\\
  &+\sum_{k=1}^{k^*(n-\Delta)-1}(n-k+2)\exp\{-(n-k+1)\inf_{\gamma\in[0,1]}E_y(R_x,R_y,\gamma)\}\nonumber\\
&\leq C_2 \exp\{-\delay [ \inf_{\gamma \in [0,1]} E_x(\Rx, \Ry,
\gamma) -\alpha]\}
\nonumber \\
&+ \sum_{l=1}^{n - \delay} (l-n-1+G(n+1-l))
 \exp\{ -(n-l+1) \inf_{\gamma \in [0,1]}
\frac{1}{1-\gamma} E_y(\Rx, \Ry, \gamma)\}\nonumber\\
&+\sum_{k=1}^{n +1- G(\delay+1)}(n-k+2)\exp\{-(n-k+1)\inf_{\gamma\in[0,1]}E_y(R_x,R_y,\gamma)\} \label{eq.largerSum}\\
&\leq C_2 \exp\{-\delay [ \inf_{\gamma \in [0,1]} E_x(\Rx, \Ry,
\gamma) -\alpha]\}
\nonumber \\
&+ (G-1)C_3
 \exp\{ -\delay \big[\inf_{\gamma \in [0,1]}
\frac{1}{1-\gamma} E_y(\Rx, \Ry, \gamma)-\alpha\big]\}\nonumber\\
&+ C_4\exp\{-\big[\delay
G\inf_{\gamma\in[0,1]}E_y(R_x,R_y,\gamma)-\alpha \big]\}\nonumber\\
&\leq C_5 \exp \Big\{  - \delay \Big[\min \Big\{ \inf_{\gamma \in
[0,1]} E_x(\Rx, \Ry, \gamma), \inf_{\gamma \in [0,1]}
\frac{1}{1-\gamma} E_y(\Rx, \Ry, \gamma) \Big\} - \alpha
\Big]\Big\}. \label{eq.finResMLSW}
\end{align}
To get (\ref{eq.largerSum}), we use the fact that $k^*(l)\geq
n+1-G(n+1-l)$ from the definition of $k^*(l)$ in (\ref{eq.kast}) to
upper bound the second term. We exploit the definition of $G$ to
convert the exponent in the third term to $\inf_{\gamma \in [0,1]}
E_x(\Rx, \Ry, \gamma)$.  Finally, to get~(\ref{eq.finResMLSW}) we
gather the constants together, sum out over the decaying
exponentials, and are limited by the smaller of the two exponents.

Note: in the proof of Theorem~\ref{thm.jointCodeML},  we regularly
double count the error events or add smaller extra probabilities to
make the summations simpler. But it should be clear that the error
exponent is not affected.

\subsection{Universal Decoding}

As discussed in Section~\ref{sec.univEnt}, we do not use a pairwise
minimum joint-entropy decoder because of polynomial term in $n$ would
multiply the exponential decay in $\delay$. Analogous to the
sequential decoder used there, we use a ``weighted suffix entropy''
decoder.  The decoding starts by first identifying candidate sequence
pairs as those that agree with the encoding bit streams up to time
$n$, i.e., $\svxBar^n \in \binX(\svx^n), \svyBar^n \in \binY(\svy^n)$.
For any one of the $|\binX(\svx^n)| |\binY(\svy^n)|$ sequence pairs in
the candidate set, i.e., $(\svxBar^n, \svyBar^n) \in \binX(\svx^n)
\times \binY(\svy^n)$ we compute $(n+1)\times (n+1)$ weighted
entropies:

\begin{eqnarray}
&&H_S(l,k,\svxBar^n, \svyBar^n)=H(\svxBar_{l}^{(n+1-l)},\svyBar_{l}^{(n+1-l)}),\ \ \ \ l=k\nonumber\\
&&H_S(l,k,\svxBar^n, \svyBar^n)=\frac{k-l}{n+1-l}H({\svxBar}_{l}^{k-1}|{\svyBar}_{l}^{k-1})+\frac{n+1-k}{n+1-l}H({\svxBar}_{k}^{n},{\svyBar}_{k}^{n}),\ \ \ \ l<k\nonumber\\
&&H_S(l,k,\svxBar^n,\svyBar^n)=\frac{l-k}{n+1-k}H({\svyBar}_{k}^{l-1}|{\svxBar}_{k}^{l-1})+\frac{n+1-l}{n+1-k}H({\svxBar}_{l}^{n},{\svyBar}_{l}^{n}),\
\ \ \ l>k.\nonumber
\end{eqnarray}

We define the \textit{score} of $({\svxBar}^n, {\svyBar}^n)$ as the
pair of integers $i_x(\svxBar^n,\svyBar^n)$, $
i_y(\svxBar^n,\svyBar^n)$ s.t.,
\begin{eqnarray}
i_x(\svxBar^n,\svyBar^n)&=&\max\{i:H_S(l,k,(\svxBar^n,\svyBar^n))<
H_S(l,k,\svxtil^n,\svytil^n) \forall k=1,2,...n+1, \forall
l=1,2,...i, \nonumber\\
&& \forall (\svxtil^n,\svytil^n)\in\binX(\svx^n) \times
\binY(\svy^n)\cap
\mathcal{F}_n(l,k,\svxBar^n,\svyBar^n) \}\\
i_y(\svxBar^n,\svyBar^n)&=&\max\{i:H_S(l,k,(\svxBar^n,\svyBar^n))<
H_S(l,k,\svxtil^n,\svytil^n) \forall l=1,2,...n+1, \forall
k=1,2,...i,\nonumber\\
&& \forall (\svxtil^n,\svytil^n)\in \binX(\svx^n) \times
\binY(\svy^n)\cap \mathcal{F}_n(l,k,\svxBar^n,\svyBar^n) \}
\end{eqnarray}
While $\mathcal{F}_n(l,k,\svx^n,\svy^n)$ is the same set as defined
in (\ref{eq.jointPart}), we repeat the definition here for
convenience,

\begin{equation}
 \mathcal{F}_n(l,k,x^n,y^n)=\{(\svxBar^n,\svytil^n)\in
\mathcal{X}^{n} \times\mathcal{Y}^{n} \; \mbox{s.t.} \;
\svxBar^{l-1} = x^{l-1},\svxBar_l \neq x_l,\svyBar^{k-1}= y^{k-1},
\svyBar_k\neq y_k\}.\nonumber
\end{equation}

The definition of
$(i_x(\svxBar^n,\svyBar^n),i_y(\svxBar^n,\svyBar^n))$ can be
visualized in the following procedure. As shown in
Fig.~\ref{fig.scoresheet},  for all $ 1\leq l,k \leq n+1$, if there
exists $({\svxBBar}^n, {\svyBBar}^n)\in
\mathcal{F}_n(l,k,(\svxBar^n,\svyBar^n))\cap \binX(\svx^n) \times
\binY(\svy^n) $ s.t.  $H_S(l,k,\svxBar^n, \svyBar^n)\geq
H_S(l,k,\svxBBar^n,\svyBBar^n)$ , then we mark $(l,k)$ on the plane
as shown in Fig.\ref{fig.scoresheet}. Eventually we pick the maximum
integer which is smaller than all marked $x$-coordinates as
$i_x(\svxBar^n,\svyBar^n)$ and the maximum integer which is smaller
than all marked $y$-coordinates as $i_y(\svxBar^n,\svyBar^n)$. The
score of $({\svxBar}^n, {\svyBar}^n)$ tells us the first
branch(either $x$ or $y$) point where  a ``better sequence pair''
(with a smaller weighted entropy) exists.

Define the set of the winners as the sequences (not sequence pair)
with the maximum   score:

$$\mathcal{W}_n^x=\{\svxBar^n\in  \binX(\svx^n) :\exists \svyBar^n\in \mathcal{B}_y(y^n), s.t.
i_x(\svxBar^n,\svyBar^n)\geq i_x(\svxtil^n,\svytil^n), \forall
(\svxtil^n,\svytil^n)\in \binX(\svx^n) \times \binY(\svy^n)\}$$
$$\mathcal{W}_n^y=\{\svyBar^n\in  \binY(\svy^n) :\exists \svxBar^n\in
\binX(\svx^n)  , s.t. i_y(\svxBar^n,\svyBar^n)\geq
i_y(\svxtil^n,\svytil^n),  \forall (\svxtil^n,\svytil^n)\in
\binX(\svx^n) \times \binY(\svy^n)\}$$

Then arbitrarily pick one sequence from $\mathcal{W}_n^x$ and one
from $\mathcal{W}_n^y$ as the decision $(\svxhat^n,\svyhat^n)$.

\setlength{\unitlength}{1mm}

 \begin{figure}
\begin{picture}(100,100)

\multiput(50,20)(5,0){12}{\multiput(0,0)(0,5){12}{\circle {1.5}}}
\put(45, 15){\vector(1,0){70}} \put(45, 15){\vector(0,1){70}}
\put(45,87){$k$} \put(117,15){$l$} \put(45, 39){\line(1,0){65}}
 \put(74, 15){\line(0,1){65}}

 \put(37,74){\rotatebox{90}{$n+1$}}

 \put(104,10){$n+1$}
 \put(48,10){$1$}
\put(37,20){\rotatebox{90}{$1$}}

\put(41, 33){\rotatebox{90}{$i_y$}}
 \put(68, 11){$i_x$}

\put(95,40){\circle*{1.5}} \put(75,45){\circle*{1.5}}
\put(105,50){\circle*{1.5}} \put(100,65){\circle*{1.5}}
\put(100,60){\circle*{1.5}} \put(95,60){\circle*{1.5}}
\put(95,65){\circle*{1.5}} \put(95,70){\circle*{1.5}}
\put(100,60){\circle*{1.5}} \put(75,65){\circle*{1.5}}
\put(95,75){\circle*{1.5}} \put(85,55){\circle*{1.5}}
\put(105,75){\circle*{1.5}}
 \end{picture}
     \caption[]{2D interpretation of the \textit{score}, $(i_x(\svxBar^n,\svyBar^n),
     i_y(\svxBar^n,\svyBar^n))$, of a sequence
     pair $(\svxBar^n,\svyBar^n)$. If there exists a sequence pair in
     $\mathcal{F}_n(l,k,\svxBar^n,\svyBar^n)$ with less or the same score, then $(l,k)$ is marked with a solid dot.
     The \textit{score} $i_x(\svxBar^n,\svyBar^n)$ is the largest
     integer which is smaller than all the $x$-coordinates of the
     marked points. Similarly for $i_y(\svxBar^n,\svyBar^n),$
     }
     \label{fig.scoresheet}
 \end{figure}
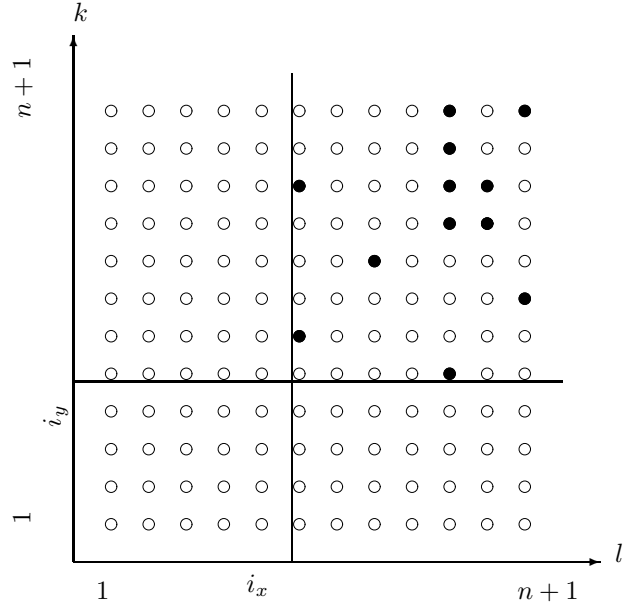

We bound the probability that there exists a sequence pair in
$\mathcal{F}_n(l,k,(\rvx^n,\rvy^n))\cap  \binX(\svx^n) \times
\binY(\svy^n)$ with smaller weighted minimum-entropy suffix score
as:
\begin{eqnarray}
 p_n(l,k)&=&\sum_{x^n}\sum_{y^n}p_{\rvx\rvy}(x^n,y^n)
P(\exists(\svxtil_{1}^{n},\svytil_{1}^{n})\in \binX(\svx^n) \times
\binY(\svy^n)\cap \mathcal{F}_n(l,k,x^n,y^n),\nonumber\\
&& s.t. H_S(l,k,\svxtil^n,\svytil^n)\leq
H_S(l,k,(x^n,y^n)))\nonumber
\end{eqnarray}
Note that the $p_n(l,k)$ here differs from the $p_n(l,k)$ defined in
the ML decoding by replacing $p_{\rvx\rvy}(\svx^n, \svy^n) \leq
p_{\rvx\rvy}(\svxtil^n, \svytil^n)$     with
$H_S(l,k,\svxtil^n,\svytil^n)\leq H_S(l,k,(x^n,y^n))$.

The following lemma, analogous  to (\ref{eq.defPn}) for ML decoding,
tells us that the ``suffix weighted entropy'' decoding rule is a
good one.

\begin{lemma} Upper bound on symbol-wise decoding error
$P_{ex}(k,k+d)$ :\label{Lemma_3_UNI_SW}

\begin{eqnarray}
\Pr[\rvxhat^{n-\delay}   \neq \rvx^{n-\delay}]  \leq
\sum_{l=1}^{n-\delay}\sum_{k=1}^{n+1}p_n(l,k)\nonumber
\end{eqnarray}

\end{lemma}

\pf According to the decoding rule, $\svxhat^{n-\delay}\neq
\svx^{n-\delay}$ implies that there exists a sequence $\svxtil^n\in
\mathcal{W}_n^x$ s.t.$\svxtil^{n-\delay}\neq x^{n-\delay}$. This
means that there exists a sequence $\svytil^n\in
 \binY(\svy^n)$, s.t. $i_x(\svxtil^n,\svytil^n)\geq
i_x(\svx^n,\svy^n)$. Suppose that $(\svxtil^n,\svytil^n)\in
\mathcal{F}_n(l,k , x^n,y^n)$, then $l\leq n-\delay $ because
$\svxtil^{n-\delay}\neq x^{n-\delay}$. By the definition of $i_x$,
we know that $H_S(l,k,\svxtil^n,\svytil^n)\leq H_S(l,k, x^n,y^n )$.
And using the union bound argument we get the desired inequality.
\hfill $\blacksquare$

We only need to bound each single error probability $p_n(l,k)$ to
finish the proof.

\begin{lemma}{Upper bound on $p_n(l,k)$, $l\leq k$:} $\forall \gamma>0$, $\exists  K_1 <
\infty$, s.t.
$$p_n(l,k)\leq \exp\{-(n-l+1) [E_{x}
(\Rx, \Ry, \lambda) - \gamma]\}$$ where  $\lambda = (k-l)/(n-l+1)
\in [0,1]$. \label{Lemma:UpperBoundon2_UN}
\end{lemma}
\pf  Here the error probability $p_n(l,k)$  can be thought as
starting from~(\ref{eq.indepBin}) with the condition $(k-l)
H(\svxtil_{l}^{k-1}|\svytil_{l}^{k-1}) + (n-k+1) H(\svxtil_k^n,
\svytil_{k}^n) < (k-l) H(\svx_{l}^{k-1}|\svy_{l}^{k-1}) + (n-k+1)
H(\svx_k^n, \svy_{k}^n)$ substituted for $p(\svxtil_l^n,
\svytil_l^n) > p(\svx_l^n, \svy_l^n)$, we get
\begin{align}
p_n(l,k) = &
\sum_{\PNK, \PKL}
\sum_{\VNK, \VKL}
\sum_{\tiny \begin{array}{c}
\svy_{l}^{k-1} \in \tclass_{\PKL},\\
\svy_{k}^n \in \tclass_{\PNK}
\end{array}}
\sum_{\tiny \begin{array}{c}
\svx_l^{k-1} \in \tclass_{\VKL}(\svy_l^{k-1}), \\
\svx_{k}^n \in \tclass_{\VNK(\svy_{k}^n)} \end{array}}
\min \Big[1,
\sum_{\tiny \begin{array}{c}\VtilNK, \VtilKL, \PtilNK \; \mbox{s.t.}\\
\minEntTil < \\\minEnt
\end{array}} \nonumber\\
& \sum_{\svytil_{k}^n \in \tclass_{\PtilNK}} \sum_{\svxtil_{l}^{k-1}
\in \tclass_{\VtilKL}(\svy_{l}^{k-1})} \sum_{\svxtil_{k}^n \in
\tclass_{\VtilNK}(\svytil_{k}^n)} \exp\{-(n-l +1) \Rx - (n-k+1)
\Ry\} \Big] p_{\rvx\rvy}(x^n, y^n) \label{eq.enumTil}
\end{align}
In~(\ref{eq.enumTil}) we enumerate all the source sequences in a way
that allows us to focus on the types of the important subsequences.
We enumerate the possibly misleading candidate sequences in terms of
their suffixes types.  We restrict the sum to those pairs
$(\svxtil^n, \svytil^n)$ that could lead to mistaken decoding,
defining the compact notation $\minEnt \defeq (k-l) H(\VKL|\PKL) +
(n-k+1) H(\PNK \times \VNK)$, which is the weighted suffix entropy
condition rewritten in terms of types.

Note that the summations within the minimization
in~(\ref{eq.enumTil}) do not depend on the arguments within these
sums.  Thus, we can bound this sum separately to get a bound on the
number of possibly misleading source pairs $(\svbxtil, \svbytil)$.
\begin{align}
& \hspace{-5em}
\sum_{\tiny \begin{array}{c}\VtilNK, \VtilKL, \PtilNK \;  \mbox{s.t.}\\
\minEntTil < \\ \minEnt \end{array}}
\sum_{\svytil_{k}^n \in \tclass_{\PtilNK}}
\sum_{\svxtil_{l}^{k-1} \in \tclass_{\VtilKL}(\svy_{l}^{k-1})}
\sum_{\svxtil_{k}^n \in \tclass_{\VtilNK}(\svytil_{k}^n)} \nonumber\\
& \leq \sum_{\tiny \begin{array}{c}\VtilNK, \VtilKL, \PtilNK
\; \mbox{s.t.}\\  \minEntTil < \\ \minEnt \end{array}}
\sum_{\svytil_{k}^n \in \tclass_{\PtilNK}}
|\tclass_{\VtilKL(\svy_{l}^{k-1})}|
|\tclass_{\VtilNK(\svytil_{k}^n)}| \label{eq.condTypeSize}\\
\leq & \sum_{\tiny \begin{array}{c}\VtilNK, \VtilKL, \PtilNK \;
\mbox{s.t.}\\
\minEntTil < \\ \minEnt \end{array}}
|\tclass_{\PtilNK}|
\exp\{(k-l)H(\VtilKL|\PKL)\}
\exp\{(n-k+1)H(\VtilNK|\PtilNK)\}
\displaybreak[2]
\label{eq.condTypeBnd}\\
\leq &\sum_{\tiny \begin{array}{c}\VtilNK, \VtilKL, \PtilNK \; \mbox{s.t.}\\
\minEntTil < \\ \minEnt \end{array}}
\exp\{(k-l)H(\VtilKL|\PKL) + (n-k+1) H(\PtilNK \times \VtilNK) \}
\displaybreak[2]
\label{eq.margTypeBnd} \displaybreak[2]\\
\leq &  \sum_{\VtilNK, \VtilKL, \PtilNK }
\exp\{(k-l)H(\VKL|\PKL) + (n-k+1) H(\PNK \times \VNK) \}
\displaybreak[2] \label{eq.scoring} \displaybreak[2] \\
\leq & \; (n-l+2)^{2 |\cX| |\cY|}
\exp\{(k-l)H(\VKL|\PKL) + (n-k+1) H(\PNK \times \VNK) \}
\label{eq.numTypes}
\end{align}
In~(\ref{eq.condTypeSize}) we sum over all $\svxtil_{l}^{k-1} \in
\tclass_{\VtilKL}(\svy_{l}^{k-1})$.
In~(\ref{eq.condTypeBnd}) we use standard bounds, e.g., $|
\tclass_{\VtilKL}(\svy_{l}^{k-1})| \leq \exp\{(k-l)
H(\VtilKL|\PKL)\}$ since $\svy_{l}^{k-1} \in \tclass_{\PKL}$.
We also sum over all $\svxtil_{k}^{n} \in
\tclass_{\VtilNK}(\svytil_{k}^{n})$ and over all
$\svytil_{k}^n \in
\tclass_{\PtilNK}$ in~(\ref{eq.condTypeBnd}).  By definition of the
decoding rule $(\svbxtil, \svbytil)$ can only lead to a decoding error
if $(k-l) H(\VtilKL|\PKL)] + (n-k+1) H(\PtilNK \times \VtilNK) <
(k-l)H(\VKL|\PKL) + (n-k+1) H(\PNK \times \VNK)$.
In~(\ref{eq.numTypes}) we apply the polynomial bound on the number of
types.

We substitute~(\ref{eq.numTypes}) into~(\ref{eq.enumTil}) and pull out
the polynomial term, giving
\begin{align}
&\hspace{-1em} p_n(l,k) \leq
(n-l+2)^{2 |\cX| |\cY|}
\sum_{\PNK, \PKL}
\sum_{\VNK, \VKL}
\sum_{\tiny \begin{array}{c}
\svy_{l}^{k-1} \in \tclass_{\PKL},\\
\svy_{k}^n \in \tclass_{\PNK}
\end{array}}
\sum_{\tiny \begin{array}{c}
\svx_l^{k-1} \in \tclass_{\VKL}(\svy_l^{k-1}), \\
\svx_{k}^n \in \tclass_{\VNK(\svy_{k}^n)} \end{array}}
\nonumber \\
&\min \Big[1, \exp\{-(k-l)[\Rx - H(\VKL|\PKL)]
- (n-k+1) [\Rx + \Ry - H(\VNK \times \PNK)] \} \Big]
\jointSource{l}{n}{l}{n} \nonumber \\
\leq &
(n-l+2)^{2 |\cX| |\cY|}
\sum_{\PNK, \PKL}
\sum_{\VNK, \VKL} \nonumber \\
& \exp\Big\{\max \Big[0, -(k-l)[\Rx - H(\VKL|\PKL)]
- (n-k+1)  [\Rx + \Ry - H(\VNK \times \PNK)] \Big]\Big\}
\nonumber \\
&\exp\left\{-(k-l)D(\VKL \times \PKL \| \PxyRV)
- (n-k+1) D(\VNK \times \PNK \| \PxyRV) \right\}
\label{eq.srcProb} \displaybreak[2]\\
\leq &
(n-l+2)^{2 |\cX| |\cY|}
\sum_{\PNK, \PKL}
\sum_{\VNK, \VKL}
\exp\Big\{-(n-l+1) \Big[\lambda D(\VKL \times \PKL \| \PxyRV)
+ \bar{\lambda} D(\VNK \times \PNK \| \PxyRV)
\nonumber \\
&+ \left|\lambda [\Rx - H(\VKL|\PKL)]
+ \bar{\lambda}  [\Rx + \Ry - H(\VNK \times \PNK)]\right|^{+}
\Big] \Big\}
\label{eq.combineDiv} \displaybreak[2]\\
\leq &
(n-l+2)^{2 |\cX| |\cY|}
\sum_{\PNK, \PKL}
\sum_{\VNK, \VKL}
\exp \Big\{-(n-l+1) \inf_{\tiny \rvxtil, \rvytil,
\rvxBar, \rvyBar}
\Big[\lambda D(p_{\rvxtil, \rvytil} \| \PxyRV)
+ \bar{\lambda} D(p_{\rvxBar, \rvyBar} \| \PxyRV) \nonumber \\
&  + \left|\lambda [\Rx - H(\rvxtil|\rvytil)]
+ \bar{\lambda}
[\Rx + \Ry - H(\rvxBar, \rvyBar)]\right|^{+} \Big] \Big\}
\label{eq.infExp} \displaybreak[2]\\
\leq &
(n-l+2)^{4 |\cX| |\cY|} \exp\{-(n-l+1)  E_{x} (\Rx, \Ry, \lambda)\}
\leq K_1 \exp\{-(n-l+1) [E_{x}
(\Rx, \Ry, \lambda) - \gamma]\} \label{eq.defExp} \displaybreak[2]\\
\end{align}
In~(\ref{eq.srcProb}) we use the memoryless property of the source,
and exponential bounds on the probability of observing
$(\svx_{l}^{k-1}, \svy_l^{k-1})$ and $(\svx_k^n, \svy_k^n)$.
In~(\ref{eq.combineDiv}) we pull out $(n-l+1)$ from all terms,
noticing that $\lambda = (k-l)/(n-l+1) \in [0,1]$ and $\bar{\lambda}
\defeq 1- \lambda = (n-k+1)/(n-l+1)$.  In~(\ref{eq.infExp}) we
minimize the exponent over all choices of distributions $p_{\rvxtil,
\rvytil}$ and $p_{\rvxBar, \rvyBar}$.  In~(\ref{eq.defExp}) we
define the universal random coding exponent $E_{x}(\Rx, \Ry,
\lambda) \defeq \inf_{\tiny \rvxtil, \rvytil, \rvxBar, \rvyBar} \{
\lambda  D(p_{\rvxtil,\rvytil} \| \PxyRV) + \bar{\lambda}
D(p_{\rvxBar, \rvyBar} \| \PxyRV) + \left|\lambda [\Rx -
H(\rvxtil|\rvytil)] + \bar{\lambda} [\Rx + \Ry - H(\rvxBar,
\rvyBar)]\right|^{+}\}$ where $0 \leq \lambda \leq 1$ and
$\bar{\lambda} = 1 - \lambda$.  We also incorporate the number of
conditional and marginal types into the polynomial bound, as well as
the sum over $k$, and then push the polynomial into the exponent
since for any polynomial $F$, $\forall E, \epsilon >0$, there exists
$C>0$, s.t. $F(\delay)e^{-\delay E}\leq Ce^{-\delay(E-\epsilon)}$ .
\hfill $\blacksquare$

A similar derivation yields a bound on $p_n(l, k)$ for $l \geq k$.

Combining Lemmas \ref{Lemma:UpperBoundon2_UN} and
\ref{Lemma_3_UNI_SW}, and then following the same derivation for ML
decoding yields Theorem~\ref{thm.jointCode}.

\section{Future Directions}

\subsection{Stationary-ergodic sources and universality}

\cite{cover:75} extends the block-coding proofs to the Slepian-Wolf
problem for stationary-ergodic sources using AEP arguments. To have a
similar extension to the streaming context, possibly additional
regularity conditions will be required so that error exponents can be
achieved. To achieve universality over sources, it is possible that
further technical restrictions will be required. For the case of
distributed Markov sources however, it seems quite clear that all the
arguments in this paper will easily generalize. In that case,
following the approach we take in \cite{SahaiUnstable}, the source can
be ``segmented'' into small blocks and the endpoints\footnote{For a
  Markov source of known order $k$, the endpoint is just $k$
  successive symbols at the end of the block.} of the blocks can be
encoded perfectly at essentially zero rate. Conditioned on these
endpoints, the blocks are then iid, with the endpoints representing a
third stream of perfectly known side-information.

\subsection{Upper bounds and demonstrating optimal delays}

This paper dealt entirely with achievability of certain error
exponents. Ideally, we would have corresponding upper bounds
demonstrating that no higher exponents are possible. In the
block-coding case, problem 3.7.1 in \cite{csiszarKorner} provides a
simple upper-bound. However, the nature of the error exponents in the
streaming case might be more complicated. \cite{Chang:06} provides an
upper bound and matching achievable scheme for point-to-point
source-coding with delay and this bound extends naturally to the case
where side-information is known at both the encoder and the decoder.
\cite{ChangISIT:06} provides an upper bound for the case of
side-information known only at the decoder, and this bound is tight
for certain symmetric cases. However, both of these extended single
encoder arguments from \cite{SahaiBlockLength} that do not immediately
generalize to the case of multiple encoders.

\subsection{Trading off error exponents for the different source terminals}
For multiple terminal systems, different error exponents can be
achieved for different users or sources. For channel coding, the
encoders can choose different distributions while generating the
randomized code book to achieve an error exponent trade-off among
different users. In \cite{Weng:05}, the error exponent region is
studied for the Gaussian multiple access channel and the broadcast
channel within the block-coding paradigm. It is unclear whether
similar tradeoffs are possible within the streaming Slepian Wolf
problems considered here since there is nothing immediately comparable
to the flexibility we have in choosing the ``input distribution'' for
channel coding problems.

\subsection{Adaptation and limited feedback}
An interesting extension is to adaptive universal streaming Slepian
Wolf encoders.  The decoders we use in this paper are based on
empirical statistics.  Therefore they can be used even if source
statistics are unknown.  The current proposal will work regardless of
source and side information statistics as long as the conditional
entropy $H(\rvx|\rvy)$ is less than the encoding rate.  Even if there
is uncertainty in statistics, the anytime nature of the coding system
should enable the system to adapt on-line to the unknown entropy rate
if some feedback channel is available.  The feedback channel would be
used to order increases (or decreases) in the binning rate.  An
increase (or decrease) could be triggered by examining the difference
between two quantities: the minimal empirical joint entropy between
the decoded sequence and observation, and the empirical joint entropy
between the particular sequence and observation yielding the
second-lowest joint entropy.  If there is a large difference between
these two entropies, we are using rate excessively, and the rate of
communication can be reduced.  If the difference is negligible, then
it's likely we are not decoding correctly.  Our target should be to
keep this difference at roughly $\epsilon$.  In the current context,
this is analogous to the rate margin by which we choose to exceed the
known conditional entropy.

\section*{Acknowledgments}
The authors wish to acknowledge a desire expressed by Zixiang Xiong
and subsequent hallway discussions during ITW 2004 that helped
precipitate the current line of research. This work was supported in
part by NSF ITR Grant No.~CNS-0326503.

\appendix

\newcommand{\pBar}{\bar{p}}

\section{Proof of Theorem \ref{THM:Universal_ML_SW}}

In this section we show that the maximum likelihood (ML) error
exponent equals the universal error exponent.  We show that for all
$\gamma$,
$$E^{ML}_x(R_x,R_y,\gamma)=E^{UN}_x(R_x,R_y,\gamma)$$
Where the ML error exponent:
\begin{eqnarray}\label{eqn:LEMMAAPPDC0_SW}
E^{ML}_x(R_x,R_y,\gamma)&=&\sup_{\rho\in[0,1]}\{\gamma
E_{x|y}(R_x,\rho)+(1-\gamma)E_{xy}(R_x,R_y,\rho)\}\nonumber\\
&=&\sup_{\rho\in[0,1]}\{\rho R^{(\gamma)} -\gamma \log(\sum_{y
}(\sum_{x }p_{\rvx\rvy}(x,y)^{\frac{1}{1+\rho}})^{1+\rho})-
(1-\gamma)(1+\rho)\log(\sum_{y}\sum_{x}p_{\rvx\rvy}(x,y)^{\frac{1}{1+\rho}})\}\nonumber\\
&=&\sup_{\rho\in[0,1]}\{E^{ML}_x(R_x,R_y,\gamma,\rho)\}\nonumber
\end{eqnarray}

Write the function inside the $\sup$ argument as
$E^{ML}_x(R_x,R_y,\gamma,\rho)$. The universal error exponent:
\begin{eqnarray}
E^{UN}_x(R_x,R_y,\gamma)&=&\inf_{ q_{xy},o_{xy}} \{\gamma D(q_{xy}||p_{\rvx\rvy})+(1-\gamma)D(o_{xy}||p_{\rvx\rvy})\nonumber\\
&&+\max\{0,\gamma (R_x-H(q_{x|y}))
+(1-\gamma)(R_x+R_y-H(o_{xy}))\}\}\nonumber\\
&=&\inf_{ q_{xy},o_{xy}} \{\gamma
D(q_{xy}||p_{\rvx\rvy})+(1-\gamma)D(o_{xy}||p_{\rvx\rvy})+\max\{0,R^{(\gamma)}-\gamma
H(q_{x|y}) -(1-\gamma)H(o_{xy})\}\}\nonumber
\end{eqnarray}
Here we define $R^{(\gamma)}=\gamma R_x +(1-\gamma)(R_x+R_y)>\gamma
H(p_{\rvx|\rvy})+(1-\gamma)H(p_{\rvx\rvy})$. For notational
simplicity, we write $q_{xy}$ and $o_{xy}$ as two arbitrary joint
distributions on $\mathcal{X}\times\mathcal{Y}$ instead of
$p_{\rvxBar\rvyBar}$ and $p_{\rvxBBar\rvyBBar}$. We still write
$p_{\rvx\rvy}$ as the distribution of the source.

Before the proof, we define a pair of distributions that we will need. \\

\begin{defn}{Tilted distribution of $p_{\rvx\rvy}$}: $p^\rho_{\rvx\rvy}$, for all $ \rho\in [-1,\infty)$

$$p^\rho_{\rvx\rvy}(x,y)=\frac{p_{\rvx\rvy}(x,y)^{\frac{1}{1+\rho}}}{\sum_t\sum_s
p_{\rvx\rvy}(s,t)^{\frac{1}{1+\rho}}}$$ The entropy of the tilted
distribution is written as $H(p^\rho_{\rvx\rvy})$. Obviously
$p^0_{\rvx\rvy}=p_{\rvx\rvy}$.\\
\end{defn}

\begin{defn}   {$\rvx-\rvy$ tilted distribution of $p_{\rvx\rvy}$}: $\pBar^\rho_{\rvx\rvy}$, for all $\rho \in
[-1,+\infty)$
\begin{eqnarray}
 \pBar^\rho_{\rvx\rvy}(x,y) &=&\frac{[\sum_s p_{\rvx\rvy}(s,y)^{\frac{1}{1+\rho}}]^{1+\rho}}{\sum_t[\sum_s p_{\rvx\rvy}(s,t)^{\frac{1}{1+\rho}}]^{1+\rho}}\times\frac{p_{\rvx\rvy}(x,y)^{\frac{1}{1+\rho}}}{\sum_s p_{\rvx\rvy}(s,y)^{\frac{1}{1+\rho}}} \nonumber\\
&=&\frac{A(y,\rho)}{B(\rho)}\times\frac{C(x,y,\rho)}{D(y,\rho)}\nonumber
\end{eqnarray}
Where
\begin{eqnarray}
A(y,\rho)&=&[\sum_s p_{\rvx\rvy}(s,y)^{\frac{1}{1+\rho}}]^{1+\rho}=D(y,\rho)^{1+\rho}\nonumber\\
B(\rho)&=& \sum_s[\sum_t p_{\rvx\rvy}(s,t)^{\frac{1}{1+\rho}}]^{1+\rho} = \sum_y A(y,\rho) \nonumber\\
C(x,y,\rho)&=&p_{\rvx\rvy}(x,y)^{\frac{1}{1+\rho}}\nonumber\\
D(y,\rho)&=&\sum_s p_{\rvx\rvy}(s,y)^{\frac{1}{1+\rho}} =\sum_x
C(x,y,\rho)\nonumber
\end{eqnarray}
\end{defn}

The marginal distribution for $\rvy$ is $\frac{A(y,\rho)}{B(\rho)}$.
Obviously $\pBar^0_{\rvx\rvy}=p_{\rvx\rvy}$. Write the conditional
distribution of $\rvx$ given $\rvy$ under distribution
$\pBar^\rho_{\rvx\rvy}$ as $\pBar^\rho_{\rvx|\rvy}$, where
$\pBar^\rho_{\rvx|\rvy}(x,y)=\frac{C(x,y,\rho)}{D(y,\rho)}$,  and
the conditional entropy of $\rvx$ given $\rvy$ under distribution
$\pBar^\rho_{\rvx\rvy}$ as
$H(\pBar^\rho_{\rvx|\rvy})$. Obviously $H(\pBar^0_{\rvx|\rvy})=H(p_{\rvx|\rvy})$.\\
The conditional  entropy of $\rvx$ given $\svy$ for the $\rvx-\rvy$
tilted distribution is
$$ H(\pBar^\rho_{\rvx|\rvy=\svy})=-\sum_x
\frac{C(x,y,\rho)}{D(y,\rho)}\log(\frac{C(x,y,\rho)}{D(y,\rho)})$$\\

We introduce $ A(y,\rho)$, $ B(\rho)$, $ C(x, y,\rho)$, $ D(y,\rho)$
to simplify the notations. Some of their properties are shown in
Lemma~\ref{LEMMAAPP1_SI}.

While tilted distributions are common optimal distributions in large
deviation theory, it is useful to contemplate why we need to introduce
these {\em two} tilted distributions. In the proof of Theorem
\ref{THM:Universal_ML_SW}, through a Lagrange multiplier argument, we
will show that $\{p^\rho_{\rvx\rvy}:\rho\in [-1,+\infty)\}$ is the
family of distributions that minimize the Kullback$-$Leibler distance
to $p_{\rvx\rvy}$ with fixed \textit{entropy} and
$\{\pBar^\rho_{\rvx\rvy}:\rho\in [-1,+\infty)\}$ is the family of
distributions that minimize the Kullback$-$Leibler distance to
$p_{\rvx\rvy}$ with fixed \textit{conditional entropy}. Using a
Lagrange multiplier argument, we parametrize the universal error
exponent $E^{UN}_x(R_x,R_y,\gamma)$ in terms of $\rho$ and show the
equivalence of the universal and maximum likelihood error exponents.

Now we are ready to prove Theorem~\ref{THM:Universal_ML_SW}:
$E^{ML}_x(R_x,R_y,\gamma)=E^{UN}_x(R_x,R_y,\gamma)$.

\pf

\subsection{case 1: $\gamma
H(p_{\rvx|\rvy})+(1-\gamma)H(p_{\rvx\rvy})< R^{(\gamma)} < \gamma
H(\pBar^1_{\rvx|\rvy} )+(1-\gamma)H(p^1_{\rvx\rvy}
)$.}\label{case:1}

First, from Lemma~\ref{LEMMA_APP10} and Lemma~\ref{LEMMA_APP11}:

 $$\frac{\partial  E^{ML}_x(R_x,R_y,\gamma,\rho)}{\partial
\rho }=R^{(\gamma)}-\gamma H(\pBar^\rho_{\rvx|\rvy}
)-(1-\gamma)H(p^\rho_{\rvx\rvy})$$

Then, using Lemma~\ref{LEMMAAPP2} and Lemma~\ref{LEMMAAPP2_SI}, we
have:

 $$\frac{\partial^2  E^{ML}_x(R_x,R_y,\gamma,\rho)}{\partial
\rho }  \leq 0$$.

So  $\rho$ maximize $E^{ML}_x(R_x,R_y,\gamma,\rho)$,  if and only
if:

\begin{eqnarray}
0=\frac{\partial E^{ML}_x(R_x,R_y,\gamma,\rho)}{\partial
\rho}=R^{(\gamma)}-\gamma H(\pBar^\rho_{\rvx|\rvy}
)-(1-\gamma)H(p^\rho_{\rvx\rvy})
\end{eqnarray}

Because $R^{(\gamma)}$ is in the interval $[\gamma
H(p_{\rvx|\rvy})+(1-\gamma)H(p_{\rvx\rvy}), \gamma
H(\pBar^1_{\rvx|\rvy})+(1-\gamma)H(p^1_{\rvx\rvy})]$ and the
entropy functions monotonically-increase over $\rho$,
we can find $\rho^*\in (0,1)$, s.t.
$$\gamma H(\pBar^{\rho^*}_{\rvx|\rvy})+(1-\gamma)H(p^{\rho^*}_{\rvx\rvy})=R^{(\gamma)}$$

Using  Lemma~\ref{LEMMA_APP8}  and Lemma~\ref{LEMMA_APP9} we get:
\begin{eqnarray}
E^{ML}_x(R_x,R_y,\gamma)&=&\gamma
D(\pBar^{\rho^*}_{\rvx\rvy}\|p_{\rvx\rvy})+(1-\gamma)
D(p^{\rho^*}_{\rvx\rvy}\|p_{\rvx\rvy})\label{eqn:ML_error_expression}
\end{eqnarray}
Where $\gamma
H(\pBar^{\rho^*}_{\rvx|\rvy})+(1-\gamma)H(p^{\rho^*}_{\rvx\rvy})=R^{(\gamma)}$
, $\rho^*$ is generally unique because
 both $H(\pBar^\rho_{\rvx|\rvy})$ and $H(p^\rho_{\rvx\rvy})$ are strictly increasing with
$\rho$.\\

Secondly
\begin{eqnarray}\label{eqn:LEMMAAPPDC2_SW}
& & E^{UN}_x(R_x,R_y,\gamma)\nonumber\\
&=&\inf_{ q_{xy},o_{xy}} \{\gamma D(q_{xy}||p_{\rvx\rvy})+(1-\gamma)D(o_{xy}||p_{\rvx\rvy})+\max\{0,R^{(\gamma)}-\gamma H(q_{x|y}) -(1-\gamma)H(o_{xy})\}\}\nonumber\\
&=&  \inf_{b} \{\inf_{q_{xy},o_{xy}:\gamma H(q_{x|y}) + (1-\gamma)H(o_{xy})=b}\{\gamma D(q_{xy}||p_{\rvx\rvy})+(1-\gamma)D(o_{xy}||p_{\rvx\rvy})+\max(0,R^{(\gamma)}-b)\}\}\nonumber\\
&=&  \inf_{b\geq \gamma H(p_{\rvx|\rvy})+(1-\gamma)H(p_{\rvx\rvy})}
\{\inf_{q_{xy},o_{xy}:\gamma H(q_{x|y}) +
(1-\gamma)H(o_{xy})=b}\{\gamma
D(q_{xy}||p_{\rvx\rvy})+(1-\gamma)D(o_{xy}||p_{\rvx\rvy})\nonumber\\
&&+\max(0,R^{(\gamma)}-b)\}\}\label{eqn:optimization_equality}
\end{eqnarray}
The last equality is true because, for $b< \gamma
H(p_{\rvx|\rvy})+(1-\gamma)H(p_{\rvx\rvy})<R^{(\gamma)}$,
\begin{eqnarray}
&&\inf_{q_{xy},o_{xy}:\gamma H(q_{x|y}) + (1-\gamma)H(o_{xy})=b}\{\gamma D(q_{xy}||p_{\rvx\rvy})+(1-\gamma)D(o_{xy}||p_{\rvx\rvy})+\max(0,R^{(\gamma)}-b)\}\}\nonumber\\
&\geq& 0 + R^{(\gamma)}-b \nonumber\\
&=&\inf_{q_{xy},o_{xy}:H(q_{x|y})=H(p_{\rvx|\rvy}),H(o_{xy})=H(p_{\rvx\rvy})}\{\gamma D(q_{xy}||p_{\rvx\rvy})+(1-\gamma)D(o_{xy}||p_{\rvx\rvy})+\max(0,R^{(\gamma)}-b)\}\}\nonumber\\
&\geq&\inf_{q_{xy},o_{xy}:H(q_{x|y})=H(p_{\rvx|\rvy}),H(o_{xy})=H(p_{\rvx\rvy})}\{\gamma D(q_{xy}||p_{\rvx\rvy})+(1-\gamma)D(o_{xy}||p_{\rvx\rvy})\nonumber\\
&&+\max(0,R^{(\gamma)}-\gamma H(p_{\rvx|\rvy})+(1-\gamma)H(p_{\rvx\rvy}))\}\}\nonumber\\
&\geq&\inf_{q_{xy},o_{xy}: \gamma H(q_{x|y})+(1-\gamma)H(o_{xy}) =\gamma H(p_{\rvx|\rvy})+(1-\gamma)H(p_{\rvx\rvy})  }\{\gamma D(q_{xy}||p_{\rvx\rvy})+(1-\gamma)D(o_{xy}||p_{\rvx\rvy})\nonumber\\
&&+\max(0,R^{(\gamma)}-\gamma
H(p_{\rvx|\rvy})+(1-\gamma)H(p_{\rvx\rvy}))\}\}\nonumber
\end{eqnarray}
Fixing $b\geq \gamma H(p_{\rvx|\rvy})+(1-\gamma)H(p_{\rvx\rvy})$, the
inner infimum in (\ref{eqn:optimization_equality}) is an
optimization problem on $q_{xy}, o_{xy}$ with equality constraints
$\sum_x\sum_y q_{xy}(x,y)=1$, $\sum_x\sum_y o_{xy}(x,y)=1$ and
$\gamma H(q_{x|y})+(1-\gamma)H(o_{xy})=b$ and the obvious inequality
constraints $ 0\leq q_{xy}(x,y)\leq 1, 0\leq o_{xy}(x,y)\leq 1,
\forall x,y$. In the following formulation of the optimization
problem, we relax one equality constraint to an inequality
constraint $\gamma H(q_{x|y})+(1-\gamma)H(o_{xy})\geq b$ to make the
optimization problem $convex$. It turns out later that the optimal
solution to the relaxed problem is also the optimal solution to the
original problem because $b\geq \gamma
H(p_{\rvx|\rvy})+(1-\gamma)H(p_{\rvx\rvy}) $. The resulting
optimization problem is:  
\begin{eqnarray}
&&\inf_{q_{xy}, o_{xy}} \{\gamma D(q_{xy}||p_{\rvx\rvy})+(1-\gamma)D(o_{xy}||p_{\rvx\rvy})\} \nonumber\\
&&\mbox{s.t.}\sum_x\sum_y  q_{xy}(x,y)=1\nonumber\\
&&\sum_x\sum_y  o_{xy}(x,y)=1\nonumber\\
&& b- \gamma H(q_{x|y})-(1-\gamma)H(o_{xy})\leq 0 \nonumber\\
&& 0\leq q_{xy}(x,y)\leq 1, \ \ \forall (x,y)\in
 \mathcal{X}\times\mathcal{Y}\nonumber\\
 && 0\leq o_{xy}(x,y)\leq 1, \ \ \forall (x,y)\in
 \mathcal{X}\times\mathcal{Y}\label{eqn:convex_opt_setup}
\end{eqnarray}
The above optimization problem is {\em convex} because the objective
function and the inequality constraint functions are convex and the
equality constraint functions are affine\cite{Boyd2004}.  The
Lagrange multiplier function for this convex optimization problem is:

\begin{eqnarray}
&&L(q_{xy},o_{xy},\rho,\mu_1,\mu_2, {\nu}_1, {\nu}_2, {\nu}_3 , {\nu}_4)\nonumber\\
&=& \gamma
D(q_{xy}||p_{\rvx\rvy})+(1-\gamma)D(o_{xy}||p_{\rvx\rvy})\nonumber\\
&&+\mu_1(\sum_x\sum_y  q_{xy}(x,y)-1) +\mu_2(\sum_x\sum_y
o_{xy}(x,y)-1)\nonumber\\
&&+\rho(b-\gamma H(q_{x|y})-(1-\gamma)H(o_{xy}))\nonumber\\
&&+\sum_x\sum_y\big\{   {\nu}_1(x,y)(-q_{xy}(x,y))+
{\nu}_2(x,y)(1-q_{xy}(x,y)) +{\nu}_3(x,y)(-o_{xy}(x,y))
+{\nu}_4(x,y)(1-o_{xy}(x,y))\big\}\nonumber\\
\end{eqnarray}
Where $\rho,\mu_1,\mu_2$ are real numbers and ${\nu}_i\in R^{
|\mathcal{X}||\mathcal{Y}|}$, $i=1,2,3,4$.

According to the KKT conditions for convex
optimization\cite{Boyd2004},  $q_{xy}, o_{xy}$  minimize the 
convex optimization problem in (\ref{eqn:convex_opt_setup}) if and
only if the following conditions are simultaneously satisfied for
some $q_{xy}$, $o_{xy}$, $\mu_1$, $\mu_2$, $\nu_1$, $\nu_2$,
$\nu_3$, $\nu_4$ and $\rho$:

\begin{eqnarray}
0&=&\frac{\partial L(q_{xy},o_{xy},\rho,\mu_1,\mu_2,{\nu}_1, {\nu}_2, {\nu}_3 , {\nu}_4)}{\partial q_{xy}(x,y)} \nonumber\\
&=& \gamma[-\log (p_{\rvx\rvy}(x,y))+(1+\rho) (1+\log(q_{xy}(x,y)))+ \rho \log(\sum_{s}q_{xy}(s,y))] +\mu_1- \nu_1(x,y)- \nu_2(x,y)\nonumber\\
0&=&\frac{\partial L(q_{xy},o_{xy},\rho,\mu_1,\mu_2,{\nu}_1, {\nu}_2, {\nu}_3 , {\nu}_4)}{\partial o_{xy}(x,y)} \nonumber\\
&=& (1-\gamma) [-\log (p_{\rvx\rvy}(x,y))+(1+\rho)
(1+\log(o_{xy}(x,y)))]+\mu_2- \nu_3(x,y)- \nu_4(x,y)
\label{eqn:multiplier1}
\end{eqnarray}
For all $x$, $y$ and
\begin{eqnarray}
&&\sum_x\sum_y  q_{xy}(x,y)=1\nonumber\\
&&\sum_x\sum_y  o_{xy}(x,y)=1\nonumber\\
&&\rho( \gamma H(q_{x|y})+(1-\gamma)H(o_{xy})-b)=0\nonumber\\
&& \rho \geq 0\nonumber\\
&& \nu_1(x,y) (-q_{xy}(x,y))=0, \ \ \  \nu_2(x,y) (1-q_{xy}(x,y))=0\
\ \ \forall x,y \nonumber\\
&& \nu_3(x,y) (-o_{xy}(x,y))=0, \ \ \  \nu_4(x,y) (1-o_{xy}(x,y))=0\
\ \ \forall x,y \nonumber\\
&&\nu_i(x,y)\geq 0, \ \ \ \forall x,y, i=1,2,3,4
\label{eqn:multiplier2}
\end{eqnarray}

Solving the above standard Lagrange multiplier equations
(\ref{eqn:multiplier1}) and (\ref{eqn:multiplier2}), we have:

\begin{eqnarray}
q_{xy}(x,y)&=&\frac{[\sum_s p_{\rvx\rvy}(s,y)^{\frac{1}{1+\rho_b}}]^{1+\rho_b}}{\sum_t[\sum_s p_{\rvx\rvy}(s,t)^{\frac{1}{1+\rho_b}}]^{1+\rho_b}}\frac{p_{\rvx\rvy}(x,y)^{\frac{1}{1+\rho_b}}}{\sum_s p_{\rvx\rvy}(s,y)^{\frac{1}{1+\rho_b}}} \nonumber\\
&=& {\pBar^{\rho_b}_{\rvx\rvy}(x,y)}\nonumber\\
o_{xy}(x,y)&=&\frac{p_{\rvx\rvy}(x,y)^{\frac{1}{1+\rho_b}}}{\sum_t\sum_s p_{\rvx\rvy}(s,t)^{\frac{1}{1+\rho_b}}} \nonumber\\
&=& {p^{\rho_b}_{\rvx\rvy}(x,y)}\nonumber\\
\nu_i(x,y)&=&0\ \ \  \forall x,y, i=1,2,3,4\nonumber\\
\rho &=&\rho_b
\end{eqnarray}
Where $\rho_b$ satisfies the following condition $$\gamma
H(\pBar^{\rho_b}_{\rvx|\rvy})+(1-\gamma)H(p^{\rho_b}_{\rvx\rvy})=b
\geq \gamma H(p_{\rvx|\rvy})+(1-\gamma)H(p_{\rvx\rvy})$$
and thus
$\rho_b\geq 0$ because both $H(\pBar^{\rho}_{\rvx|\rvy})$ and
$H(p^{\rho}_{\rvx\rvy})$ are monotonically increasing with $\rho$ as
shown in Lemma~\ref{LEMMAAPP2} and Lemma~\ref{LEMMAAPP2_SI}.

Notice that all the KKT conditions are simultaneously satisfied with
the inequality constraint $\gamma H(q_{x|y})+(1-\gamma)H(o_{xy})\geq
b$ being met with equality. Thus, the relaxed optimization problem has
the same optimal solution as the original problem as promised. The
optimal $q_{xy}$ and $o_{xy}$ are the $\rvx-\rvy$ tilted distribution
$\pBar^{\rho_b}_{\rvx\rvy}$ and standard tilted distribution
$p^{\rho_b}_{\rvx\rvy}$ of $p_{\rvx\rvy}$ with the same parameter
$\rho_b\geq 0$. chosen s.t.
$$\gamma H(\pBar^{\rho_b}_{\rvx|\rvy})+(1-\gamma)H(p^{\rho_b}_{\rvx\rvy})=b$$
 Now we have :
\begin{eqnarray}
&&E^{UN}_x(R_x,R_y,\gamma)\nonumber\\
&=&\inf_{b\geq \gamma H(p_{\rvx|\rvy})+(1-\gamma)H(p_{\rvx\rvy})} \{\inf_{q_{xy},o_{xy}:\gamma H(q_{x|y}) + (1-\gamma)H(o_{xy})=b} \{\gamma D(q_{xy}||p_{\rvx\rvy})+(1-\gamma)D(o_{xy}||p_{\rvx\rvy})+\max(0,R^{(\gamma)}-b)\}\}\nonumber\\
&=&\inf_{b\geq \gamma H(p_{\rvx|\rvy})+(1-\gamma)H(p_{\rvx\rvy})}\{\gamma D(\pBar^{\rho_b}_{\rvx\rvy}||p_{\rvx\rvy})+(1-\gamma )D(p^{\rho_b}_{\rvx\rvy}||p_{\rvx\rvy})+\max(0,R^{(\gamma)}-b)\}\nonumber\\
&=& \min [\inf_{\rho\geq 0: R^{(\gamma)} \geq \gamma H(\pBar^\rho_{\rvx|\rvy})+(1-\gamma)H(p^\rho_{\rvx\rvy})}\{\gamma D(\pBar^\rho_{\rvx\rvy}||p_{\rvx\rvy})+(1-\gamma )D(p_{\rvx\rvy_{\rho}}||p_{\rvx\rvy})+R^{(\gamma)}-\gamma H(\pBar^\rho_{\rvx|\rvy})-(1-\gamma)H(p^\rho_{\rvx\rvy})\},\nonumber\\
&&\inf_{\rho \geq 0: R^{(\gamma)} \leq \gamma
H(\pBar^\rho_{\rvx|\rvy})+(1-\gamma)H(p^\rho_{\rvx\rvy})}\{\gamma
D(\pBar^\rho_{\rvx\rvy}||p_{\rvx\rvy})+(1-\gamma
)D(p_{\rvx\rvy_{\rho}}||p_{\rvx\rvy})\}]\label{eqn:big_two_conditions}
\end{eqnarray}
Notice that $H(p^\rho_{\rvx\rvy})$, $H(\pBar^\rho_{\rvx|\rvy})$,
$D(\pBar^\rho_{\rvx\rvy}||p_{\rvx\rvy})$ and
$D(p^\rho_{\rvx\rvy}||p_{\rvx\rvy})$ are all strictly increasing with
$\rho>0$ as shown in Lemma~\ref{LEMMAAPP2_SI},
Lemma~\ref{LEMMAAPP3_SI}, Lemma~\ref{LEMMAAPP2} and
Lemma~\ref{LEMMAAPP3} later in this appendix. We have:
\begin{eqnarray}
& & \inf_{\rho \geq 0: R^{(\gamma)} \leq \gamma
H(\pBar^\rho_{\rvx|\rvy})+(1-\gamma)H(p^\rho_{\rvx\rvy})}\{\gamma
D(\pBar^\rho_{\rvx\rvy}||p_{\rvx\rvy})+(1-\gamma
)D(p^\rho_{\rvx\rvy}||p_{\rvx\rvy})\} \nonumber\\
&=&\gamma
D(\pBar^{\rho^*}_{\rvx\rvy}||p_{\rvx\rvy})+(1-\gamma
)D(p^{\rho^*}_{\rvx\rvy}||p_{\rvx\rvy})\label{eqn:condition1}
\end{eqnarray}
where $R^{(\gamma)} =\gamma
H(\pBar^{\rho^*}_{\rvx|\rvy})+(1-\gamma)H(p^{\rho^*}_{\rvx\rvy})$.
Applying the results in Lemma~\ref{LEMMAAPP4_SI} and
Lemma~\ref{LEMMAAPP4}, we get:
\begin{eqnarray}
&&\inf_{\rho \geq 0: R^{(\gamma)} \geq \gamma H(\pBar^\rho_{\rvx|\rvy})+(1-\gamma)H(p^\rho_{\rvx\rvy})}\{\gamma D(\pBar^\rho_{\rvx\rvy}||p_{\rvx\rvy})+(1-\gamma )D(p^\rho_{\rvx\rvy}||p_{\rvx\rvy})+R^{(\gamma)}-\gamma H(\pBar^\rho_{\rvx|\rvy})-(1-\gamma)H(p^\rho_{\rvx\rvy})\}\nonumber\\
&&=\gamma D(\pBar^\rho_{\rvx\rvy}||p_{\rvx\rvy})+(1-\gamma
)D(p^\rho_{\rvx\rvy}||p_{\rvx\rvy}) +R^{(\gamma)}-\gamma
H(\pBar^\rho_{\rvx|\rvy})-(1-\gamma)H(p^\rho_{\rvx\rvy})|_{\rho=\rho^*}\nonumber\\
&&=\gamma D(\pBar^{\rho^*}_{\rvx\rvy}||p_{\rvx\rvy})+(1-\gamma
)D(p^{\rho^*}_{\rvx\rvy}||p_{\rvx\rvy})\label{eqn:condition2}
\end{eqnarray} This is true because for $\rho : R^{(\gamma)} \geq
\gamma H(\pBar^\rho_{\rvx|\rvy})+(1-\gamma)H(p^\rho_{\rvx\rvy})$,
we know $\rho\leq 1$ because of the range of $R^{(\gamma)}$:
$R^{(\gamma)} < \gamma H(\pBar^1_{\rvx|\rvy}
)+(1-\gamma)H(p^1_{\rvx\rvy} )$. Substituting (\ref{eqn:condition1})
and (\ref{eqn:condition2}) into (\ref{eqn:big_two_conditions}), we
get
\begin{eqnarray}
E^{UN}_x(R_x,R_y,\gamma)&=&\gamma
D(\pBar^{\rho^*}_{\rvx\rvy}||p_{\rvx\rvy})+(1-\gamma
)D(p^{\rho^*}_{\rvx\rvy}||p_{\rvx\rvy})\nonumber\\
&& \mbox{where }  \ \ R^{(\gamma)} =\gamma
H(\pBar^{\rho^*}_{\rvx|\rvy})+(1-\gamma)H(p^{\rho^*}_{\rvx\rvy})
\end{eqnarray}
So for $\gamma H(p_{\rvx|\rvy})+(1-\gamma)H(p_{\rvx\rvy})\leq
R^{(\gamma)} \leq \gamma
H(\pBar^1_{\rvx|\rvy})+(1-\gamma)H(p^1_{\rvx\rvy})$, from
(\ref{eqn:ML_error_expression}) we have the desired property:
$$E^{ML}_x(R_x,R_y,\gamma)=E^{UN}_x(R_x,R_y,\gamma)$$

\subsection{case 2: $ R^{(\gamma)} \geq \gamma
H(\pBar^1_{\rvx|\rvy})+(1-\gamma)H(p^1_{\rvx\rvy})$.}\label{case:2}

In this case, for all $0\leq \rho\leq 1$
$$\frac{\partial  E^{ML}_x(R_x,R_y,\gamma,\rho)}{\partial
\rho }=R^{(\gamma)}-\gamma H(\pBar^\rho_{\rvx|\rvy}
)-(1-\gamma)H(p^\rho_{\rvx\rvy})\geq R^{(\gamma)}-\gamma
H(\pBar^1_{\rvx|\rvy} )-(1-\gamma)H(p^1_{\rvx\rvy})\geq 0$$

So $\rho$ takes value $1$ to maximize the error exponent
$E^{ML}_x(R_x,R_y,\gamma,\rho)$, thus
\begin{eqnarray}
E^{ML}_x(R_x,R_y,\gamma)=R^{(\gamma)} -\gamma
\log(\sum_{y}(\sum_{x}p_{\rvx\rvy}(x,y)^{\frac{1}{2}})^{2})-
2(1-\gamma)\log(\sum_{y}\sum_{x}p_{\rvx\rvy}(x,y)^{\frac{1}{2}})
\end{eqnarray}

Using the same convex optimization techniques as case \ref{case:1}, we
notice the fact that $\rho^*\geq 1$ for $R^{(\gamma)}
=\gamma
H(\pBar^{\rho^*}_{\rvx|\rvy})+(1-\gamma)H(p^{\rho^*}_{\rvx\rvy})$.
Then applying Lemma~\ref{LEMMAAPP4_SI} and Lemma~\ref{LEMMAAPP4}, we
have:
\begin{eqnarray}
 &&\inf_{\rho\geq 0: R^{(\gamma)} \geq \gamma H(\pBar^\rho_{\rvx|\rvy})+(1-\gamma)H(p^\rho_{\rvx\rvy})}\{\gamma D(\pBar^\rho_{\rvx\rvy}||p_{\rvx\rvy})+(1-\gamma )D(p^\rho_{\rvx\rvy}||p_{\rvx\rvy})+R^{(\gamma)}-\gamma H(\pBar^\rho_{\rvx|\rvy})-(1-\gamma)H(p_{\rvx\rvy_\rho})\},\nonumber\\
&&=\gamma D(\pBar^1_{\rvx\rvy}||p_{\rvx\rvy})+(1-\gamma
)D(p^1_{\rvx\rvy}||p_{\rvx\rvy})+R^{(\gamma)}-\gamma
H(\pBar^{1}_{\rvx|\rvy})-(1-\gamma)H(p^1_{\rvx\rvy})\nonumber
\end{eqnarray}
And
\begin{eqnarray}
&&\inf_{\rho \geq 0: R^{(\gamma)} \leq \gamma H(\pBar^\rho_{\rvx|\rvy})+(1-\gamma)H(p^\rho_{\rvx\rvy})}\{\gamma D(\pBar^\rho_{\rvx\rvy}||p_{\rvx\rvy})+(1-\gamma )D(p^\rho_{\rvx\rvy}||p_{\rvx\rvy})\}]\nonumber\\
&&=\gamma D(\pBar^{\rho^*}_{\rvx\rvy}||p_{\rvx\rvy})+(1-\gamma
)D(p^{\rho^*}_{\rvx\rvy}||p_{\rvx\rvy})\nonumber\\
&&=\gamma D(\pBar^{\rho^*}_{\rvx\rvy}||p_{\rvx\rvy})+(1-\gamma
)D(p^{\rho^*}_{\rvx\rvy}||p_{\rvx\rvy})+R^{(\gamma)}-\gamma
H(\pBar^{\rho^*}_{\rvx|\rvy})-(1-\gamma)H(p^{\rho^*}_{\rvx\rvy})\nonumber\\
&&\leq \gamma D(\pBar^1_{\rvx\rvy}||p_{\rvx\rvy})+(1-\gamma
)D(p^1_{\rvx\rvy}||p_{\rvx\rvy})+R^{(\gamma)}-\gamma
H(\pBar^{1}_{\rvx|\rvy})-(1-\gamma)H(p^1_{\rvx\rvy})\nonumber
 \end{eqnarray}

Finally:
\begin{eqnarray}
&&E^{UN}_x(R_x,R_y,\gamma)\nonumber\\
&=&\inf_{b\geq \gamma H(p_{\rvx|\rvy})+(1-\gamma)H(p_{\rvx\rvy})} \{\inf_{q_{xy},o_{xy}:\gamma H(q_{x|y}) + (1-\gamma)H(o_{xy})=b} \{\gamma D(q_{xy}||p_{\rvx\rvy})+(1-\gamma)D(o_{xy}||p_{\rvx\rvy})+\max(0,R^{(\gamma)}-b)\}\}\nonumber\\
&=&\inf_{b\geq \gamma H(p_{\rvx|\rvy})+(1-\gamma)H(p_{\rvx\rvy})}\{\gamma D(\pBar^{\rho_b}_{\rvx\rvy}||p_{\rvx\rvy})+(1-\gamma )D(p^{\rho_b}_{\rvx\rvy}||p_{\rvx\rvy})+\max(0,R^{(\gamma)}-b)\}\nonumber\\
&=&\min [\inf_{\rho\geq 0: R^{(\gamma)} \geq \gamma H(\pBar^\rho_{\rvx|\rvy})+(1-\gamma)H(p^\rho_{\rvx\rvy})}\{\gamma D(\pBar^\rho_{\rvx\rvy}||p_{\rvx\rvy})+(1-\gamma )D(p^\rho_{\rvx\rvy}||p_{\rvx\rvy})+R^{(\gamma)}-\gamma H(\pBar^\rho_{\rvx|\rvy})-(1-\gamma)H(p^\rho_{\rvx\rvy})\},\nonumber\\
&&\inf_{\rho \geq 0: R^{(\gamma)} \leq \gamma H(\pBar^\rho_{\rvx|\rvy})+(1-\gamma)H(p^\rho_{\rvx\rvy})}\{\gamma D(\pBar^\rho_{\rvx\rvy}||p_{\rvx\rvy})+(1-\gamma )D(p^\rho_{\rvx\rvy}||p_{\rvx\rvy})\}]\nonumber\\
&=&\gamma D(\pBar^1_{\rvx\rvy}||p_{\rvx\rvy})+(1-\gamma
)D(p^1_{\rvx\rvy}||p_{\rvx\rvy})+R^{(\gamma)}-\gamma
H(\pBar^1_{\rvx|\rvy})-(1-\gamma)H(p^1_{\rvx\rvy})\nonumber\\
&=&R^{(\gamma)} -\gamma \log(\sum_{y }(\sum_{x
}p_{\rvx\rvy}(x,y)^{\frac{1}{2}})^{2})- 2(1-\gamma)\log(\sum_{y
}\sum_{x }p_{\rvx\rvy}(x,y)^{\frac{1}{2}})
\end{eqnarray}
The last equality is true by setting $\rho =1$ in
Lemma~\ref{LEMMA_APP8} and Lemma~\ref{LEMMA_APP9}.

Again,  $E^{ML}_x(R_x,R_y,\gamma)=E^{UN}_x(R_x,R_y,\gamma)$, thus we finish the
proof.\hfill$\blacksquare$\\

\subsection{Technical Lemmas}
Some technical lemmas we used in the above proof of
Theorem~\ref{THM:Universal_ML_SW} are now discussed:

\begin{lemma}\label{LEMMAAPP2}
$\frac{\partial H(p^\rho_{\rvx\rvy})}{\partial \rho}\geq0$
\end{lemma}
\pf  From the definition of the tilted distribution we have the
following observation:

$\log(p^\rho_{\rvx\rvy}(x_1,y_1))-\log
(p^\rho_{\rvx\rvy}(x_2,y_2))=\log(p_{\rvx\rvy}(x_1,y_1)^{\frac{1}{1+
 \rho}})-\log(p_{\rvx\rvy}(x_2,y_2)^{\frac{1}{1+
 \rho}})$\\ Using the above equality, we first derive the derivative
 of the tilted distribution,  for all  $x,y$

\begin{eqnarray}
 \frac{\partial p^\rho_{\rvx\rvy}(x,y)  }{\partial \rho}
 &=&\frac{-1}{(1+\rho)^2}
 \frac{ p_{\rvx\rvy}(x,y)^{\frac{1}{1+
 \rho}}\log(p_{\rvx\rvy}(x,y)) (\sum_t\sum_s
p_{\rvx\rvy}(s,t)^{\frac{1}{1+\rho}})}{(\sum_t\sum_s
p_{\rvx\rvy}(s,t)^{\frac{1}{1+\rho}})^2}
\nonumber\\
&&-\frac{-1}{(1+\rho)^2}
 \frac{p_{\rvx\rvy}(x,y)^{\frac{1}{1+
 \rho}} (\sum_t\sum_s
p_{\rvx\rvy}(s,t)^{\frac{1}{1+\rho}}
\log(p_{\rvx\rvy}(s,t)))}{(\sum_t\sum_s
p_{\rvx\rvy}(s,t)^{\frac{1}{1+\rho}})^2}
\nonumber\\
&=&\frac{-1}{1+\rho}
 p^\rho_{\rvx\rvy}(x,y)[ \log(p_{\rvx\rvy}(x,y)^{\frac{1}{1+
 \rho}})-\sum_t\sum_s
p^\rho_{\rvx\rvy}(s,t)\log(p_{\rvx\rvy}(s,t)^{\frac{1}{1+
 \rho}})]
\nonumber\\
&=&\frac{-1}{1+\rho}
 p^\rho_{\rvx\rvy}(x,y)[ \log(p^\rho_{\rvx\rvy}(x,y))-\sum_t\sum_s
p^\rho_{\rvx\rvy}(s,t)\log(p^\rho_{\rvx\rvy}(s,t))]\nonumber\\
&=&-\frac{p^\rho_{\rvx\rvy}(x,y)}{1+\rho}[\log(p^\rho_{\rvx\rvy}(x,y))+H(p^\rho_{\rvx\rvy})]
\end{eqnarray}
Then:

\begin{eqnarray}
\frac{\partial H(p^\rho_{\rvx\rvy})}{\partial \rho}&=&-\frac{\partial \sum_{x,y} p^\rho_{\rvx\rvy}(x,y) \log( p^\rho_{\rvx\rvy}(x,y))}{\partial \rho}\nonumber\\
&=&-\sum_{x,y} (1+\log(p^\rho_{\rvx\rvy}(x,y)))\frac{\partial p^\rho_{\rvx\rvy}(x,y)}{\partial \rho}\nonumber \\
&=&\sum_{x,y} (1+\log(p^\rho_{\rvx\rvy}(x,y)))\frac{p^\rho_{\rvx\rvy}(x,y)}{1+\rho}(\log(p^\rho_{\rvx\rvy}(x,y))+H(p^\rho_{\rvx\rvy}))\nonumber\\
&=&\frac{1}{1+\rho}\sum_{x,y} p^\rho_{\rvx\rvy}(x,y) \log(p^\rho_{\rvx\rvy}(x,y)) (\log(p^\rho_{\rvx\rvy}(x,y))+H(p^\rho_{\rvx\rvy}))\nonumber\\
&=&\frac{1}{1+\rho}[\sum_{x,y}p^\rho_{\rvx\rvy}(x,y) (\log(p^\rho_{\rvx\rvy}(x,y)))^2-H(p^\rho_{\rvx\rvy})^2]\nonumber\\
&=&\frac{1}{1+\rho}[\sum_{x,y} p^\rho_{\rvx\rvy}(x,y) (\log(p^\rho_{\rvx\rvy}(x,y)))^2\sum_{x,y} p^\rho_{\rvx\rvy}(x,y)-H(p^\rho_{\rvx\rvy})^2]\nonumber\\
&\geq_{(a)}&\frac{1}{1+\rho}[(\sum_{x,y}p^\rho_{\rvx\rvy}(x,y) \log(p^\rho_{\rvx\rvy}(x,y)))^2-H(p^\rho_{\rvx\rvy})^2]\nonumber\\
&=& 0
\end{eqnarray}
where (a) is true by the Cauchy-Schwartz inequality. \hfill$\blacksquare$\\

\begin{lemma}\label{LEMMAAPP3}
$\frac{\partial D(p^\rho_{\rvx\rvy}\|P)}{\partial
\rho}=\rho\frac{\partial H(p^\rho_{\rvx\rvy})}{\partial \rho} $
\end{lemma}
\pf  As shown in Lemma~\ref{LEMMA_APP8} and Lemma~\ref{LEMMA_APP10}
respectively:
$$D(p^\rho_{\rvx\rvy}\|p_{\rvx\rvy})=\rho
H(p^\rho_{\rvx\rvy})-(1+\rho) \log(\sum_{x,y
}p_{\rvx\rvy}(x,y)^{\frac{1}{1+\rho}})$$
$$  H(p^\rho_{\rvx\rvy})=\frac{\partial (1+\rho)\log(\sum_{y}\sum_{x}p_{\rvx\rvy}(x,y)^{\frac{1}{1+\rho}})}{\partial \rho} $$

We have:
\begin{eqnarray}
\frac{\partial D(p^\rho_{\rvx\rvy}\|p_{\rvx\rvy})}{\partial \rho}&=&
H(p^\rho_{\rvx\rvy}) +\rho\frac{\partial
H(p^\rho_{\rvx\rvy})}{\partial \rho}-\frac{\partial (1+\rho)\log(\sum_{y}\sum_{x}p_{\rvx\rvy}(x,y)^{\frac{1}{1+\rho}})}{\partial \rho}\nonumber\\
&=&  H(p^\rho_{\rvx\rvy}) +\rho\frac{\partial
H(p^\rho_{\rvx\rvy})}{\partial \rho}-H(p^\rho_{\rvx\rvy})  \nonumber\\
&=&\rho\frac{\partial H(p^\rho_{\rvx\rvy})}{\partial \rho}
\end{eqnarray}
 \hfill$\blacksquare$\\

\begin{lemma}\label{LEMMAAPP4}
$sign\frac{\partial
[D(p^\rho_{\rvx\rvy}\|p_{\rvx\rvy})-H(p^\rho_{\rvx\rvy})]}{\partial
\rho}=sign(\rho-1)$.

\end{lemma}
\pf  Combining the results of the previous two lemmas, we have:
\begin{eqnarray}
&&\frac{\partial
D(p^\rho_{\rvx\rvy}\|p_{\rvx\rvy})-H(p^\rho_{\rvx\rvy})}{\partial
\rho}=(\rho-1)\frac{\partial H(p^\rho_{\rvx\rvy})}{\partial
\rho}=sign(\rho-1)\nonumber
\end{eqnarray} \hfill$\blacksquare$\\

\begin{lemma}\label{LEMMAAPP1_SI} Properties of
$\frac{\partial A(y,\rho)}{\partial \rho}$, $\frac{\partial
B(\rho)}{\partial \rho}$,  $\frac{\partial C(x, y,\rho)}{\partial
\rho}$, $\frac{\partial D(y,\rho)}{\partial \rho}$ and
$\frac{\partial H(\pBar^\rho_{\rvx|\rvy=\svy})}{\partial \rho}$
\end{lemma}

First,
\begin{eqnarray}
\frac{\partial C(x, y,\rho)}{\partial \rho}&=&\frac{\partial p_{\rvx\rvy}(x,y)^{\frac{1}{1+\rho}}}{\partial \rho}= -\frac{1}{1+\rho}p_{\rvx\rvy}(x,y)^{\frac{1}{1+\rho}} \log(p_{\rvx\rvy}(x,y)^{\frac{1}{1+\rho}})\nonumber\\
&=&-\frac{C(x,y,\rho)}{1+\rho}\log(C(x,y,\rho))\nonumber\\
\frac{\partial D(y,\rho)}{\partial \rho}&=&\frac{\partial \sum_ s
p_{\rvx\rvy}(s,y)^{\frac{1}{1+\rho}}}{\partial
\rho}=-\frac{1}{1+\rho}\sum_s p_{\rvx\rvy}(s,y)^{\frac{1}{1+\rho}}
\log(p_{\rvx\rvy}(s,y)^{\frac{1}{1+\rho}})\nonumber\\&=&-\frac{\sum_x
C(x,y,\rho)\log(C(x,y,\rho))}{1+\rho}
\end{eqnarray}

For a differentiable function $f(\rho)$,
$$\frac{\partial f(\rho)^{1+\rho}}{\partial \rho}=f(\rho)^{1+\rho}\log(f(\rho))+ (1+\rho)f(\rho)^\rho\frac{\partial f(\rho)}{\partial
\rho}$$ So
\begin{eqnarray}
\frac{\partial A(y,\rho)}{\partial \rho}&=&\frac{\partial D(y,\rho)^{1+\rho}}{\partial \rho}= D(y,\rho)^{1+\rho}\log( D(y,\rho))+ (1+\rho) D(y,\rho)^\rho\frac{\partial  D(y,\rho)}{\partial \rho}\nonumber\\
&=&D(y,\rho)^{1+\rho}(\log(D(y,\rho))- \sum_x \frac{C(x,y,\rho)}{D(y,\rho)}\log(C(x,y,\rho)))\nonumber\\
&=&D(y,\rho)^{1+\rho} (- \sum_x \frac{C(x,y,\rho)}{D(y,\rho)}\log(\frac{C(x,y,\rho)}{D(y,\rho))}))\nonumber\\
&=& A(y,\rho) H(\pBar^\rho_{\rvx|\rvy=\svy})\nonumber\\
\frac{\partial B(\rho)}{\partial \rho}&=&\sum_y \frac{\partial
A(y,\rho)}{\partial \rho}=\sum_y
A(y,\rho)H(\pBar^\rho_{\rvx|\rvy=\svy})=B(\rho)\sum_y
\frac{A(y,\rho)}{B(\rho)}H(\pBar^\rho_{\rvx|\rvy=\svy})
=B(\rho)H(\pBar^\rho_{\rvx|\rvy})\nonumber
\end{eqnarray}
And last:

\begin{eqnarray}
& & \frac{\partial H(\pBar^\rho_{\rvx|\rvy=\svy})}{\partial
\rho} \nonumber\\
&=&-\sum_x [\frac{\frac{\partial C(x,y,\rho)}{\partial \rho}}{D(y,\rho)}-\frac{C(x,y,\rho)\frac{\partial D(y,\rho)}{\partial \rho}}{D(y,\rho)^2}][1+\log(\frac{C(x,y,\rho)}{D(y,\rho)})]\nonumber\\
 &=&-\sum_x
 [\frac{-\frac{C(x,y,\rho)}{1+\rho}\log(C(x,y,\rho))}{D(y,\rho)}+\frac{C(x,y,\rho)\frac{\sum_s
C(s,y,\rho)\log(C(s,y,\rho))}{1+\rho}}{D(y,\rho)^2}][1+\log(\frac{C(x,y,\rho)}{D(y,\rho)})]\nonumber\\
 &=&\frac{1}{1+\rho}\sum_x
 [\pBar^\rho_{\rvx|\rvy}(x,y)\log(C(x,y,\rho))-\pBar^\rho_{\rvx|\rvy}(x,y) \sum_s
\pBar^\rho_{\rvx|\rvy}(s,y)\log(C(s,y,\rho))][1+\log(\pBar^\rho_{\rvx|\rvy}(x,y))]\nonumber\\
 &=&\frac{1}{1+\rho}\sum_x
 \pBar^\rho_{\rvx|\rvy}(x,y)[\log(\pBar^\rho_{\rvx|\rvy}(x,y))- \sum_s
\pBar^\rho_{\rvx|\rvy}(s,y)\log(\pBar^\rho_{\rvx|\rvy}(s,y))][1+\log(\pBar^\rho_{\rvx|\rvy}(x,y))]\nonumber\\
 &=&\frac{1}{1+\rho}\sum_x
 \pBar^\rho_{\rvx|\rvy}(x,y)\log(\pBar^\rho_{\rvx|\rvy}(x,y))[\log(\pBar^\rho_{\rvx|\rvy}(x,y))- \sum_s
\pBar^\rho_{\rvx|\rvy}(s,y)\log(\pBar^\rho_{\rvx|\rvy}(s,y))]\nonumber\\
 &=&\frac{1}{1+\rho}\sum_x
 \pBar^\rho_{\rvx|\rvy}(x,y)\log(\pBar^\rho_{\rvx|\rvy}(x,y))\log(\pBar^\rho_{\rvx|\rvy}(x,y)) -\frac{1}{1+
 \rho} [\sum_x
\pBar^\rho_{\rvx|\rvy}(x,y)\log(\pBar^\rho_{\rvx|\rvy}(x,y))]^2\nonumber\\
&\geq& 0
\end{eqnarray}
The inequality is true by the Cauchy-Schwartz inequality and by
noticing that $\sum_x \pBar^\rho_{\rvx|\rvy}(x,y)=1$.  \hfill$\blacksquare$

These properties will again be used in the proofs in the following
lemmas.

\begin{lemma}\label{LEMMAAPP2_SI}
$\frac{\partial H(\pBar^\rho_{\rvx|\rvy})}{\partial \rho}\geq0$
\end{lemma}
\pf
\begin{eqnarray}
\frac{\partial \frac{A(y,\rho)}{B(\rho)}}{\partial \rho}&=&\frac{1}{B(\rho)^2}(\frac{\partial A(y,\rho)}{\partial \rho}B(\rho)-\frac{\partial B(\rho)}{\partial \rho}A(y, \rho))\nonumber\\
&=&\frac{1}{B(\rho)^2}( A(y,\rho)H(\pBar^\rho_{\rvx|\rvy=\svy})B(\rho)- H(\pBar^\rho_{\rvx|\rvy})B(\rho)A(y,\rho))\nonumber\\
&=&\frac{A(y,\rho)}{B(\rho)}( H(\pBar^\rho_{\rvx|\rvy=\svy})-
H(\pBar^\rho_{\rvx|\rvy}))\nonumber
\end{eqnarray}
Now,

\begin{eqnarray}
\frac{\partial H(\pBar^\rho_{\rvx|\rvy})}{\partial\rho}&=& \frac{\partial}{\partial\rho}\sum_y\frac{A(y,\rho)}{B(\rho)}\sum_x \frac{C(x,y,\rho)}{D(y,\rho)}[-\log(\frac{C(x,y,\rho)}{D(y,\rho)})] \nonumber\\
&=& \frac{\partial}{\partial\rho}\sum_y\frac{A(y,\rho)}{B(\rho)}H(\pBar^\rho_{\rvx|\rvy=\svy}) \nonumber\\
&=& \sum_y\frac{A(y,\rho)}{B(\rho)}\frac{\partial H(\pBar^\rho_{\rvx|\rvy=\svy})}{\partial\rho} + \sum_y  \frac{\partial \frac{A(y,\rho)}{B(\rho)}}{\partial\rho}H(\pBar^\rho_{\rvx|\rvy=\svy})\nonumber\\
&\geq& \sum_y  \frac{\partial \frac{A(y,\rho)}{B(\rho)}}{\partial\rho}H(\pBar^\rho_{\rvx|\rvy=\svy})\nonumber\\
&=& \sum_y \frac{A(y,\rho)}{B(\rho)}( H(\pBar^\rho_{\rvx|\rvy=\svy})-H(\pBar^\rho_{\rvx|\rvy})) H(\pBar^\rho_{\rvx|\rvy=\svy})\nonumber\\
&=& \sum_y \frac{A(y,\rho)}{B(\rho)} H(\pBar^\rho_{\rvx|\rvy=\svy})^2-H(\pBar^\rho_{\rvx|\rvy}) ^2\nonumber\\
&=& (\sum_y \frac{A(y,\rho)}{B(\rho)} H(\pBar^\rho_{\rvx|\rvy=\svy})^2)(\sum_y \frac{A(y,\rho)}{B(\rho)})-H(\pBar^\rho_{\rvx|\rvy}) ^2\nonumber\\
&\geq_{(a)}& (\sum_y \frac{A(y,\rho)}{B(\rho)} H(\pBar^\rho_{\rvx|\rvy=\svy}))^2-H(\pBar^\rho_{\rvx|\rvy}) ^2\nonumber\\
&=&0
\end{eqnarray}
where (a) is again true by the Cauchy-Schwartz inequality. \hfill$\blacksquare$

\begin{lemma}\label{LEMMAAPP3_SI}
$\frac{\partial D(\pBar^\rho_{\rvx\rvy}\|p_{\rvx\rvy})}{\partial
\rho}=\rho \frac{\partial H(\pBar^\rho_{\rvx|\rvy})}{\partial \rho}$
\end{lemma}
\pf As shown in Lemma~\ref{LEMMA_APP9} and Lemma~\ref{LEMMA_APP11}
respectively:
$$D(\pBar^\rho_{\rvx\rvy}\|p_{\rvx\rvy})=\rho H(\pBar^\rho_{\rvx|\rvy})- \log(\sum_{y
}(\sum_{x }p_{\rvx\rvy}(x,y)^{\frac{1}{1+\rho}})^{1+\rho})$$
$$  H(\pBar^\rho_{\rvx|\rvy})=\frac{\partial \log(\sum_{y
}(\sum_{x }p_{\rvx\rvy}(x,y)^{\frac{1}{1+\rho}})^{1+\rho})}{\partial
\rho} $$

We have:
\begin{eqnarray}
\frac{\partial D(\pBar^\rho_{\rvx\rvy}\|p_{\rvx\rvy})}{\partial
\rho}&=&   H(\pBar^\rho_{\rvx|\rvy}) +\rho\frac{\partial
H(\pBar^\rho_{\rvx|\rvy})}{\partial \rho}-\frac{\partial
\log(\sum_{y
}(\sum_{x }p_{\rvx\rvy}(x,y)^{\frac{1}{1+\rho}})^{1+\rho})}{\partial \rho }\nonumber\\
&=&  H(\pBar^\rho_{\rvx|\rvy}) +\rho\frac{\partial
H(\pBar^\rho_{\rvx|\rvy})}{\partial \rho}-H(\pBar^\rho_{\rvx|\rvy})  \nonumber\\
&=&\rho\frac{\partial H(\pBar^\rho_{\rvx|\rvy})}{\partial \rho}
\end{eqnarray} \hfill$\blacksquare$

\begin{lemma}\label{LEMMAAPP4_SI}
$sign\frac{\partial
[D(\pBar^\rho_{\rvx\rvy}\|p_{\rvx\rvy})-H(\pBar^\rho_{\rvx|\rvy})]}{\partial
\rho}=sign(\rho-1)$.
\end{lemma}
\pf Using the previous lemma, we get:
\begin{eqnarray}
&&\frac{\partial
D(\pBar^\rho_{\rvx\rvy}\|p_{\rvx\rvy})-H(\pBar^\rho_{\rvx|\rvy})}{\partial
\rho}=(\rho-1)\frac{\partial H(\pBar^\rho_{\rvx|\rvy})}{\partial
\rho}\nonumber
\end{eqnarray}
Then by Lemma~\ref{LEMMAAPP2_SI}, we get the
conclusion.\hfill$\blacksquare$\\

\begin{lemma}\label{LEMMA_APP8}
 $$\rho H(p^\rho_{\rvx\rvy})-(1+\rho)\log(\sum_{y}\sum_{x}p_{\rvx\rvy}(x,y)^{\frac{1}{1+\rho}})=D(p^\rho_{\rvx\rvy}\|p_{\rvx\rvy})$$
\end{lemma}
\pf  By noticing that
$\log(p_{\rvx\rvy}(x,y))=(1+\rho)[\log(p^\rho_{\rvx\rvy}(x,y))+\log(\sum_{s,t}
p_{\rvx\rvy}(s,t)^{\frac{1}{1+\rho}})]$. We have:
\begin{eqnarray}
 D(p^\rho_{\rvx\rvy}\|p_{\rvx\rvy})&=&-H(p^\rho_{\rvx\rvy})-\sum_{x,y}p^\rho_{\rvx\rvy}(x,y)\log(p_{\rvx\rvy}(x,y))\nonumber\\
&=&-H(p^\rho_{\rvx\rvy})-\sum_{x,y}p^\rho_{\rvx\rvy}(x,y)(1+\rho)[\log(p^\rho_{\rvx\rvy}(x,y))+\log(\sum_{s,t}
p_{\rvx\rvy}(s,t)^{\frac{1}{1+\rho}})]\nonumber\\
&=&-H(p^\rho_{\rvx\rvy})+(1+\rho)H(p^\rho_{\rvx\rvy})-(1+\rho)\sum_{x,y}p^\rho_{\rvx\rvy}(x,y)\log(\sum_{s,t}
p_{\rvx\rvy}(s,t)^{\frac{1}{1+\rho}})\nonumber\\
&=&\rho
H(p^\rho_{\rvx\rvy})-(1+\rho)\log(\sum_{s,t}p_{\rvx\rvy}(s,t)^{\frac{1}{1+\rho}})\end{eqnarray}

\hfill$\blacksquare$

\begin{lemma}\label{LEMMA_APP9}
 $$\rho H(\pBar^\rho_{\rvx|\rvy})- \log(\sum_{y
}(\sum_{x
}p_{\rvx\rvy}(x,y)^{\frac{1}{1+\rho}})^{1+\rho})=D(\pBar^\rho_{\rvx\rvy}\|p_{\rvx\rvy})$$
\end{lemma}
\pf
\begin{eqnarray}
D(\pBar^\rho_{\rvx\rvy}\|p_{\rvx\rvy})&=&\sum_y\sum_x \frac{A(y,\rho)}{B(\rho)}\frac{C(x,y,\rho)}{D(y,\rho)}\log(\frac{\frac{A(y,\rho)}{B(\rho)}\frac{C(x,y,\rho)}{D(y,\rho)}}{p_{\rvx\rvy}(x,y)})\nonumber\\
&=& \sum_y\sum_x \frac{A(y,\rho)}{B(\rho)}\frac{C(x,y,\rho)}{D(y,\rho)}[\log(\frac{A(y,\rho)}{B(\rho)})+\log(\frac{C(x,y,\rho)}{D(y,\rho)})-\log(p_{\rvx\rvy}(x,y))]\nonumber\\
&=& -\log(B(\rho)) - H(\pBar^\rho_{\rvx|\rvy}) + \sum_y\sum_x \frac{A(y,\rho)}{B(\rho)}\frac{C(x,y,\rho)}{D(y,\rho)}[\log(D(y,\rho)^{1+\rho})-\log(C(x,y,\rho)^{1+\rho})]\nonumber\\
&=& -\log(B(\rho)) - H(\pBar^\rho_{\rvx|\rvy}) +(1+\rho) H(\pBar^\rho_{\rvx|\rvy})\nonumber\\
&=& - \log(\sum_{y }(\sum_{x
}p_{\rvx\rvy}(x,y)^{\frac{1}{1+\rho}})^{1+\rho}) + \rho
H(\pBar^\rho_{\rvx|\rvy}) \nonumber
\end{eqnarray}
 \hfill$\blacksquare$

\begin{lemma}\label{LEMMA_APP10}
$$  H(p^\rho_{\rvx\rvy})=\frac{\partial (1+\rho)\log(\sum_{y}\sum_{x}p_{\rvx\rvy}(x,y)^{\frac{1}{1+\rho}})}{\partial \rho} $$
\end{lemma}
\pf

\begin{eqnarray}
 & & \frac{\partial
(1+\rho)\log(\sum_{y}\sum_{x}p_{\rvx\rvy}(x,y)^{\frac{1}{1+\rho}})}{\partial
\rho} \nonumber\\
 &=&\log(\sum_{t}\sum_{s}p_{\rvx\rvy}(s,t)^{\frac{1}{1+\rho}})-
\sum_{y}\sum_{x}\frac{p_{\rvx\rvy}(x,y)^{\frac{1}{1+\rho}}}{\sum_{t}\sum_{s}p_{\rvx\rvy}(s,t)^{\frac{1}{1+\rho}}}\log(p_{\rvx\rvy}(x,y)^{\frac{1}{1+\rho}})\nonumber\\
&=& -
\sum_{y}\sum_{x}\frac{p_{\rvx\rvy}(x,y)^{\frac{1}{1+\rho}}}{\sum_{t}\sum_{s}p_{\rvx\rvy}(s,t)^{\frac{1}{1+\rho}}}\log(\frac{p_{\rvx\rvy}(x,y)^{\frac{1}{1+\rho}}}{\sum_{t}\sum_{s}p_{\rvx\rvy}(s,t)^{\frac{1}{1+\rho}}})\nonumber\nonumber\\
&=&H(p^\rho_{\rvx\rvy})
\end{eqnarray}

\hfill$\blacksquare$
\begin{lemma}\label{LEMMA_APP11}
$$  H(\pBar^\rho_{\rvx|\rvy})=\frac{\partial \log(\sum_{y
}(\sum_{x }p_{\rvx\rvy}(x,y)^{\frac{1}{1+\rho}})^{1+\rho})}{\partial
\rho} $$\\
\end{lemma}
\pf Notice that $B(\rho)=\sum_{y }(\sum_{x
}p_{\rvx\rvy}(x,y)^{\frac{1}{1+\rho}})^{1+\rho}$, and $
\frac{\partial B(\rho)}{\partial \rho}
=B(\rho)H(\pBar^\rho_{\rvx|\rvy})$ as shown in
Lemma~\ref{LEMMAAPP1_SI}. It is clear that:

\begin{eqnarray}
\frac{\partial \log(\sum_{y }(\sum_{x
}p_{\rvx\rvy}(x,y)^{\frac{1}{1+\rho}})^{1+\rho})}{\partial
\rho}&=&\frac{\partial \log(B(\rho))}{\partial \rho}\nonumber\\
&=& \frac{1}{B(\rho)}\frac{\partial B(\rho)}{\partial
\rho}\nonumber\\
&=&H(\pBar^\rho_{\rvx|\rvy})
\end{eqnarray}
\hfill$\blacksquare$


\bibliographystyle{IEEEtran}
\bibliography{IEEEabrv,references}

\begin{thebibliography}{10}
\providecommand{\url}[1]{#1}
\csname url@rmstyle\endcsname
\providecommand{\newblock}{\relax}
\providecommand{\bibinfo}[2]{#2}
\providecommand\BIBentrySTDinterwordspacing{\spaceskip=0pt\relax}
\providecommand\BIBentryALTinterwordstretchfactor{4}
\providecommand\BIBentryALTinterwordspacing{\spaceskip=\fontdimen2\font plus
\BIBentryALTinterwordstretchfactor\fontdimen3\font minus
  \fontdimen4\font\relax}
\providecommand\BIBforeignlanguage[2]{{%
\expandafter\ifx\csname l@#1\endcsname\relax
\typeout{** WARNING: IEEEtran.bst: No hyphenation pattern has been}%
\typeout{** loaded for the language `#1'. Using the pattern for}%
\typeout{** the default language instead.}%
\else
\language=\csname l@#1\endcsname
\fi
#2}}

\bibitem{slepianWolf:73}
D.~Slepian and J.~K. Wolf, ``Noiseless coding of correlated information
  sources,'' \emph{IEEE Trans.\ Inform.\ Theory}, vol.~19, pp. 471--480, July
  1973.

\bibitem{koulgiEtAl:03}
P.~Koulgi, E.~Tuncel, S.~Regunathan, and K.~Rose, ``On zero-error coding of
  correlated sources,'' \emph{IEEE Trans.\ Inform.\ Theory}, vol.~49, pp.
  2856--2873, Nov. 2003.

\bibitem{sahaiSimsek:04}
A.~Sahai and T.~\c{S}im\c{s}ek, ``On the variable-delay reliability function of
  discrete memoryless channels with access to noisy feedback,'' in \emph{IEEE
  Information Theory Workshop, San Antonio, Texas}, 2004.

\bibitem{draperSahai:06}
S.~C. Draper and A.~Sahai, ``Noisy feedback improves communication
  reliability,'' in \emph{Proc.\ Int.\ Symp.\ Inform.\ Theory}, 2006.

\bibitem{draperAllerton:04}
S.~C. Draper, ``Universal incremental slepian-wolf coding,'' in \emph{Proc.\
  42nd Allerton Conf.\ on Communication, Control and Computing}, Oct. 2004.

\bibitem{Chang:06}
C.~Chang and A.~Sahai, ``The error exponent with delay for lossless source
  coding,'' \emph{IEEE Information Theory Workshop}, March 2006.

\bibitem{jelinek:68}
F.~Jelinek, ``Buffer overflow in variable length coding of fixed rate
  sources,'' \emph{IEEE Trans.\ Inform.\ Theory}, vol.~14, pp. 490--501, May
  1968.

\bibitem{csiszarKorner}
I.~Csisz{\'a}r and J.~K{\"o}rner, \emph{Information Theory, Coding Theorems for
  Discrete Memoryless Systems}.\hskip 1em plus 0.5em minus 0.4em\relax
  Akad{\'e}miai Kiad{\'o}, 1981.

\bibitem{gallagerTech:76}
R.~G. Gallager, ``Source coding with side information and universal coding,''
  Mass.~Instit.~Tech., Tech. Rep. LIDS-P-937, 1976.

\bibitem{Forney:74}
G.~Forney, ``Convolutional codes iii. sequential decoding,'' \emph{Information
  and Control}, vol.~25, no.~3, pp. 267--297, 1974.

\bibitem{lapidothNarayan:98}
A.~Lapidoth and P.~Narayan, ``Reliable communication under channel
  uncertainty,'' \emph{IEEE Trans.\ Inform.\ Theory}, vol.~44, pp. 2148--2177,
  Oct. 1998.

\bibitem{cover:75}
T.~M. Cover, ``A proof of the data compression theorem of {S}lepian and {W}olf
  for ergodic sources,'' \emph{IEEE Trans.\ Inform.\ Theory}, vol.~21, pp.
  226--228, Mar. 1975.

\bibitem{SahaiUnstable}
A.~Sahai and S.~Mitter, ``Source coding and channel requirements for unstable
  processes,'' \emph{Submitted to IEEE Trans.\ Inform.\ Theory}, 2006.

\bibitem{ChangISIT:06}
C.~Chang and A.~Sahai, ``Upper bound on error exponents with delay for lossless
  source coding with side-information,'' \emph{Proc.\ Int.\ Symp.\ Inform.\
  Theory}, July 2006.

\bibitem{SahaiBlockLength}
A.~Sahai, ``Why block length and delay are not the same thing,''
  \emph{Submitted to IEEE Trans.\ Inform.\ Theory}, 2006.

\bibitem{Weng:05}
L.~Weng, S.~Pradhan, and A.~Anastasopoulos, ``Error exponent regions for
  gaussian broadcast and mulitple access channels,'' \emph{submitted to
  Transactions on Information Theory}, 2005.

\bibitem{Boyd2004}
S.~Boyd and L.~Vandenberghe, \emph{Convex Optimization}.\hskip 1em plus 0.5em
  minus 0.4em\relax Cambridge University Press, 2004.

\end{thebibliography}

\end{document}